\documentclass[reprint,prd,twocolumn,superscriptaddress,nofootinbib,preprintnumbers]{revtex4-2}
\usepackage{amsmath}
\usepackage{amssymb}
\usepackage{bigstrut}
\usepackage{bm,color}
\usepackage{booktabs}
\usepackage{csquotes}
\usepackage{fontawesome}
\usepackage{graphicx}
\usepackage{hyperref}
\usepackage{ifthen}
\usepackage{multirow}
\usepackage{nicefrac}
\usepackage{float}
\usepackage{empheq}

\usepackage{savesym} % use savesym to work around command conflict between siunitx and tablenum
\savesymbol{tablenum}
\usepackage{siunitx}
\restoresymbol{SIX}{tablenum}
\usepackage{orcidlink}
\usepackage{physics}
\savesymbol{op}

\usepackage{textcomp}
\usepackage{threeparttable}
\usepackage{ulem}
\usepackage{verbatim}
\usepackage{xcolor}
\usepackage{xspace}
\usepackage{bbm}
\usepackage{lineno}
\usepackage{makecell}
\usepackage{threeparttable}

\hypersetup{
    colorlinks,
    linkcolor={red!50!black},
    citecolor={blue!50!black},
    urlcolor={blue!80!black}
}

\MakeOuterQuote{"}

\graphicspath{{figures/}{./}{../..}} %Setting the graphicspath

\newcommand{\healpix}{{\sc HEALpix}\xspace}

\newcommand{\vanish}[1]{}
\newcommand{\lcdm}{\ensuremath{\Lambda {\rm CDM}}\xspace}
\newcommand{\planck}{Planck\xspace}
\DeclareMathAlphabet{\mathbbgreek}{U}{bbold}{m}{n}

\newcommand{\op}[1]{\mathbb{#1}}

\newcommand{\Len}[1][]{\op L\ifthenelse{\equal{#1}{}}{}{(#1)}}
\newcommand{\Cflen}[1][]{\op{\widetilde{C}}_{f}\ifthenelse{\equal{#1}{}}{}{(#1)}}
\newcommand{\Cf}[1][]{\op C_{f}\ifthenelse{\equal{#1}{}}{}{(#1)}}
\newcommand{\Cphi}[1][]{\op C_{\phi}\ifthenelse{\equal{#1}{}}{}{(#1)}}

\newcommand{\opG}[1][]{\op G\ifthenelse{\equal{#1}{}}{}{(#1)}}
\newcommand{\D}[1][]{\op D\ifthenelse{\equal{#1}{}}{}{(#1)}}

\newcommand{\psipol}{\psi_{\rm pol}}
\newcommand{\epsQ}{\epsilon_{\rm Q}}
\newcommand{\epsU}{\epsilon_{\rm U}}

\DeclareUnicodeCharacter{2212}{-}
\newcommand{\smap}[1][]{s^{\rm \scriptscriptstyle MAP}\ifthenelse{\equal{#1}{}}{}{_#1}}
\date{\today}
\begin{document}
%%%%%%%%%%%%%%%%%%%%% Title, etc. %%%%%%%%%%%%%%%%%%%%%
\title{Cosmology From CMB Lensing and Delensed EE Power Spectra Using 2019-2020 SPT-3G Polarization Data}
% Authors
\author{F.~Ge\,\orcidlink{0000-0002-3833-8133}}
\email{fge@ucdavis.edu}
\affiliation{Department of Physics \& Astronomy, University of California, One Shields Avenue, Davis, CA 95616, USA}
\author{M.~Millea\,\orcidlink{0000-0001-7317-0551}}
\affiliation{Department of Physics \& Astronomy, University of California, One Shields Avenue, Davis, CA 95616, USA}
\author{E.~Camphuis\,\orcidlink{0000-0003-3483-8461}}
\affiliation{Sorbonne Universit\'e, CNRS, UMR 7095, Institut d'Astrophysique de Paris, 98 bis bd Arago, 75014 Paris, France}
\author{C.~Daley\,\orcidlink{0000-0002-3760-2086}}
\affiliation{Department of Astronomy, University of Illinois Urbana-Champaign, 1002 West Green Street, Urbana, IL, 61801, USA}
\affiliation{Universit\'e Paris-Saclay, Universit\'e Paris Cit\'e, CEA, CNRS, AIM, 91191, Gif-sur-Yvette, France}
\author{N.~Huang}
\affiliation{Department of Physics, University of California, Berkeley, CA, 94720, USA}
\author{Y.~Omori}
\affiliation{Department of Astronomy and Astrophysics, University of Chicago, 5640 South Ellis Avenue, Chicago, IL, 60637, USA}
\affiliation{Kavli Institute for Cosmological Physics, University of Chicago, 5640 South Ellis Avenue, Chicago, IL, 60637, USA}
\author{W.~Quan}
\affiliation{Department of Physics, University of Chicago, 5640 South Ellis Avenue, Chicago, IL, 60637, USA}
\affiliation{Kavli Institute for Cosmological Physics, University of Chicago, 5640 South Ellis Avenue, Chicago, IL, 60637, USA}
\author{E.~Anderes}
\affiliation{Department of Statistics, University of California, One Shields Avenue, Davis, CA 95616, USA}
\author{A.~J.~Anderson\,\orcidlink{0000-0002-4435-4623}}
\affiliation{Fermi National Accelerator Laboratory, MS209, P.O. Box 500, Batavia, IL, 60510, USA}
\affiliation{Kavli Institute for Cosmological Physics, University of Chicago, 5640 South Ellis Avenue, Chicago, IL, 60637, USA}
\affiliation{Department of Astronomy and Astrophysics, University of Chicago, 5640 South Ellis Avenue, Chicago, IL, 60637, USA}
\author{B.~Ansarinejad}
\affiliation{School of Physics, University of Melbourne, Parkville, VIC 3010, Australia}
\author{M.~Archipley\,\orcidlink{0000-0002-0517-9842}}
\affiliation{Kavli Institute for Cosmological Physics, University of Chicago, 5640 South Ellis Avenue, Chicago, IL, 60637, USA}
\affiliation{Department of Astronomy and Astrophysics, University of Chicago, 5640 South Ellis Avenue, Chicago, IL, 60637, USA}
\author{L.~Balkenhol\,\orcidlink{0000-0001-6899-1873}}
\affiliation{Sorbonne Universit\'e, CNRS, UMR 7095, Institut d'Astrophysique de Paris, 98 bis bd Arago, 75014 Paris, France}
\author{K.~Benabed}
\affiliation{Sorbonne Universit\'e, CNRS, UMR 7095, Institut d'Astrophysique de Paris, 98 bis bd Arago, 75014 Paris, France}
\author{A.~N.~Bender\,\orcidlink{0000-0001-5868-0748}}
\affiliation{High-Energy Physics Division, Argonne National Laboratory, 9700 South Cass Avenue., Lemont, IL, 60439, USA}
\affiliation{Kavli Institute for Cosmological Physics, University of Chicago, 5640 South Ellis Avenue, Chicago, IL, 60637, USA}
\affiliation{Department of Astronomy and Astrophysics, University of Chicago, 5640 South Ellis Avenue, Chicago, IL, 60637, USA}
\author{B.~A.~Benson\,\orcidlink{0000-0002-5108-6823}}
\affiliation{Fermi National Accelerator Laboratory, MS209, P.O. Box 500, Batavia, IL, 60510, USA}
\affiliation{Kavli Institute for Cosmological Physics, University of Chicago, 5640 South Ellis Avenue, Chicago, IL, 60637, USA}
\affiliation{Department of Astronomy and Astrophysics, University of Chicago, 5640 South Ellis Avenue, Chicago, IL, 60637, USA}
\author{F.~Bianchini\,\orcidlink{0000-0003-4847-3483}}
\affiliation{Kavli Institute for Particle Astrophysics and Cosmology, Stanford University, 452 Lomita Mall, Stanford, CA, 94305, USA}
\affiliation{Department of Physics, Stanford University, 382 Via Pueblo Mall, Stanford, CA, 94305, USA}
\affiliation{SLAC National Accelerator Laboratory, 2575 Sand Hill Road, Menlo Park, CA, 94025, USA}
\author{L.~E.~Bleem\,\orcidlink{0000-0001-7665-5079}}
\affiliation{High-Energy Physics Division, Argonne National Laboratory, 9700 South Cass Avenue., Lemont, IL, 60439, USA}
\affiliation{Kavli Institute for Cosmological Physics, University of Chicago, 5640 South Ellis Avenue, Chicago, IL, 60637, USA}
\author{F.~R.~Bouchet\,\orcidlink{0000-0002-8051-2924}}
\affiliation{Sorbonne Universit\'e, CNRS, UMR 7095, Institut d'Astrophysique de Paris, 98 bis bd Arago, 75014 Paris, France}
\author{L.~Bryant}
\affiliation{Enrico Fermi Institute, University of Chicago, 5640 South Ellis Avenue, Chicago, IL, 60637, USA}
\author{J.~E.~Carlstrom}
\affiliation{Kavli Institute for Cosmological Physics, University of Chicago, 5640 South Ellis Avenue, Chicago, IL, 60637, USA}
\affiliation{Enrico Fermi Institute, University of Chicago, 5640 South Ellis Avenue, Chicago, IL, 60637, USA}
\affiliation{Department of Physics, University of Chicago, 5640 South Ellis Avenue, Chicago, IL, 60637, USA}
\affiliation{High-Energy Physics Division, Argonne National Laboratory, 9700 South Cass Avenue., Lemont, IL, 60439, USA}
\affiliation{Department of Astronomy and Astrophysics, University of Chicago, 5640 South Ellis Avenue, Chicago, IL, 60637, USA}
\author{C.~L.~Chang}
\affiliation{High-Energy Physics Division, Argonne National Laboratory, 9700 South Cass Avenue., Lemont, IL, 60439, USA}
\affiliation{Kavli Institute for Cosmological Physics, University of Chicago, 5640 South Ellis Avenue, Chicago, IL, 60637, USA}
\affiliation{Department of Astronomy and Astrophysics, University of Chicago, 5640 South Ellis Avenue, Chicago, IL, 60637, USA}
\author{P.~Chaubal}
\affiliation{School of Physics, University of Melbourne, Parkville, VIC 3010, Australia}
\author{G.~Chen}
\affiliation{University of Chicago, 5640 South Ellis Avenue, Chicago, IL, 60637, USA}
\author{P.~M.~Chichura\,\orcidlink{0000-0002-5397-9035}}
\affiliation{Department of Physics, University of Chicago, 5640 South Ellis Avenue, Chicago, IL, 60637, USA}
\affiliation{Kavli Institute for Cosmological Physics, University of Chicago, 5640 South Ellis Avenue, Chicago, IL, 60637, USA}
\author{A.~Chokshi}
\affiliation{University of Chicago, 5640 South Ellis Avenue, Chicago, IL, 60637, USA}
\author{T.-L.~Chou\,\orcidlink{0000-0002-3091-8790}}
\affiliation{Department of Astronomy and Astrophysics, University of Chicago, 5640 South Ellis Avenue, Chicago, IL, 60637, USA}
\affiliation{Kavli Institute for Cosmological Physics, University of Chicago, 5640 South Ellis Avenue, Chicago, IL, 60637, USA}
\author{A.~Coerver}
\affiliation{Department of Physics, University of California, Berkeley, CA, 94720, USA}
\author{T.~M.~Crawford\,\orcidlink{0000-0001-9000-5013}}
\affiliation{Kavli Institute for Cosmological Physics, University of Chicago, 5640 South Ellis Avenue, Chicago, IL, 60637, USA}
\affiliation{Department of Astronomy and Astrophysics, University of Chicago, 5640 South Ellis Avenue, Chicago, IL, 60637, USA}
\author{T.~de~Haan}
\affiliation{High Energy Accelerator Research Organization (KEK), Tsukuba, Ibaraki 305-0801, Japan}
\author{K.~R.~Dibert}
\affiliation{Department of Astronomy and Astrophysics, University of Chicago, 5640 South Ellis Avenue, Chicago, IL, 60637, USA}
\affiliation{Kavli Institute for Cosmological Physics, University of Chicago, 5640 South Ellis Avenue, Chicago, IL, 60637, USA}
\author{M.~A.~Dobbs}
\affiliation{Department of Physics and McGill Space Institute, McGill University, 3600 Rue University, Montreal, Quebec H3A 2T8, Canada}
\affiliation{Canadian Institute for Advanced Research, CIFAR Program in Gravity and the Extreme Universe, Toronto, ON, M5G 1Z8, Canada}
\author{M.~Doohan}
\affiliation{School of Physics, University of Melbourne, Parkville, VIC 3010, Australia}
\author{A.~Doussot}
\affiliation{Sorbonne Universit\'e, CNRS, UMR 7095, Institut d'Astrophysique de Paris, 98 bis bd Arago, 75014 Paris, France}
\author{D.~Dutcher\,\orcidlink{0000-0002-9962-2058}}
\affiliation{Joseph Henry Laboratories of Physics, Jadwin Hall, Princeton University, Princeton, NJ 08544, USA}
\author{W.~Everett}
\affiliation{Department of Astrophysical and Planetary Sciences, University of Colorado, Boulder, CO, 80309, USA}
\author{C.~Feng}
\affiliation{Department of Physics, University of Illinois Urbana-Champaign, 1110 West Green Street, Urbana, IL, 61801, USA}
\author{K.~R.~Ferguson\,\orcidlink{0000-0002-4928-8813}}
\affiliation{Department of Physics and Astronomy, University of California, Los Angeles, CA, 90095, USA}
\affiliation{Department of Physics and Astronomy, Michigan State University, East Lansing, MI 48824, USA}
\author{K.~Fichman}
\affiliation{Department of Physics, University of Chicago, 5640 South Ellis Avenue, Chicago, IL, 60637, USA}
\affiliation{Kavli Institute for Cosmological Physics, University of Chicago, 5640 South Ellis Avenue, Chicago, IL, 60637, USA}
\author{A.~Foster\,\orcidlink{0000-0002-7145-1824}}
\affiliation{Joseph Henry Laboratories of Physics, Jadwin Hall, Princeton University, Princeton, NJ 08544, USA}
\author{S.~Galli}
\affiliation{Sorbonne Universit\'e, CNRS, UMR 7095, Institut d'Astrophysique de Paris, 98 bis bd Arago, 75014 Paris, France}
\author{A.~E.~Gambrel}
\affiliation{Kavli Institute for Cosmological Physics, University of Chicago, 5640 South Ellis Avenue, Chicago, IL, 60637, USA}
\author{R.~W.~Gardner}
\affiliation{Enrico Fermi Institute, University of Chicago, 5640 South Ellis Avenue, Chicago, IL, 60637, USA}
\author{N.~Goeckner-Wald}
\affiliation{Department of Physics, Stanford University, 382 Via Pueblo Mall, Stanford, CA, 94305, USA}
\affiliation{Kavli Institute for Particle Astrophysics and Cosmology, Stanford University, 452 Lomita Mall, Stanford, CA, 94305, USA}
\author{R.~Gualtieri\,\orcidlink{0000-0003-4245-2315}}
\affiliation{Department of Physics and Astronomy, Northwestern University, 633 Clark St, Evanston, IL, 60208, USA}
\author{F.~Guidi\,\orcidlink{0000-0001-7593-3962}}
\affiliation{Sorbonne Universit\'e, CNRS, UMR 7095, Institut d'Astrophysique de Paris, 98 bis bd Arago, 75014 Paris, France}
\author{S.~Guns}
\affiliation{Department of Physics, University of California, Berkeley, CA, 94720, USA}
\author{N.~W.~Halverson}
\affiliation{CASA, Department of Astrophysical and Planetary Sciences, University of Colorado, Boulder, CO, 80309, USA }
\affiliation{Department of Physics, University of Colorado, Boulder, CO, 80309, USA}
\author{E.~Hivon\,\orcidlink{0000-0003-1880-2733}}
\affiliation{Sorbonne Universit\'e, CNRS, UMR 7095, Institut d'Astrophysique de Paris, 98 bis bd Arago, 75014 Paris, France}
\author{G.~P.~Holder\,\orcidlink{0000-0002-0463-6394}}
\affiliation{Department of Physics, University of Illinois Urbana-Champaign, 1110 West Green Street, Urbana, IL, 61801, USA}
\author{W.~L.~Holzapfel}
\affiliation{Department of Physics, University of California, Berkeley, CA, 94720, USA}
\author{J.~C.~Hood}
\affiliation{Kavli Institute for Cosmological Physics, University of Chicago, 5640 South Ellis Avenue, Chicago, IL, 60637, USA}
\author{D.~Howe}
\affiliation{University of Chicago, 5640 South Ellis Avenue, Chicago, IL, 60637, USA}
\author{A.~Hryciuk}
\affiliation{Department of Physics, University of Chicago, 5640 South Ellis Avenue, Chicago, IL, 60637, USA}
\affiliation{Kavli Institute for Cosmological Physics, University of Chicago, 5640 South Ellis Avenue, Chicago, IL, 60637, USA}
\author{F.~K\'eruzor\'e}
\affiliation{High-Energy Physics Division, Argonne National Laboratory, 9700 South Cass Avenue., Lemont, IL, 60439, USA}
\author{A.~R.~Khalife\,\orcidlink{0000-0002-8388-4950}}
\affiliation{Sorbonne Universit\'e, CNRS, UMR 7095, Institut d'Astrophysique de Paris, 98 bis bd Arago, 75014 Paris, France}
\author{L.~Knox}
\affiliation{Department of Physics \& Astronomy, University of California, One Shields Avenue, Davis, CA 95616, USA}
\author{M.~Korman}
\affiliation{Department of Physics, Case Western Reserve University, Cleveland, OH, 44106, USA}
\author{K.~Kornoelje}
\affiliation{Department of Astronomy and Astrophysics, University of Chicago, 5640 South Ellis Avenue, Chicago, IL, 60637, USA}
\affiliation{Kavli Institute for Cosmological Physics, University of Chicago, 5640 South Ellis Avenue, Chicago, IL, 60637, USA}
\author{C.-L.~Kuo}
\affiliation{Kavli Institute for Particle Astrophysics and Cosmology, Stanford University, 452 Lomita Mall, Stanford, CA, 94305, USA}
\affiliation{Department of Physics, Stanford University, 382 Via Pueblo Mall, Stanford, CA, 94305, USA}
\affiliation{SLAC National Accelerator Laboratory, 2575 Sand Hill Road, Menlo Park, CA, 94025, USA}
\author{A.~T.~Lee}
\affiliation{Department of Physics, University of California, Berkeley, CA, 94720, USA}
\affiliation{Physics Division, Lawrence Berkeley National Laboratory, Berkeley, CA, 94720, USA}
\author{K.~Levy}
\affiliation{School of Physics, University of Melbourne, Parkville, VIC 3010, Australia}
\author{A.~E.~Lowitz}
\affiliation{Kavli Institute for Cosmological Physics, University of Chicago, 5640 South Ellis Avenue, Chicago, IL, 60637, USA}
\author{C.~Lu}
\affiliation{Department of Physics, University of Illinois Urbana-Champaign, 1110 West Green Street, Urbana, IL, 61801, USA}
\author{A.~Maniyar}
\affiliation{Kavli Institute for Particle Astrophysics and Cosmology, Stanford University, 452 Lomita Mall, Stanford, CA, 94305, USA}
\affiliation{Department of Physics, Stanford University, 382 Via Pueblo Mall, Stanford, CA, 94305, USA}
\affiliation{SLAC National Accelerator Laboratory, 2575 Sand Hill Road, Menlo Park, CA, 94025, USA}
\author{E.~S.~Martsen}
\affiliation{Department of Astronomy and Astrophysics, University of Chicago, 5640 South Ellis Avenue, Chicago, IL, 60637, USA}
\affiliation{Kavli Institute for Cosmological Physics, University of Chicago, 5640 South Ellis Avenue, Chicago, IL, 60637, USA}
\author{F.~Menanteau}
\affiliation{Department of Astronomy, University of Illinois Urbana-Champaign, 1002 West Green Street, Urbana, IL, 61801, USA}
\affiliation{Center for AstroPhysical Surveys, National Center for Supercomputing Applications, Urbana, IL, 61801, USA}
\author{J.~Montgomery}
\affiliation{Department of Physics and McGill Space Institute, McGill University, 3600 Rue University, Montreal, Quebec H3A 2T8, Canada}
\author{Y.~Nakato}
\affiliation{Department of Physics, Stanford University, 382 Via Pueblo Mall, Stanford, CA, 94305, USA}
\author{T.~Natoli}
\affiliation{Kavli Institute for Cosmological Physics, University of Chicago, 5640 South Ellis Avenue, Chicago, IL, 60637, USA}
\author{G.~I.~Noble\,\orcidlink{0000-0002-5254-243X}}
\affiliation{Dunlap Institute for Astronomy \& Astrophysics, University of Toronto, 50 St. George Street, Toronto, ON, M5S 3H4, Canada}
\affiliation{David A. Dunlap Department of Astronomy \& Astrophysics, University of Toronto, 50 St. George Street, Toronto, ON, M5S 3H4, Canada}
\author{Z.~Pan\,\orcidlink{0000-0002-6164-9861}}
\affiliation{High-Energy Physics Division, Argonne National Laboratory, 9700 South Cass Avenue., Lemont, IL, 60439, USA}
\affiliation{Kavli Institute for Cosmological Physics, University of Chicago, 5640 South Ellis Avenue, Chicago, IL, 60637, USA}
\affiliation{Department of Physics, University of Chicago, 5640 South Ellis Avenue, Chicago, IL, 60637, USA}
\author{P.~Paschos}
\affiliation{Enrico Fermi Institute, University of Chicago, 5640 South Ellis Avenue, Chicago, IL, 60637, USA}
\author{K.~A.~Phadke\,\orcidlink{0000-0001-7946-557X}}
\affiliation{Department of Astronomy, University of Illinois Urbana-Champaign, 1002 West Green Street, Urbana, IL, 61801, USA}
\affiliation{Center for AstroPhysical Surveys, National Center for Supercomputing Applications, Urbana, IL, 61801, USA}
\author{A.~W.~Pollak}
\affiliation{University of Chicago, 5640 South Ellis Avenue, Chicago, IL, 60637, USA}
\author{K.~Prabhu}
\affiliation{Department of Physics \& Astronomy, University of California, One Shields Avenue, Davis, CA 95616, USA}
%\author{S.~Raghunathan\,\orcidlink{0000-0003-1405-378X}}
%\affiliation{Center for AstroPhysical Surveys, National Center for Supercomputing Applications, Urbana, IL, 61801, USA}
\author{M.~Rahimi}
\affiliation{School of Physics, University of Melbourne, Parkville, VIC 3010, Australia}
\author{A.~Rahlin\,\orcidlink{0000-0003-3953-1776}}
\affiliation{Department of Astronomy and Astrophysics, University of Chicago, 5640 South Ellis Avenue, Chicago, IL, 60637, USA}
\affiliation{Kavli Institute for Cosmological Physics, University of Chicago, 5640 South Ellis Avenue, Chicago, IL, 60637, USA}
\author{C.~L.~Reichardt\,\orcidlink{0000-0003-2226-9169}}
\affiliation{School of Physics, University of Melbourne, Parkville, VIC 3010, Australia}
\author{D.~Riebel}
\affiliation{University of Chicago, 5640 South Ellis Avenue, Chicago, IL, 60637, USA}
\author{M.~Rouble}
\affiliation{Department of Physics and McGill Space Institute, McGill University, 3600 Rue University, Montreal, Quebec H3A 2T8, Canada}
\author{J.~E.~Ruhl}
\affiliation{Department of Physics, Case Western Reserve University, Cleveland, OH, 44106, USA}
\author{E.~Schiappucci}
\affiliation{School of Physics, University of Melbourne, Parkville, VIC 3010, Australia}
\author{J.~A.~Sobrin\,\orcidlink{0000-0001-6155-5315}}
\affiliation{Fermi National Accelerator Laboratory, MS209, P.O. Box 500, Batavia, IL, 60510, USA}
\affiliation{Kavli Institute for Cosmological Physics, University of Chicago, 5640 South Ellis Avenue, Chicago, IL, 60637, USA}
\author{A.~A.~Stark}
\affiliation{Harvard-Smithsonian Center for Astrophysics, 60 Garden Street, Cambridge, MA, 02138, USA}
\author{J.~Stephen}
\affiliation{Enrico Fermi Institute, University of Chicago, 5640 South Ellis Avenue, Chicago, IL, 60637, USA}
\author{C.~Tandoi}
\affiliation{Department of Astronomy, University of Illinois Urbana-Champaign, 1002 West Green Street, Urbana, IL, 61801, USA}
\author{B.~Thorne}
\affiliation{Department of Physics \& Astronomy, University of California, One Shields Avenue, Davis, CA 95616, USA}
\author{C.~Trendafilova}
\affiliation{Center for AstroPhysical Surveys, National Center for Supercomputing Applications, Urbana, IL, 61801, USA}
\author{C.~Umilta\,\orcidlink{0000-0002-6805-6188}}
\affiliation{Department of Physics, University of Illinois Urbana-Champaign, 1110 West Green Street, Urbana, IL, 61801, USA}
\author{J.~D.~Vieira}
\affiliation{Department of Astronomy, University of Illinois Urbana-Champaign, 1002 West Green Street, Urbana, IL, 61801, USA}
\affiliation{Department of Physics, University of Illinois Urbana-Champaign, 1110 West Green Street, Urbana, IL, 61801, USA}
\affiliation{Center for AstroPhysical Surveys, National Center for Supercomputing Applications, Urbana, IL, 61801, USA}
\author{A.~Vitrier}
\affiliation{Sorbonne Universit\'e, CNRS, UMR 7095, Institut d'Astrophysique de Paris, 98 bis bd Arago, 75014 Paris, France}
\author{Y.~Wan}
\affiliation{Department of Astronomy, University of Illinois Urbana-Champaign, 1002 West Green Street, Urbana, IL, 61801, USA}
\affiliation{Center for AstroPhysical Surveys, National Center for Supercomputing Applications, Urbana, IL, 61801, USA}
\author{N.~Whitehorn\,\orcidlink{0000-0002-3157-0407}}
\affiliation{Department of Physics and Astronomy, Michigan State University, East Lansing, MI 48824, USA}
\author{W.~L.~K.~Wu\,\orcidlink{0000-0001-5411-6920}}
\affiliation{Kavli Institute for Particle Astrophysics and Cosmology, Stanford University, 452 Lomita Mall, Stanford, CA, 94305, USA}
\affiliation{SLAC National Accelerator Laboratory, 2575 Sand Hill Road, Menlo Park, CA, 94025, USA}
\author{M.~R.~Young}
\affiliation{Fermi National Accelerator Laboratory, MS209, P.O. Box 500, Batavia, IL, 60510, USA}
\affiliation{Kavli Institute for Cosmological Physics, University of Chicago, 5640 South Ellis Avenue, Chicago, IL, 60637, USA}
\author{J.~A.~Zebrowski}
\affiliation{Kavli Institute for Cosmological Physics, University of Chicago, 5640 South Ellis Avenue, Chicago, IL, 60637, USA}
\affiliation{Department of Astronomy and Astrophysics, University of Chicago, 5640 South Ellis Avenue, Chicago, IL, 60637, USA}
\affiliation{Fermi National Accelerator Laboratory, MS209, P.O. Box 500, Batavia, IL, 60510, USA}
\collaboration{SPT-3G Collaboration}
\noaffiliation

\begin{abstract}
    \vspace{1cm}

    From CMB polarization data alone we reconstruct the CMB lensing power spectrum, comparable in overall constraining power to previous temperature-based  reconstructions, and an {\it unlensed} $E$-mode power spectrum, with clear detections of the third through tenth acoustic peaks. The observations, taken in 2019 and 2020 with the South Pole Telescope (SPT) and the SPT-3G camera, cover 1500\,deg$^2$ at 95, 150, and 220 GHz with arcminute resolution and roughly 4.9\,$\mu$K-arcmin coadded noise in polarization. The power spectrum estimates, together with systematic parameter estimates and a joint covariance matrix, follow from a Bayesian analysis using the Marginal Unbiased Score Expansion (MUSE) method. The $E$-mode spectrum at $\ell\,{>}\,2000$ and lensing spectrum at $L\,{>}\,350$ are the most precise to date. Assuming the \lcdm model, and using only these SPT data and priors on $\tau$ and absolute calibration from Planck, we find $H_0\,{=}\,66.81\,{\pm}\,0.81$\,km/s/Mpc, comparable in precision to the Planck determination and in $5.4\,\sigma$ tension with the most precise $H_0$ inference derived via the distance ladder. We also find $S_8 \equiv \sigma_8 (\Omega_{\rm m}/0.3)^{0.5}\,{=}\,0.850\,{\pm}\,0.017$, providing further independent evidence of a slight tension with low-redshift structure probes. The \lcdm model provides a good simultaneous fit to the combined \planck, ACT, and SPT data, and thus passes a powerful test. Combining these CMB datasets with BAO observations, we explore extensions to the \lcdm model.
    We find that the effective number of neutrino species, spatial curvature, and primordial helium fraction are consistent with standard model values, and that the 95\% confidence upper limit on the neutrino mass sum is 0.075\,eV, close to the minimum sum expected from observations of solar and atmospheric neutrino oscillations. The SPT data are consistent with the somewhat weak (${<}\,3\,\sigma$) preference for excess lensing power seen in Planck and ACT data relative to predictions of the \lcdm model given the combined \planck, ACT, and BAO data sets.
    We also detect at greater than $3\,\sigma$ the influence of non-linear evolution in the CMB lensing power spectrum and discuss it in the context of the $S_8$ tension.
    Forthcoming SPT-3G analyses will feature deeper and wider observations in temperature and polarization, providing even tighter constraints and more powerful tests of the \lcdm model.
    \vspace{1cm}
\end{abstract}


\maketitle

%\tableofcontents

\section{Introduction}

Since the successful completion of the \planck\ satellite's sky surveys \citep{Planck:2018nkj}, the frontier in the study of the cosmic microwave background (CMB) anisotropies has been its polarization. Here we present the strongest cosmological constraints to date derived from polarization data alone, constraints that are, by some measures, comparable to those from temperature data alone. Low-noise polarization measurements allow for new tests of the standard cosmological model, as the cosmological parameter inference relies on different signals, and they allow for additional robustness to many sources of potential systematic error.  

The data we use come from the SPT-3G camera \citep{sobrin.etal22} installed on the South Pole Telescope (SPT) in early 2017. By October of 2024, SPT-3G had been used to survey 10,000\,deg$^2$, with the bulk of the Austral winter observing time spent on a 1500\,deg$^2$ patch. The resulting data allow for powerful tests of the standard cosmological model, \lcdm, via the primary and CMB lensing signals \citep{SPT-3G:2024qkd}, improved constraints on primordial gravitational waves via de-lensing of the degree-scale observations of the BICEP/Keck Collaboration \citep{BK21}, and many other science applications \citep{Guns21,Chaubal22,Ferguson22,Schiappucci23,Ansarinejad24,Tandoi24,raghunathan2024}.

This paper is the first of a series of papers on CMB power spectra, the CMB lensing power spectrum, and cosmological parameter estimates inferred from the 2019 and 2020 Austral winter observations of the 1500\,deg$^2$ patch. The temperature and polarization maps used in these analyses are the result of four times as much observing time and twice as many detectors as were used for similar analyses that used a half season of data taken in 2018 with a partial SPT-3G focal plane \citep{SPT-Dutcher:2021eoc,SPT-Balkenhol:2021wgf,SPT-Balkenhol:2022hvq,SPT-Pan:2023jql}. 

This paper is distinct from the others in the coming series by its focus on polarization data only, and by its use of a novel algorithm for optimally and simultaneously estimating the CMB lensing power spectrum, the {\it unlensed}\footnote{We note a subtle distinction between a de-lensed spectrum, which contains a residual lensing component which must be modeled, and our inference of an unlensed spectrum, which contains no such residual, independent of cosmological model. Although of minor impact, this explains our usage of "unlensed" throughout the text.} CMB $E$-mode polarization power spectrum ($EE$), and systematics parameters. This algorithm, the Marginal Unbiased Score Expansion (MUSE), was developed by \citet{Millea:2021had} and \citet{millea2022a}, with its precursors discussed in \cite{Seljak:2017rmr, Horowitz:2018tbe}. This is the first time the MUSE method has been applied to real data. We present the analysis in some detail and emphasize its advantages, including straightforward incorporation of systematic effects into a statistical model, a lensing potential map with lower noise than if it were produced by a quadratic estimator \citep[e.g.][]{Hu:2001kj}, and an unlensed $EE$ power spectrum with reduced sample variance relative to any inference of the lensed $EE$ power spectrum \citep{Hotinli:2021umk}. 

De-lensed power spectra from quadratic estimators have been estimated previously from data by \citet{planck2018lensing} and \citet{ACT:2020goa}. The methodology for optimal lensing reconstruction has been principally developed in \cite{hirata2003a,carron2017,Legrand:2021qdu,Legrand:2023jne,millea2019,millea2020,Millea:2021had}; and an application to SPTpol data was described in \cite{millea2021}.
With the SPT-3G data we analyze here, we are crossing the threshold of polarization noise level below which optimal methods lead to significant improvements in lensing reconstruction.

We focus solely on polarization anisotropies in order to avoid modeling extragalactic foregrounds and their non-Gaussian properties. Such modeling would likely be necessary for a similar analysis which included temperature anisotropies, where the relative impact of foregrounds is larger. Due to the low noise and high angular resolution of the SPT-3G measurements, this initial conservative approach still leaves us with an inferred CMB lensing power spectrum with high signal-to-noise ratio. In the part of sky we observe, over the angular scales used in the analysis, galactic foregrounds are also negligibly small \citep{SPT-Dutcher:2021eoc, SPT-3G:2022hvq}. 

We compare with other CMB measurements in both temperature and polarization, including those from the \planck Collaboration \citep{Planck2018Overview20} and the Atacama Cosmology Telescope (ACT) Collaboration \citep{ACT:2023kun, ACT:2023dou, aiola2020, ACT:2020frw}. We check for consistency of our results in the context of the \lcdm model. We also determine cosmological parameters assuming the \lcdm model and extensions from SPT data alone and in combination with these other datasets. Although the SPT observations we use here cover much less sky relative to Planck and ACT, the very low noise (4.9\,$\mu$K-arcmin in coadded polarization maps) and arc-minute angular resolution enable measurements of the $EE$ power spectrum and CMB lensing potential power spectrum ($\phi\phi$) that are the most precise measurements ever made at $\ell\,{>}\,2000$ for $EE$ and $L\,{>}\,350$ for $\phi\phi$.

We discuss how the signals in our data allow for the determination of cosmological parameters assuming the \lcdm model. The origins of these determinations have interesting differences compared to those from \planck CMB data. For example, the Hubble constant ($H_0$) inference depends heavily on the lensing information; without it, the error increases by more than 10 times. Consistency between the \planck and SPT inferences of $H_0$ places further constraints on attempts to solve the $H_0$ tension \citep{DiValentino:2021izs} with changes to cosmological models \citep{Schoneberg:2021qvd, Khalife:2023qbu}.

We pay particular attention to the implications of our data for the "negative neutrino mass" problem and its relation to excess lensing power pointed out by \citet{Craig:2024tky}, and to the suggestion from \citet{Amon:2022azi} that the $\sigma_8$ tension is associated with non-linear evolution, either due to mismodeling the \lcdm predictions or by neglect of some new physics that modifies evolution on non-linear scales.  
 
The rest of this paper is constructed as follows.  In Sect.~\ref{sec:data_processing} we describe the process for making CMB maps from real and mock observations. We give an overview of MUSE and our data modeling in Sect.~\ref{sec:methodology}. We present our pipeline and data validation processes, the blinding procedure, and results in Sect.~\ref{sec:validation}. We show the resulting bandpowers and cosmological parameter constraints in Sect.~\ref{sec:results}. 
We conclude in Sect.~\ref{sec:conclusions}. Throughout this work we use a series of abbreviations for external datasets tabulated in Tab.~\ref{table:dataset}.

\vspace{0.3cm}
\section{Data and Mapmaking}
\label{sec:data_processing}

We use data taken with the SPT-3G camera during the 2019 and 2020 winter observing seasons. The 1500\,deg$^2$ SPT-3G winter footprint extends from $-\ang{42}$ to $-\ang{70}$ in declination and $-\ang{50}$ to $\ang{50}$ in right ascension. The footprint is divided into four subfields with declination centers of $-\ang{44.75}, -\ang{52.25}, -\ang{59.75}$, and $-\ang{67.25}$, and that allows for fitting different systematics parameters between patches that improve detector response consistency and linearity across the entire field. Observations are performed in two-hour periods and map a single sub-field at a time. During one observation, the telescope performs 72 constant-elevation (and thus constant-declination) scans with a scan speed of 1\,deg/s in right ascension and an elevation step of 12.5\,arcmin between scans. The subfields are observed 1034, 902, 772, and 578 times respectively across the two seasons, which produces approximately uniform map noise levels across subfields.

Each of the $\,{\sim}\,$16,000 detectors in the SPT-3G camera samples the sky as the telescope scans, producing one-dimensional ``timestreams'' that record sky brightness as a function of right ascension. The timestreams used in this analysis have on-sky sample rates between 0.3\,arcmin/sample and 0.6\,arcmin/sample depending on declination, corresponding to an effective Nyquist frequency well beyond our maximum multipole of interest. 

The timestreams are low-pass filtered to reduce aliasing when later binning into maps, and high-pass filtered to remove large-scale instrumental and atmospheric noise. The high-pass filter is performed by deprojecting a set of Legendre polynomials up to 30th-order and a set of sines and cosines up to an equivalent angular scale of $\ell\,{\leq}\,300$. 

The filtering described above can produce undesired "ringing" features in the maps around the locations of point sources (a term we will loosely use here to describe both emissive galaxies and the Sunyaev-Zeldovich effect from galaxy clusters, the latter which can be spatially extended).
To reduce these features, when fitting the deprojection coefficients, we ignore regions around point sources, specifically around all emissive sources with 150\,GHz flux ${>}\,6\,{\rm mJy}$ and all galaxy clusters with a signal-to-noise ${>}\,10$, for a total of 2655 objects. This procedure reduces ringing but does not mask the point sources, which instead happens at the map level later in the pipeline. 

After filtering, we bin the timestream samples using inverse-variance weights into $0.5625^\prime$ pixels given a Lambert projection centered on a right ascension of $\ang{0}$ and a declination of $-\ang{59.55}$. Weights are determined from the detector noise in the range $320\,{<}\,{\ell}\,{<}\,4000$. The pixel size was chosen so that the 1500\,deg$^2$ patch fits into maps with a number of pixels on each side that are powers of two, which maximizes the efficiency of FFTs. For this resolution, this yields $4096\,{\times}\,8192$ maps.

After maps have been made for each observation in the 2019 and 2020 winter observing seasons, we sum these maps into a full-depth ``coadd'' with a small number of data cuts.
In addition to full-depth coadds, we also construct 500 "sign-flip" noise realizations by summing the observations with a random half of them multiplied by $-1$, canceling the signal but leaving the statistical properties of the noise unchanged. We use these to quantify the noise covariance, and to directly add to simulated signal maps to produce highly realistic simulations of the full dataset. We estimate that the noise realizations are correlated with each other at the percent level due to the finite number of observations in the dataset, but this is below the level which can significantly impact this analysis. 

Next, we apply an isotropic anti-aliasing filter to the $0.5625^\prime$ maps and rebin them to the final $2.25^\prime$ resolution, $1024\,{\times}\,2048$ pixel analysis maps. We note that the reason for not binning the timestreams directly into the final $2.25^\prime$ resolution is exactly to be able to apply this anti-aliasing filter, as otherwise nothing would prevent aliasing in the scan-perpendicular direction. The final Nyquist frequency for these maps corresponds to $\ell\,{=}\,4800$, high enough to accommodate the $\ell_{\rm max}\,{=}\,4000$ of our analysis while at the same time minimizing the pixel count and hence the computational cost of MUSE. 

Finally, we apply the pipeline masks described in Sect.~\ref{sec:model_summary}. These processed and masked $1024\,{\times}\,2048$-pixel maps of $Q$ and $U$ stokes polarization at 95, 150, and 220\,GHz are the basic data, $d$, which is input to our analysis pipeline.

\vspace{0.3cm}
\section{Methodology: Bayesian inference with MUSE}
\label{sec:methodology}

\subsection{Bandpower and systematics inference}

We perform a map-level simultaneous Bayesian inference of
the gravitational lensing potential bandpowers, the unlensed CMB $EE$ bandpowers and systematic parameters. This procedure is described in \citet{Millea:2021had} and \citet{millea2022a}, and summarized below with a few additions specific to this work. The central piece of the analysis is the joint posterior probability function, 
\begin{align}
    \mathcal{P}(f,\phi,\theta\,|\,d)
\end{align}
where $f$ refers to the unlensed CMB maps (here just polarization), $\phi$ to the gravitational lensing potential, $\theta$ to the  bandpower and systematics parameter vector, and $d$ to the maps described in the previous section. The exact details of our modeling and construction of this function are given in the next section. Here, we first discuss how we use it to perform inference. 

The goal of our analysis is to marginalize over the ``latent'' $f$ and $\phi$ variables to produce the marginal posterior on just the parameters of interest, $\theta$:
\begin{align}
    \mathcal{P}(\theta\,|\,d) = \int {\rm d}f \, {\rm d}\phi \, \mathcal{P}(f,\phi,\theta\,|\,d)
    \label{eq:postmarg}
\end{align}
The constraints on $\theta$ described by this function are formally optimal, satisfying our goal of an optimal analysis. This optimality is achieved because, implicitly, the marginalization sources information from moments of the data at all orders, unlike e.g. quadratic estimators, which only use second-order moments. In particular, this means our results use information from all combinations of the data such as $EE$, $EB$, $BB$, $EEE$, $EEB$, etc...

Various techniques exist for performing the difficult high-dimensional integration in Eqn.~\eqref{eq:postmarg}. A popular but computationally expensive choice is to use Hamiltonian Monte Carlo to yield the exact answer, up to sampling errors. MUSE instead performs an approximate but much faster marginalization, which yields a multi-variate Gaussian approximation to the marginal posterior,
\begin{align}
    -2\log \mathcal{P}(\theta\,|\,d) \approx (\theta - \hat\theta)^\dagger (\Sigma_{\rm MUSE})^{-1} (\theta - \hat\theta) + C
\end{align}
for posterior mean $\hat\theta$, covariance $\Sigma_{\rm MUSE}$, and unimportant constant, $C$. The Gaussian approximation is well-motivated due to the central limit theorem as long as many modes of $f$ and $\phi$ contribute to the constraint on each $\theta$, which is well-satisfied here.

The marginal posterior mean, $\hat \theta$, is given by solving the following equation for $\theta$:

\begin{align}
    \smap[i](\theta,d) = \Big \langle \smap[i](\theta,d^\prime) \Big \rangle_{d^\prime \sim \mathcal{P}(d^\prime\,|\,\theta)},
    \label{eq:muse_estimate}
\end{align}
where $\smap[i](\theta,d)$ is the gradient of the joint posterior with respect to parameter $\theta_i$, evaluated at the {\it maximum a posteriori} (``MAP'') estimate of $z\,{\equiv}\,(f,\phi)$, i.e.
\begin{align}
    \smap[i](\theta,d) = \frac{d}{d\theta_i} \log \mathcal{P}(\hat z(d,\theta),\theta\,|\,d)\Big\vert_{\theta}
    \label{eq:smap}
\end{align}
with
\begin{align}
    \hat z(d, \theta) = \underset{z}{\rm arg\,max} \, \mathcal{P}(z,\theta\,|\,d).
    \label{eq:zhat}
\end{align}

One way to understand this definition is to consider $\smap[i]$ as a data compression, which, for a given value of $\theta$, takes as input our ${\sim}$\,million-dimensional data vector and returns another vector of the same dimensionality as $\theta$, which in our case is around a hundred bandpower and systematics parameters. The MUSE algorithm then proceeds by picking a starting guess for $\theta$ and compressing the data according to this, generating a suite of data simulations given the same $\theta$ and compressing them as well, and finally iteratively searching for the value of $\theta$ which make the data and simulations most similar, as compared via this data compression (Eqn.~\ref{eq:muse_estimate}). Intuitively, this yields the parameters $\theta$ which are most likely to have actually generated the real data.

More formally, \citet{Millea:2021had} show that following this procedure leads to an asymptotically unbiased estimate of parameters, $\hat \theta$, for any Gaussian or non-Gaussian latent space, $z$. For a Gaussian latent space, the data compression is a sufficient statistic, and therefore $\hat \theta$ gives the exact marginal posterior mean. For non-Gaussian latent spaces such as the one here, the data compression can be considered ``lossy'', but this leads only to a possible loss of optimality, with the asymptotic bias remaining zero. For a lensing analysis similar to the one performed here, \citet{Millea:2021had} bounds any loss of optimality to less than 10\% of the statistical uncertainty. 

The covariance, $\Sigma_{\rm MUSE}$, is computed from:
\begin{align}
    \Sigma_{\rm MUSE}^{-1} = H^\dagger \, J^{-1} \, H + \Sigma_{\rm prior}^{-1}
    \label{eq:muse_cov}
\end{align}
where 
\begin{align}
    J_{ij} &= {\rm cov}\Big(\smap[i](\hat\theta,d), \smap[j](\hat\theta,d)\Big)_{d\sim\mathcal{P}(d\,|\,\hat\theta)} \label{eq:J} \\
    H_{ij} &= \left. \frac{d}{d\theta_j} \left[ \Big\langle \smap[i](\hat\theta,d) \Big\rangle_{d\sim\mathcal{P}(d\,|\,\theta)} \right] \right|_{\theta=\hat\theta}, \label{eq:H}
\end{align}
and $\Sigma_{\rm prior}$ is the covariance of any assumed prior. Note that this is not a pure Monte Carlo covariance of the MUSE estimate, which would be prohibitive computationally, but rather a cheaper Monte Carlo covariance of just $\smap$, with analytic propagation to the resulting covariance in parameters. All derivatives in practice are computed with automatic differentiation, and Monte Carlo error stemming from $J$ is reduced using a method described in App.~\ref{app:shrinkage}.

One aforementioned aspect of MUSE which is key for this analysis is that it is unbiased even for ``lossy'' data compressions. This holds as long as the simulation-generating distribution, $\mathcal{P}(d\,|\,\theta)$, is an accurate representation of the real data. There is not actually {\it any} requirement on the form of $\smap[i]$ for MUSE to remain asymptotically unbiased and for the covariance expressions to be valid. The default choice for $\smap[i]$ given in Eqn.~\eqref{eq:smap} is simply the one which makes MUSE exact for Gaussian problems. 

Here, we keep this choice largely intact, but modify the joint posterior appearing in Eqns.~\eqref{eq:smap} and \eqref{eq:zhat}, $\mathcal{P}(f,\phi,\theta\,|\,d)$, to be slightly different from the likelihood function used to generate the simulations, $\mathcal{P}(d\,|\,\theta)$. Normally, the two distributions would be related by:
\begin{align}
    \mathcal{P}(d\,|\,\theta) = \frac{\mathcal{P}(d)}{ \mathcal{P}(\theta)} \int {\rm d}f \, {\rm d}\phi \, \mathcal{P}(f,\phi,\theta\,|\,d). 
\end{align}
Instead, here we have one statistical model denoted as the {\it posterior model}, $\mathcal{P}(f,\phi,\theta\,|\,d)$, and another called the {\it simulation model}, $\mathcal{P}(d\,|\,\theta)$.

In the analysis, the posterior model is used in computationally costly high-dimensional maximizations, while the simulation model is only ever used to generate a few hundred samples from $\mathcal{P}(d\,|\,\theta)$. This hints at the motivation for this choice of separate models, namely to allow defining a simulation model which fully meets our accuracy requirements, while keeping a simpler "surrogate" posterior model which is much faster to compute and makes the analysis more tractable. With an accurate simulation model, the MUSE analysis will remain unbiased, and we find little loss of optimality due to approximations in the posterior model.

We note that our analysis can also be considered a form of ``simulation-based inference'' (SBI). This term is used to describe a class of algorithms which leverage the ability to generate samples from $\mathcal{P}(d\,|\,\theta)$ directly into the ability to perform inference, and has been of increasing interest in various domains in cosmology \citep[see e.g.][]{hahn2017,levasseur2017,coogan2020,legin2022,lin2023,lemos2023,karchev2023}. In typical high-dimensional SBI applications to-date, an important first step is to obtain a suitable data compression function to reduce the dimensionality of the problem, often by training a neural network on the generated samples themselves \citep{charnock2018}. The compressed data is then used to infer parameters with something like Approximate Bayesian Computation (ABC) \citep{rubin1984,beaumont2002}, or by training a neural network surrogate model for the likelihood which can be inverted to obtain the posterior \citep[see][]{cramer2019}. 

MUSE is similar in that it also involves a data compression function, but rather than learned with a neural network, the generic and semi-analytic $\smap[i]$ is used instead. Furthermore, rather than using something like ABC or training surrogate models for inference, we exploit our access to gradients and Jacobians through the posterior and simulation-generating function to derive semi-analytic expressions for the posterior mean and covariance. In this way, MUSE can be considered one of the possible augmentations of SBI suggested by \cite{cramer2019} which use additional tractable quantities from the simulator. 

\subsection{Cosmological parameter inference}
\label{sec:cosmo_inference}

The output of MUSE is a parameter mean, $\hat \theta$, and its covariance, $\Sigma_{\rm MUSE}$. The inferred parameters for this analysis include the $\phi\phi$ and unlensed $EE$ spectra binned into bandpowers, and systematics parameters. The systematics-marginalized CMB bandpowers and covariance can be obtained by simply dropping the systematics from the estimate and dropping the corresponding block from the covariance (which is equivalent to marginalizing them). That is, part of the output of MUSE is automatically a set of "CMB-compressed" bandpowers such as those which have been derived for other experiments \citep{dunkley2013,prince2019,prince2024}. We can then estimate cosmological parameters, $\gamma$, as usual, by forming a cosmological parameter likelihood,
\begin{equation}
    -2\,\mathrm{log}\,\mathcal{P}(\hat\theta \, |  \, \gamma) = \left[\theta(\gamma) - \hat\theta\right]^{\dagger} \Sigma^{-1}_{\rm MUSE} \left[\theta(\gamma) - \hat\theta\right],
\end{equation}
where $\theta(\gamma)$ is our theory $\phi\phi$ and unlensed $EE$ spectra binned according to our specified bandpower window functions (App.~\ref{sec:lensedcmb}). For our main results we sample $\mathcal{P}(\hat\theta \, |  \, \gamma)$ with \texttt{Cobaya}\footnote{\url{https://cobaya.readthedocs.io/en/latest/}} \citep{Torrado:2020dgo} and use \texttt{CAMB}\footnote{\url{https://camb.info/}} \citep{Lewis:1999bs} to compute theory spectra. To get the accurate theory CMB spectra at large $\ell$, we set \texttt{lens\_potential\_accuracy=4} and \texttt{lens\_margin=1250} in \texttt{CAMB}. We also set \texttt{halofit\_version=mead2020} for the non-linear correction in \texttt{CAMB}. The convergence criterion of MCMC is set to $R-1<0.01$ of the Gelman-Rubin statistic.

\vspace{1cm}
\subsection{Map-level model summary}
\label{sec:model_summary}

The posterior and simulation models are specified in Eqns.~(\ref{eq:model_start}-\ref{eq:model_end}), with a summary given below and full details given in Appendix~\ref{app:model}.

\begin{widetext}
\begin{align}
{\rm Simulation} & {\rm \;model:} \nonumber \label{eq:model_start} \\
f &\sim \mathcal{N}(0, \mathbb{C}_f^{\rm curv\,sky}(A_b^{\rm EE})) \\
\phi &\sim \mathcal{N}(0, \mathbb{C}_\phi^{\rm curv\,sky}(A_b^{\phi\phi})) \\
n^\nu &\sim \{n_{\rm signflips}^\nu\} \\
d^{\nu,i} &= \mathbb{M}_{\rm fourier} \cdot \mathbb{M}_{\rm trough} \cdot \mathbb{M}_{\rm pix} \cdot \big(\mathbb{PWF} \cdot \mathbb{TF}^\nu \cdot \mathbb{R}(\psi_{\rm pol}^\nu) \cdot A_{\rm cal}^{\nu,i} \cdot \mathbb{B}(\beta_n, \beta_{\rm pol}^\nu) \cdot \mathbb{G} \cdot \mathbb{P} \cdot \mathbb{L}(\phi) \cdot f \nonumber \\ & \hspace{11cm} + \epsilon_{\rm Q}^{\nu,i} \cdot t_{\rm Q}^\nu + \epsilon_{\rm U}^{\nu,i} \cdot t_{\rm U}^\nu + n^\nu \big) \label{eq:sim_data_model}  \\
\nonumber\\
{\rm Posterior} & {\rm \;model:} \nonumber \\
f &\sim \mathcal{N}(0, \mathbb{C}_f^{\rm flat\,sky}(A_b^{\rm EE})) \\
\phi &\sim \mathcal{N}(0, \mathbb{C}_\phi^{\rm flat\,sky}(A_b^{\phi\phi})) \\
\mu^{\nu,i} &= \mathbb{M}_{\rm fourier} \cdot \mathbb{M}_{\rm trough} \cdot \mathbb{M}_{\rm pix} \cdot \big(\mathbb{PWF} \cdot \mathbb{TF}^\nu \cdot \mathbb{R}(\psi_{\rm pol}^\nu) \cdot A_{\rm cal}^{\nu,i} \cdot \mathbb{B}(\beta_n, \beta_{\rm pol}^\nu) \cdot \mathbb{L}(\phi) \cdot f + \epsilon_{\rm Q}^{\nu,i} \cdot t_{\rm Q}^\nu + \epsilon_{\rm U}^{\nu,i} \cdot t_{\rm U}^\nu \big) \label{eq:post_mu_model}\\
d^\nu &\sim \mathcal{N}(\mu^\nu, \mathbb{C}_n^{\nu}) \label{eq:model_end} \\
&\Rightarrow -2\log\mathcal{P}(f,\phi,\theta\,|\,d) = f^\dagger \mathbb{C}_f^{-1} f + \phi^\dagger \mathbb{C}_\phi^{-1} \phi + \sum_\nu (d^\nu - \mu^\nu)^\dagger (\mathbb{C}_n^\nu)^{-1} (d^\nu - \mu^\nu) -2\log\mathcal{P}(\phi) \label{eq:posterior_function} \\
&{\rm where} \; -2\log\mathcal{P}(\phi) = \big \lVert \mathbb{M}_{\rm pix} \nabla^2 \phi \big \rVert^2 / 10^{-8} \label{eq:super_sample_prior}
\end{align}
\end{widetext}

Reading these models top-to-bottom describes how to produce a random sample from the model, with ``$\sim$'' used to represent generation of a random sample from the distribution on the RHS and $\mathcal{N}(\mu,\Sigma)$ a multi-variate Gaussian with mean and covariance $\mu$ and $\Sigma$. 

To summarize, in the simulation model, we generate curved-sky unlensed CMB maps ($f$) and lensing potential maps ($\phi$) from appropriate covariance operators ($\mathbb{C}$), which are parameterized by the bandpower amplitudes which we will infer ($A_b^{\rm EE}$ and $A_b^{\rm \phi\phi}$). We apply the lensing operation ($\mathbb{L}$) to produce lensed CMB maps and project to the flat-sky Lambert pixelization of the data ($\mathbb{P}$). We apply an additional small deflection ($\mathbb{G}$) to account for relativistic aberration due to our proper motion relative to the CMB rest frame. We apply the beam convolution operator ($\mathbb{B}(\beta_n, \beta^\nu_{\rm pol})$), which is parameterized by a set of beam eigenmode amplitudes ($\beta_n$) and the polarization fraction of the beam sidelobes at each frequency ($\beta^\nu_{\rm pol}$). We rotate the global polarization angle of the maps ($\mathbb{R}(\psipol)$) by an angle $\psipol$ to model systematic errors in the overall polarization angle of our detectors. In each subfield, $i$, we scale the maps by $A^{\nu,i}_{\rm cal}$, to model the uncertain absolute calibration of the instrument. We apply the model transfer function ($\mathbb{TF}$), and apply the pixel window function ($\mathbb{PWF}$). We add in temperature to polarization monopole leakage $Q$ and $U$ templates determined directly from our temperature observations ($t_{\rm Q}$ and $t_{\rm U}$) with subfield-dependent amplitudes ($\epsQ^i$ and $\epsU^i$), as well as add in a sample of the noise ($n^\nu$) from one of the sign-flip noise realizations. Finally, we apply our chosen pixel, ``trough'', and Fourier masks (see App.~\ref{sec:masking}) to produce a simulated data vector, $d^\nu$. 

The posterior model is nearly the same, with the only differences being that the covariances, $\mathbb{C}$, assume flat-sky statistics, the noise is generated from a covariance and is unmasked, and aberration is ignored. For the posterior model, it is possible to calculate the posterior probability function, given in Eqn.~\eqref{eq:posterior_function}. Here, we also choose to include a "super-sample" prior, $\mathcal{P}(\phi)$, in order to reduce the lensing mean-field (before its impact is automatically removed entirely in the process of obtaining the MUSE estimate). 

The final inferred parameter vector is
\begin{align}
    \theta = \{A_b^{\phi\phi}, A_b^{\rm EE}, \beta_n, \beta^\nu_{\rm pol}, A^{\nu,i}_{\rm cal}, \psipol, \epsQ^i, \epsU^i\},
\end{align}
containing 77 bandpower parameters and 47 systematics parameters. The MUSE estimate is a joint best-fit and a covariance between all of them.

We note that the approximation of leaving the noise unmasked, while seemingly benign, is actually critical to obtaining a sufficiently fast posterior function since it avoids an additional large matrix inversion, while having little impact on the final results (see also Sec.~3.7 of \citep{millea2021}). We also note that the ability to directly use the sign-flip noise realizations in the simulation model is extremely advantageous. Without this, a Bayesian analysis would depend on an accurate statistical model of all moments of the noise (power spectrum, bispectrum, etc...), and would require both sampling and evaluating the likelihood of a noise map given such a model. Here, noise modeling is essentially a non-issue since MUSE requires only sampling the noise, and the sign-flip realizations provide these samples and empirically match the true noise distribution. 

There are a few effects which we considered but ultimately chose not to include the model. The first effect is polarized foregrounds. The dominant expected foreground comes from polarized point sources. For the same point source flux cut used here, \citet{SPT-Dutcher:2021eoc} construct a set of priors on the foreground amplitudes based on measurements of temperature foregrounds scaled to polarization with very conservative priors on the polarization fraction. Based on this, we find that the maximum possible contamination has power two orders of magnitude lower than our noise spectra at all frequencies. That the foreground contribution is small is also consistent with \cite{qu2024}, in which it is found that even for much deeper observations than ours, the bias to lensing from polarized foregrounds is fairly marginal. As mentioned, this finding that the foreground contribution is negligible in polarization for this analysis was key in our decision to focus only on polarization in this paper. 

The second effect we ignore is the bias due to spatial correlation between the lensing potential and the location of our masked point source holes \citep{lembo2022}. Using the \textsc{Agora} simulations \citep{omori2024}, which include a model for these spatial correlations, we find that for our point source masking threshold, the maximum bias to $C_L^{\phi\phi}$ would be ${\sim}\,0.5\%$. However, \citet{lembo2022} also show that this maximum bias is highly suppressed in an optimally filtered $\phi$ map such as the MAP $\phi$ maps which enter our analysis. We thus conclude this effect is also negligible here.

\begin{figure*}
    \centering
    \includegraphics[width=\textwidth]{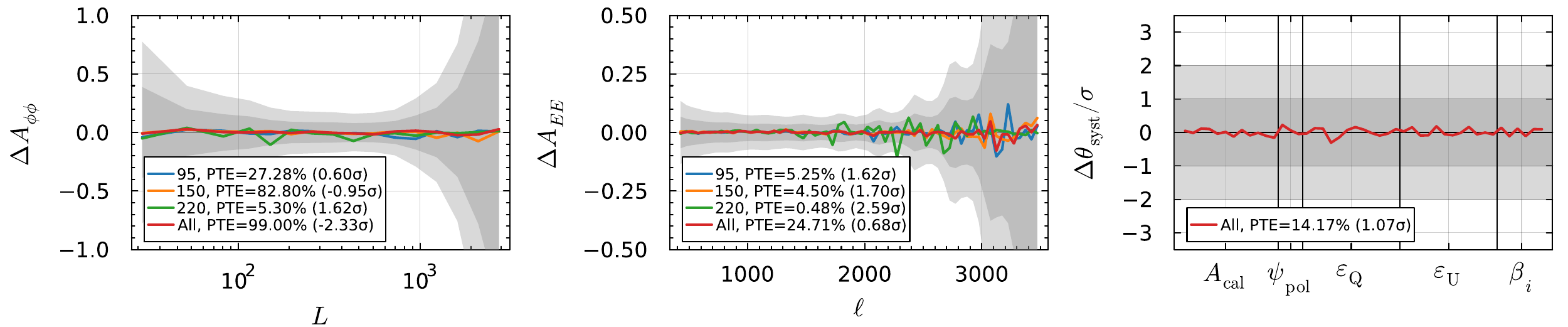}
    \caption{Verification that our pipeline recovers unbiased bandpower and systematics parameters using 100 mock simulations. The colored lines are the average MUSE parameter estimates over simulations after subtracting the input simulation truth (in the final panel we also divide by the errorbars for one simulation due to the otherwise large dynamic range). The first two panels show $\phi\phi$ and unlensed $EE$ bandpower amplitude parameters (and also consider single-frequency runs) and the last panel groups together all the systematics parameters (see Sec.~\ref{sec:model_summary} and App.~\ref{app:model} for description of the systematics considered). The PTEs of the colored lines relative to the expected scatter and 2-tailed numbers of $\sigma$ are shown in the legends. To get the PTEs, we Monte Carlo sample the expected distribution of $\chi^2$ values, accounting both for the look-elsewhere effect of 4 tests (3 single-frequency runs and 1 all-frequency run) and the fact that we compute the expected scatter itself from 100 mocks. For reference, gray bands are the all-frequency 1 and $2\,\sigma$ errors (note that this is not the expected scatter of any of the lines, rather is a way to judge the size of any potential bias).}
    \label{fig:bias_sims_single_freq}
\end{figure*}

\vspace{0.3cm}
\section{Validation of Pipeline and Data}
\label{sec:validation}

Before computing our main results, we conducted a set of validation tests which we present in this section. To mitigate and quantify confirmation bias, we performed these tests ``blind'', that is, without comparison to data from other experiments or to theory models derived from these experiments. After unblinding, we allowed ourselves to make further changes to the pipeline before producing our final baseline results, but committed to reporting in this paper the parameters that we initially found before any such changes. This motivated and focused the validation work, and allows us to quantify any level of confirmation bias; the initial unblinded results could not be impacted by confirmation bias, only the shifts after unblinding could be. A full discussion of changes made after unblinding is at the end of this section (Sec.~\ref{sec:post-unblind}). 

Our chosen set of rules for blinding were: 1) do not plot our bandpowers on the same plot as bandpowers or theory curves from any other previous analysis or experiment (including from SPT itself), or show bandpowers relative to any theory curves, 2) if we do plot bandpowers, remove all tick marks and tick labels and do not look at numerical values of bandpowers, 3) do not look at absolute cosmological parameters derived from bandpowers (internal null parameter differences, e.g. the parameters from 95\,GHz minus parameters from 150\,GHz were allowed).

This procedure left us blind to all but very gross departures from \lcdm and did not allow us to know if bandpowers were consistent with any other experiment before unblinding, while not being overly burdensome. 

We now describe the set of validation tests of our pipeline and data which we performed while blinded.

\subsection{Pipeline tests}
\label{subsec:pipeline_test}

\subsubsection{Simulated Data}
\label{sec:mock_making}

To validate our full bandpower and cosmological parameter estimation pipeline, we use an ensemble of highly realistic simulations of the full dataset which we call "mock simulations". The signal part of the mock simulations consists of a lensed CMB realization generated assuming a fiducial \planck \lcdm cosmology and a Gaussian foreground model based on \textsc{Agora} simulations \citep{omori2024} that is calibrated against data at very high-$\ell$. These maps are processed through a procedure called mock observation, which generates simulated timestreams and simulates the observation and processing of the timestreams into maps at each frequency, using the identical code which processes the real data.\footnote{To reduce computational cost and because many observations have identical filtering, we use only 5\% of the total observations in this step.} The noise part of the mock simulations comes from sign-flipped noise realizations.

\subsubsection{End-to-end pipeline tests}
\label{sec:endtoend}

To check that our pipeline is unbiased, we start with an end-to-end test of the pipeline on a set of 100 mock simulations. First, we compute MUSE estimates of bandpowers and systematic parameters on each mock simulation using all three 95+150+220 GHz polarization maps (hereafter, "all-frequency"). Then we repeat the same procedures on single-frequency polarization maps with the corresponding systematic parameters fixed to the all-frequency results (this is how we handle systematics in the single-frequency runs on real data as well).

In the left two panels of Fig.~\ref{fig:bias_sims_single_freq}, we show the mean bandpowers over 100 mock simulations compared to the input truth. The PTE of the difference between the mean and input truth and its conversion to a number of $\sigma$ is given in the legend.\footnote{All PTEs we quote are corrected for look-elsewhere in the case that there are multiple tests. We will specify if the conversion to number of $\sigma$ is 2-tailed, in which case negative values correspond to a PTE better than the mean expectation; otherwise the quoted $\sigma$ number is 1-tailed.} No bias is detected at $3\,\sigma$ significance from the 100 mock simulations. The scatter of the mean bandpowers is at the level of 10\% of the statistical uncertainty, as expected from a test with 100 realizations. In the right panel of Fig.~\ref{fig:bias_sims_single_freq}, we also show the mean systematics parameters over 100 all-frequency mock simulations, here scaled by their 1\,$\sigma$ uncertainty. These are also recovered with no significant bias. Single-frequency runs do not vary systematics parameters, hence those estimates are not shown.

\begin{figure}
    \centering
    \includegraphics[width=\columnwidth]{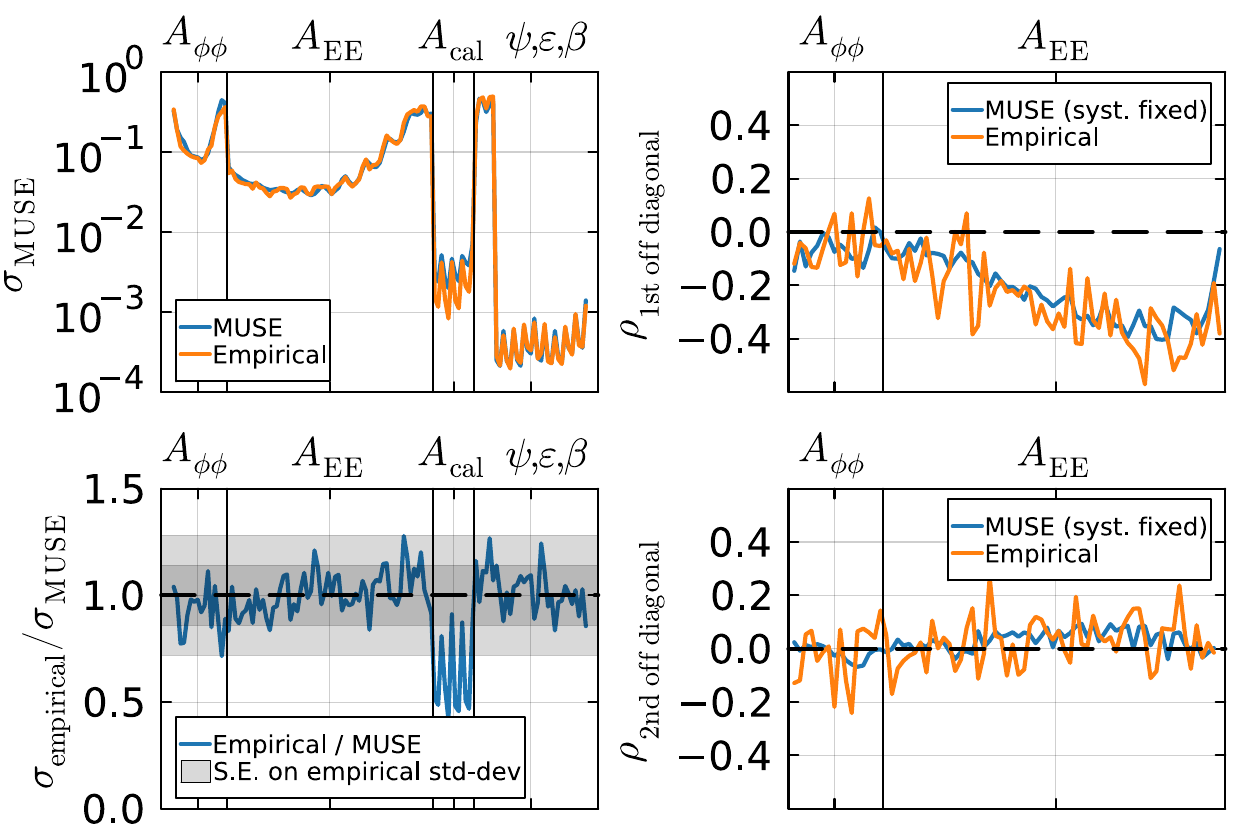}
    \caption{Verification of the accuracy of the all-frequency MUSE covariance, $\Sigma_{\rm MUSE}$. {\it (Top left)} Comparison of the square root of the diagonal elements between the MUSE covariance, $\sigma_{\rm MUSE}\,{\equiv}\,\sqrt{}{\rm diag}(\Sigma_{\rm MUSE})$, and the empirical covariance from running the estimate on mock simulations. {\it (Bottom left)} The ratio of the two curves in the upper panel. The shaded band shows the 1 and $2\,\sigma$ standard error on $\sigma_{\rm empirical}$ from the 100 simulations used. These are expected to match for fully likelihood-dominated parameters, and do so for all parameters except the partially prior-dominated $A_{\rm cal}$ parameters. {\it (Top and bottom right)} Comparisons of the first and second off-diagonal elements of the MUSE and empirical correlation matrices. In this case, systematics are assumed fixed in the MUSE covariance, since they are fixed in the simulations.}
    \label{MUSE_empirical_cov}
\end{figure}

We next compare the empirical covariance of the 100 all-frequency estimates with the MUSE covariance given in Eqn.~\eqref{eq:muse_cov}. The left column of Fig.~\ref{MUSE_empirical_cov} compares the square root of the diagonal covariance entries. We see good consistency between the two except for $A_{\rm cal}$. This is because the mock simulations were made without varying systematic parameters within their prior, while the MUSE covariance takes the prior uncertainty of all parameters into account. This difference is particularly evident for prior-dominated parameters such as $A_{\rm cal}$.

In the right column of Fig.~\ref{MUSE_empirical_cov}, we compare the first and second off-diagonal elements in the bandpower covariance matrices, this time fixing the systematics in $\Sigma_{\rm MUSE}$ for a more like-to-like comparison. We see good agreement between the empirical covariance and the MUSE covariance. The anti-correlation in the first off-diagonal elements is caused by projection effects, masking, delensing and the transfer function, where the projection effects explain about half of the anti-correlation. Here the projection effects refers to our use of the flat-sky approximation in the posterior model. This effect has been found in the previous SPT-3G analysis \citep{SPT-3G:2022hvq}.

\begin{figure}
    \centering
    \includegraphics[width=0.9\columnwidth]{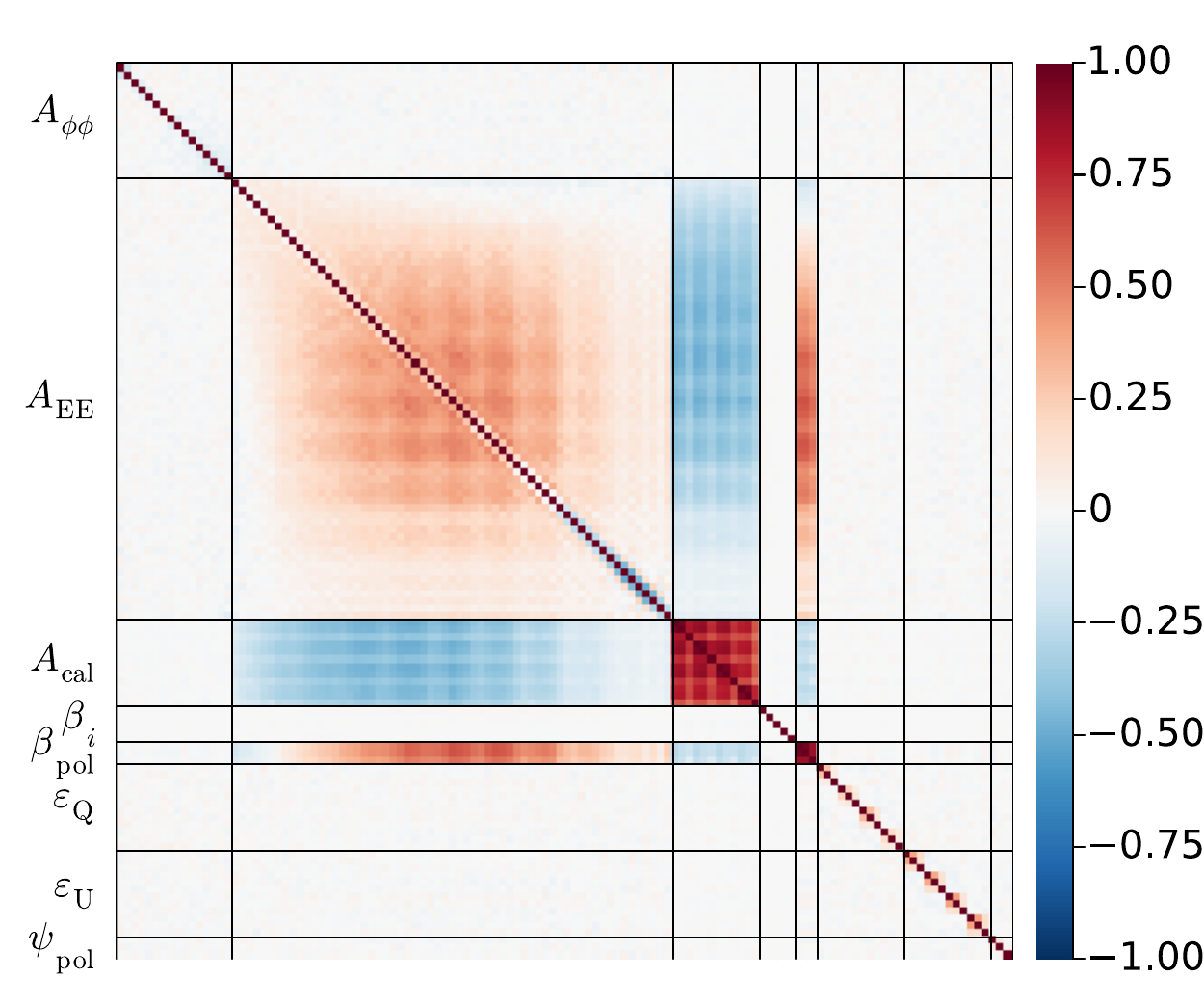}
    \caption{The posterior correlation matrix of $\phi\phi$ bandpowers, unlensed $EE$ bandpowers, and systematics parameters, given all-frequency data.}
    \label{fig:muse_bp_forr}
\end{figure}

In Fig.~\ref{fig:muse_bp_forr}, we show the MUSE covariance with its diagonal entries rescaled to unity (i.e. the correlation matrix between all bandpowers and systematics parameters). We find no significant correlation between lensing bandpowers and any other parameter, {except for $A_{\rm \phi\phi}$ at the lowest $L$ bin, which shows correlations with $A_{\rm EE}$ at $\lesssim 10\%$ level and with $\beta_{\rm pol}$ at $\sim12\%$ level. This suggests} a robust determination of the lensing spectrum with little uncertainty from systematic effects. The correlations within the unlensed $EE$ block are due to masking, delensing, projection and correlation with systematics parameters. The $EE$ bandpowers are significantly correlated with both calibration uncertainty and the sidelobe polarization fraction, both of which add appreciable systematic uncertainty to our final $EE$ estimates. 

We next obtain $\Lambda$CDM cosmological parameter estimates from the 100 all-frequency mock simulations. Unlike the main results which use \texttt{Cobaya} and \texttt{CAMB} to perform parameter estimation, here, for speed, we have trained and verified the accuracy of an emulator using \texttt{Capse.jl} \citep{Bonici:2023xjk}, and sample using Hamiltonian Monte Carlo. 

\begin{figure}
    \centering
    \includegraphics[width=\columnwidth]{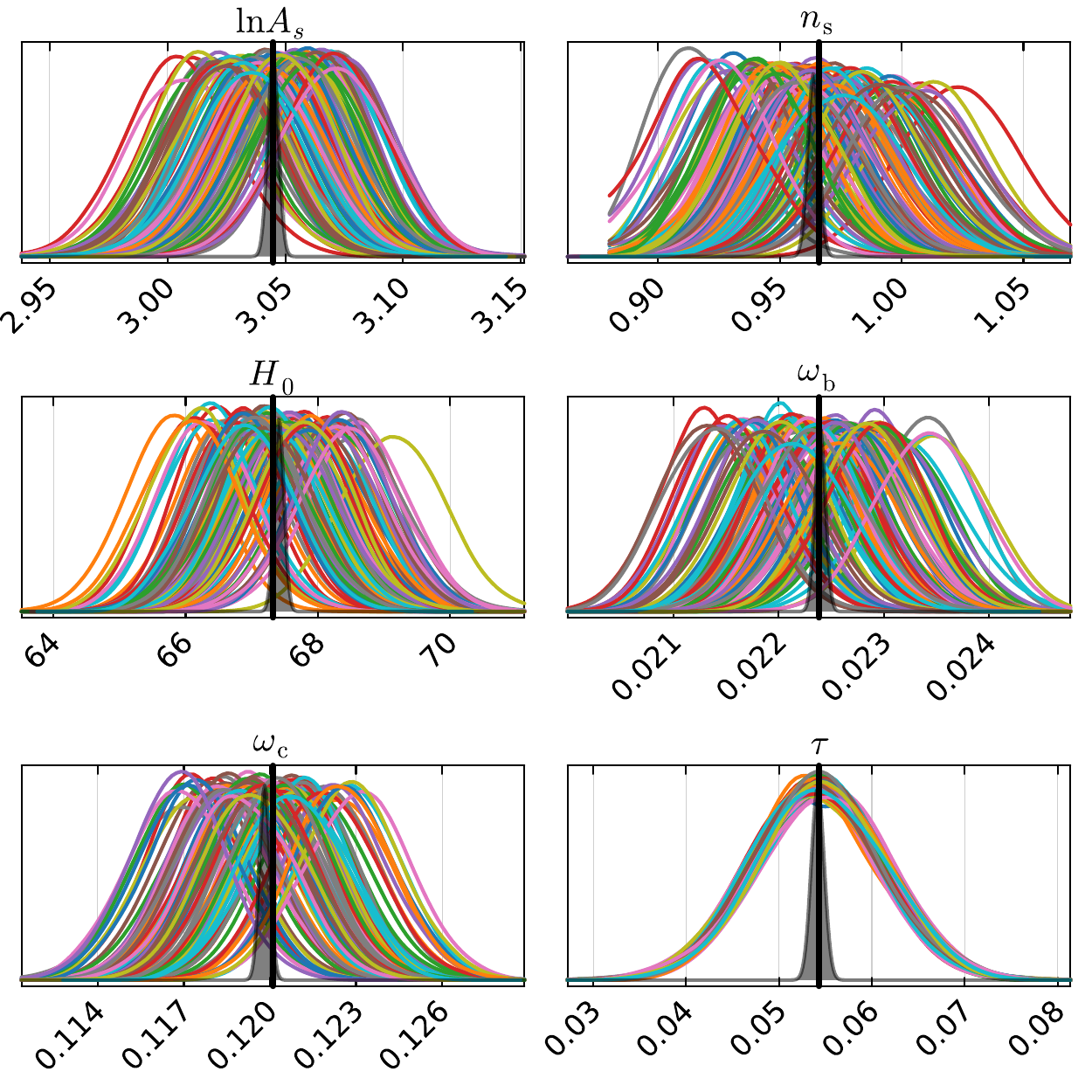}
    \caption{Verification that our pipeline recovers unbiased cosmological parameters using 100 mock simulations. We infer  $\Lambda$CDM parameters from each simulated all-frequency dataset, and plot the one-dimensional marginalized posteriors above as different colored lines for each simulation. The shaded gray region in each panel is the product over all posteriors, which should cover the simulation truth, denoted as vertical black lines. Note that the smaller scatter for $\tau$ is because this parameter is mainly constrained by our prior. Across the other 5 parameters, the shaded region is consistent with the input truth with significance equivalent to $1.1\,\sigma$, suggesting no detection of any bias at the level afforded by our 100 simulations.}
    \label{fig:coverage_test_lcdm}
\end{figure}

The one-dimensional cosmological parameter posterior distributions inferred from each mock simulation and the product over all of these posteriors is shown in Fig.~\ref{fig:coverage_test_lcdm}. It is expected that the product over all posteriors should cover the input truth, which is found to be the case. The non-detection of any bias with 100 sims suggests that, at 68\% confidence, we have bounded any possible pipeline-induced systematic error to be less than ${\lesssim}\,0.1\,\sigma$ for each cosmological parameter. 

To summarize, we have validated the pipeline using 100 highly realistic mock simulations. With 100 simulations, we are able to detect biases at the level of 10\% of the statistical uncertainty. We do not detect any such biases to within $3\,\sigma$ of the expected scatter on the mean in either our estimates of $\phi\phi$ lensing spectra, unlensed $EE$ bandpowers, systematics, or in the cosmological parameters derived from them. We find good agreement of the bandpower covariance matrices between the one derived from mock simulations and the one calculated in MUSE using Eqn.~\eqref{eq:muse_cov}.

\subsection{Data tests}
\label{subsec:data_test}
\begin{figure*}
    \centering
    \includegraphics[width=\textwidth]{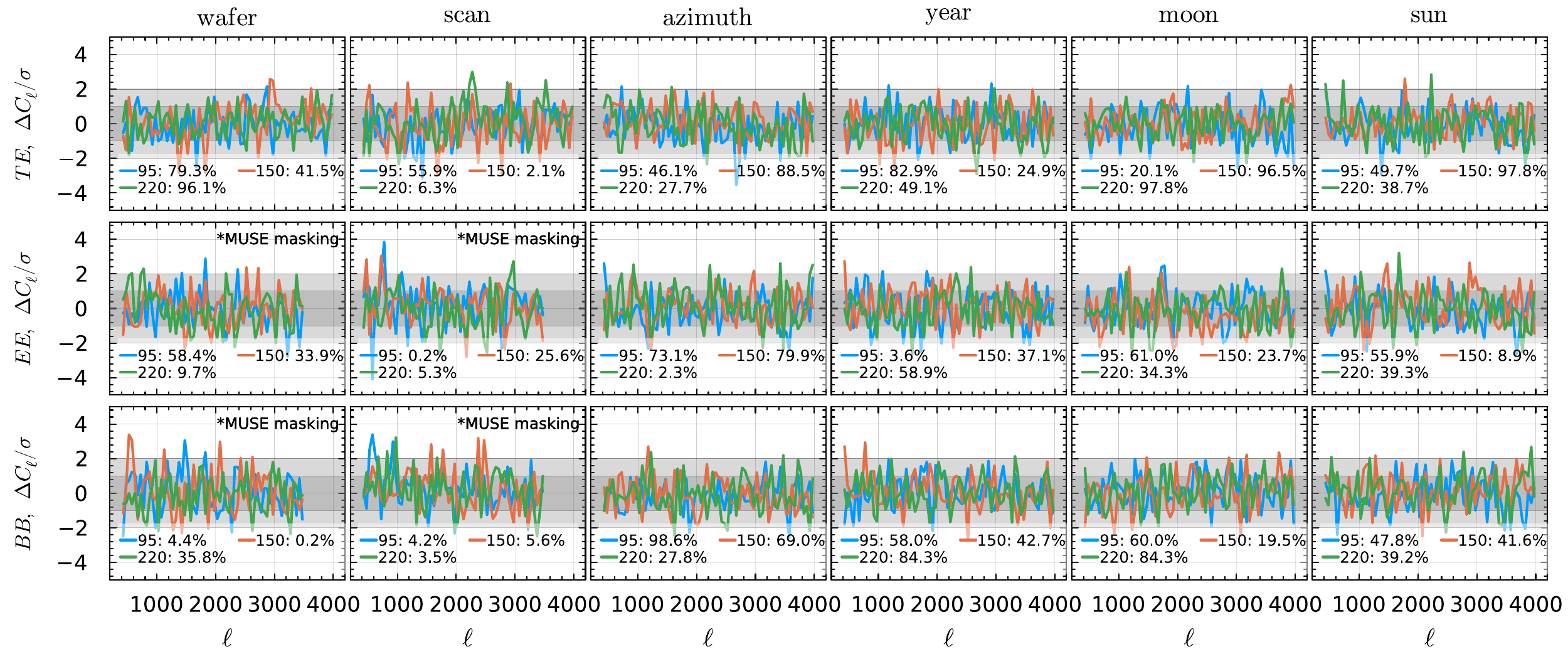}
    \caption{Spectra of null maps computed from various data splits divided by the expected $1\,\sigma$ scatter due to noise. Detector wafer and scan direction tests also have had a small expected contribution from signal leakage subtracted, and in $EE$ and $BB$ have been computed with the exact pixel and Fourier space masking used in the main results pipeline (indicated as "MUSE masking"; the other tests are not expected to depend significantly on masking). Legends give the PTE for each test without correcting for the look-elsewhere effect. The $2\,\sigma$ threshold when considering the look-elsewhere effect is a PTE$>0.07\,\%$ on any individual test, which is passed in all cases.}
    \label{fig:null_tests}
\end{figure*}
Next, we turn to tests of the data itself. These include  intra-frequency null tests using different splits of the data, inter-frequency comparison of $\phi\phi$ and unlensed $EE$ bandpower results, pixel-level $\chi^2$ tests of our model fits, and internal comparison of our systematics estimates with those derived with more traditional methods. We report the results of these tests in this section.

\subsubsection{Intra-frequency null tests}
\label{sec:nulltests}

Null tests involve splitting the full dataset into two parts and differencing them such that the signal cancels, then checking whether the remaining portion is consistent with the expected statistical properties of noise. Finding something outside of this expectation could indicate the presence of a systematic, which might then also be present in the full dataset coadd at some level. By canceling the signal and thus the sample variance from the signal, these tests can probe systematics to levels well below even what could impact the final results, which are themselves limited by sample variance. Since temperature data does enter our analysis via leakage templates, we also present null TE spectra to more thoroughly vet the full temperature and polarization dataset. Further details of the null tests will be presented in a future work. 

We consider 6 types of data splits: sun, moon, azimuth, year, scan, and wafer. The sun and moon tests are designed to check whether a portion of the dataset is significantly contaminated by the radiation from the sun or moon. In these tests, one part of the dataset comprises the data we took when the sun or moon was above the horizon, and the other part below the horizon. The azimuth test is designed to check whether a portion of the dataset is contaminated by thermal radiation from a building near the telescope. One part of the dataset comprises the data we took when the telescope's azimuth angle was within 90 degrees of that of the building, and the other part azimuth angles outside that range. The year test is designed to check whether the data we took during the 2019 season are significantly different from the data we took in 2020. The scan test is designed to check whether the data we took when the telescope was scanning in one direction are significantly different from the data we took when the scanning was in the opposite direction. Lastly, the wafer test is designed to check whether the data from one half of the full detector array are significantly different from the data from the other half.

For the scan and wafer tests, we do not expect the null spectra to perfectly cancel the signal due to details of our mapmaking pipeline. For the wafer test, this is because we filter time-ordered data from different detector wafers differently. For the scan test, this is mainly because we do not correct for each individual detector's time constant effect when converting its time-ordered data into map pixel values. The time constant effects refers to the fact that each detector does not respond to changing radiation from the sky instantaneously. As a result, a map made from left-going scans contain a version of the CMB shifted to the left, and vice-versa for right-going scans. 
For these two tests, we calculate the expected null spectra through simulations of the mapmaking pipeline and subtract it. This is computed assuming a fiducial cosmology, but because these expected null spectra are very small compared to the signal, it does not significantly imprint model-dependence to these null tests or degrade their usefulness.

For each null test type, we divide each of the two halves of the data into 25 smaller parts and call them bundles. We create 25 null maps by differencing bundles from each half one-by-one, then compute all 300 cross-spectra from these 25 null maps and bin them into the same bandpowers as the analysis. The average of these is what we call a null spectrum, and the standard error is our estimate of the expected scatter of the null spectrum.

The maps used for these tests are curved-sky projections of the same timestreams which enter the main analysis, with spectra computed with \texttt{PolSpice} \citep{Chon:2003gx}. For convenience, most of the tests have been computed using slightly different masks than the ones used to produce our main results. These tests, which we do not expect to depend significantly on the masking choice, use a pixel mask which includes slightly more sky, and a harmonic space mask which zeros all $a_{\ell m}$ for $\ell \in [500, 680]$ and $m \in [350, 425]$. This harmonic mask removes contamination from a narrowband signal around 1.1 Hz, which is similar, but not exactly the same as the ``trough'' mask used to remove the contamination in the MUSE pipeline. As the 1.1 Hz signal presents itself mainly in detector wafer and scan direction tests, we have recomputed these with the exact MUSE pixel and Fourier masking to verify the MUSE masking is sufficient to remove it.

After masking, computing null spectra, and subtracting expected contributions as necessary, we calculate a $\chi^2$ with respect to the expected scatter, assuming no correlation between neighboring bandpowers. To compute a threshold probability-to-exceed (PTE) for the observed $\chi^2$, we Monte Carlo sample the expected distribution of $\chi^2$ values, accounting both for the look-elsewhere effect of 54 tests and the fact that we compute the expected scatter itself from a finite 300 simulations. 
We consider the test passed if the PTE is ${>}\,5\%$ (corresponding to a ${<}\,2\,\sigma$ deviation) with respect to this Monte Carlo distribution. This corresponds to a PTE\,${>}\,0.07\%$ on any single individual test amongst our suite (this is roughly $5\%/54$ with an additional correction for the slightly noisy expected scatter).

A summary of all tests and their PTEs is given in Fig.~\ref{fig:null_tests}. All tests pass at our threshold of $0.07\%$, and no obvious features are otherwise visible. 

\subsubsection{Inter-frequency agreement}

We next check for inter-frequency agreement of the $\phi\phi$ and unlensed $EE$ bandpowers inferred from single-frequency maps. We follow the same procedure for inference as described in Sect.~\ref{sec:methodology}, except that the data, $d$, includes only a single-frequency map and the systematic parameters are fixed to the value derived from the all-frequency run; only the bandpowers $A_b^{\rm EE}$ and $A_b^{\rm \phi\phi}$ are free parameters. To quantify the expected scatter, we run this identical procedure on 100 mock simulations. 

\begin{figure}[!tphb]
    \centering
    \includegraphics[width=\columnwidth]{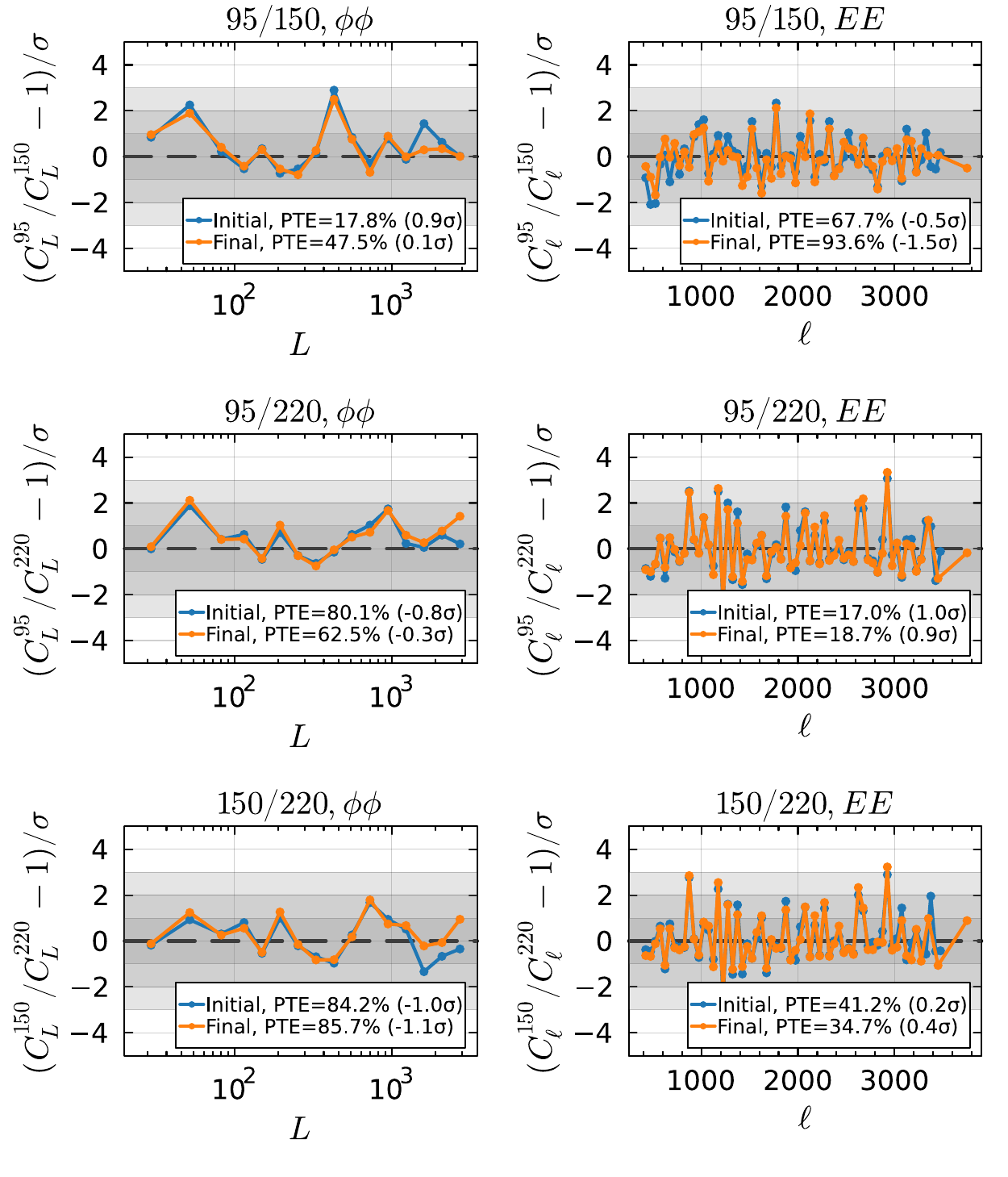}
    \caption{Tests of the inter-frequency bandpower agreement of our data. Each panel shows the ratio of $\phi\phi$ or unlensed $EE$ bandpowers from single-frequency MUSE runs at different frequencies, normalized by the $1\,\sigma$ expected scatter as computed from simulations. Legends show the $\chi^2$ and 2-tailed effective number of $\sigma$ fluctuation of this $\chi^2$ given the number of degrees of freedom. The blue line shows the result using our model before unblinding, which yielded acceptable $\chi^2$ in all cases. The orange line shows the result after the post-unblinding changes to our model, which notably slightly improves the agreement between 95\,GHz and 150\,GHz.}
    \label{fig:data_inter_freq}
\end{figure}

In Fig.~\ref{fig:data_inter_freq}, we show the ratios of bandpowers inferred from maps at different frequencies relative to the expected scatter. The blue lines show our result prior to unblinding, which showed good overall agreement. The legends give 2-tailed effective number of $\sigma$ for the $\chi^2$ values, and all are consistent with statistical fluctuations to within ${\pm}\,2\,\sigma$. We also note that with model changes after unblinding, shown as orange lines, agreement improved slightly, particularly in the comparison between 95\,GHz and 150\,GHz.

\subsubsection{Goodness-of-fit}
\label{sec:chi2s}

We next compare the pixel-level goodness-of-fit of our model to the observed data. To do so, we define $\chi^2\,{=}\,{-}\,2\log\mathcal{P}(\hat z(\hat \theta, d), \hat \theta\,|\,d)$, that is, minus twice the posterior probability function evaluated at the MUSE estimate of parameters, $\hat \theta$, and the MAP given that $\hat \theta$. Because of our decoupling of the posterior and simulation models, we cannot compute the expected $\chi^2$ simply from counting degrees of freedom. Instead, we use the distribution of $\chi^2$ obtained from mock simulations to quantify the expected distribution.

Tab.~\ref{tab:chi2s} gives the data $\chi^2$ values observed for single-frequency and all-frequency runs before unblinding, as compared to this expectation. We see that the values are on the order of $10^8$, despite that the number of pixels in $d$ is close to $10^7$ for the all-frequency case, confirming the fact that the expectation is not given simply by counting degrees of freedom. That the single-frequency expected $\chi^2$ are larger than the all-frequencies case is due to the fixing of systematic parameters in those cases. Regardless, with the expected distribution mapped out by simulations, we find that our observed values in all cases are consistent with a statistical fluctuation within $3\,\sigma$. This suggests our model is sufficient to describe the observed data.

\begin{table}[tphb]
\centering
\begin{tabular}{c|c|c|c|c}
\toprule
& $\chi^2_{\rm obs}/10^6$ & $\chi^2_{\rm expected}/10^6$ & PTE & $n_\sigma$ \\
\midrule
95+150+220 & 93.33 & $93.50\,{\pm}\,0.12$ & 4.0\% & 2.1 \\ 
95 & 108.21 & $108.28\,{\pm}\,0.18$ & 37.0\% & 0.9 \\ 
150 & 109.32 & $109.51\,{\pm}\,0.19$ & 7.0\% & 1.8 \\ 
220 & 100.97 & $101.00\,{\pm}\,0.04$ & 24.0\% & 1.2 \\ \bottomrule
\end{tabular}
\caption{The $\chi^2$ of the best-fit pixel-level model against the data, as compared to the expectation computed from mock simulations, as described in Sec.~\ref{sec:chi2s}. The number of $\sigma$ is 1-tailed. We find values within $3\,\sigma$ of the expectation for individual frequency fits, as well as for our baseline combination of all frequencies.}
\label{tab:chi2s}
\end{table}

\subsubsection{Agreement of systematics with alternative estimates}

Our analysis infers systematic parameters jointly with bandpowers by incorporating systematics into a map-level model for the data. In this section, we compare our systematics estimates to independent determinations from alternate methods which have traditionally been used to estimate systematics in CMB analyses.

In this comparison, we consider the global polarization rotation angle, $\psi_{\rm pol}$, the amplitude of monopole temperature-to-polarization leakage, $\epsilon_Q^i$ and $\epsilon_U^i$, and the calibration at each frequency, $A_{\rm cal}^{\nu,i}$. Details of how these are estimated with alternate methods are given in Appendix~\ref{app:model}. Briefly here, $\psi_{\rm pol}$ can be estimated by assuming the sky $EB$ spectrum is zero, and searching for the angle correction which achieves this in the data. $\epsilon_Q^i$ and $\epsilon_U^i$ can be estimated by assuming that the sky $TQ$ and $TU$ spectra are zero and finding the deprojection of leakage templates which achieves this. $A_{\rm cal}^{\nu,i}$ can be estimated taking ratios of power spectra to a known calibrated spectrum and averaging over very broad multipole ranges. Here, $A_{\rm cal}^{150}$ is fit via ratio to \planck, which performs its own absolute calibration, and $A_{\rm cal}^{95}$ and $A_{\rm cal}^{220}$ are fit from internal comparison to the \planck-calibrated SPT 150\,GHz spectrum. 

The source of MUSE estimates of these systematics is generally the same as described above, but happens implicitly and jointly inside the global MUSE fit, driven by our model definition (e.g. the model assumes the CMB $EB$ spectrum is zero and allows a global rotation controlled by $\psi_{\rm pol}$ to absorb any $EB$ power observed in the data during the fit).

The best-fit systematic parameters inferred with MUSE from the SPT data are summarized in Tab.~\ref{tab:muse_sys}. Broadly speaking, we find around $1\%$ temperature to polarization leakage, a little less than half of a degree of global polarization angle miscalibration, and beam sidelobes (see Appendix~\ref{sec:beams}) which are polarized at around the $50\%$ level.

\begin{figure}
    \centering
    \includegraphics[clip,width=\columnwidth]{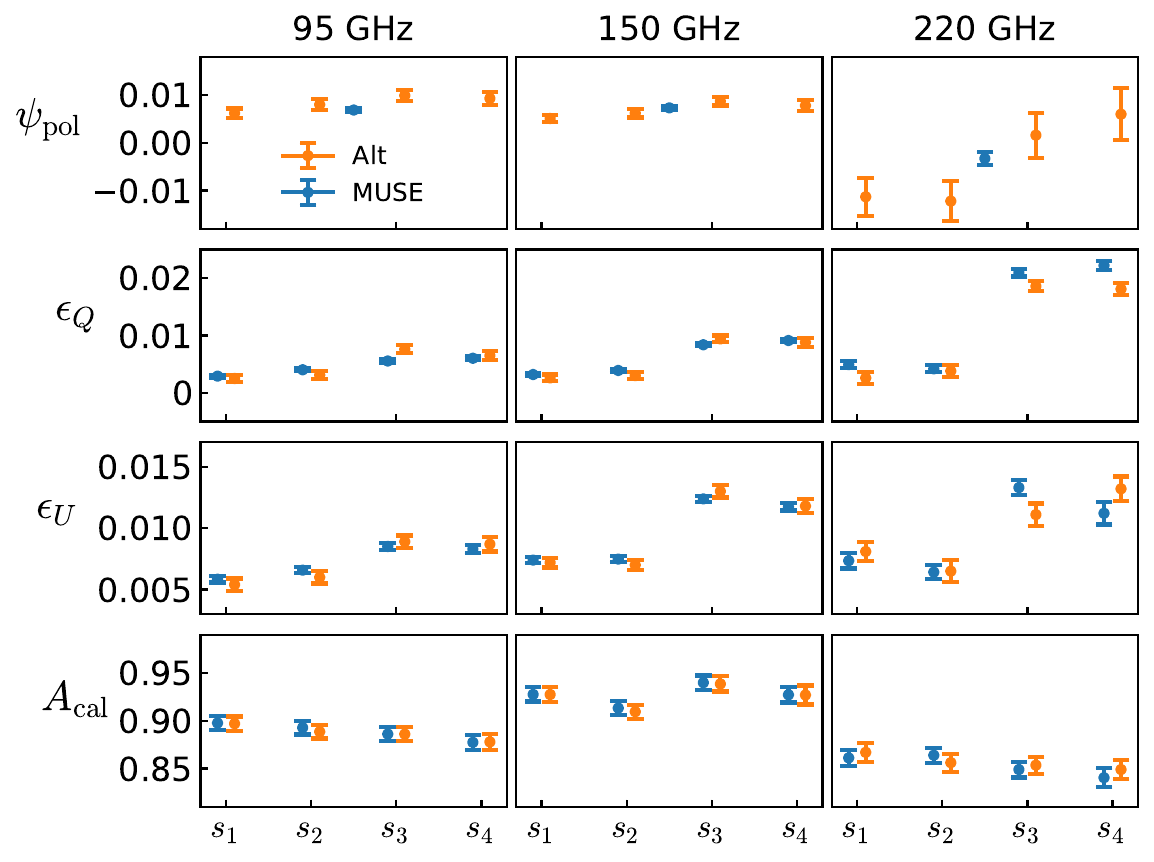}
    \caption{Agreement of polarization angle rotation angle ($\psi_{\rm pol}$), monopole T$\rightarrow$P leakage ($\epsQ$ and $\epsU$) and calibration estimates ($A_{\rm cal}$), between MUSE estimates and alternative methods which have been previously used. The groups of 4 points refer to the 4 different subfields in cases where those are estimated separately. Details of the alternative estimates are given in App.~\ref{app:model}.
    }
    \label{fig:systematics_comparison_pipelines}
\end{figure}

In Fig.~\ref{fig:systematics_comparison_pipelines}, we compare to systematic parameters derived from the alternative methods. Up to small differences expected due to slightly different choices of masking and filtering made in these alternative estimates, we find good agreement between our two sets of estimates. We also note that in general the MUSE systematic estimate is slightly tighter, likely due to simultaneously fitting for all systematics and increased optimality of the Bayesian approach. 

\begin{table}
\centering
\begin{tabular}{r|rrr}
  \toprule
  \multicolumn{4}{c}{\textbf{Posterior}} \\\midrule
  \ & {95\,GHz} & {150\,GHz} & {220\,GHz} \\\midrule
  $A_{\rm cal}^1$ & 0.8977±0.0074 & 0.9276±0.0074 & 0.8612±0.0084 \\
  $A_{\rm cal}^2$ & 0.8928±0.0072 & 0.9133±0.0071 & 0.864±0.0079 \\
  $A_{\rm cal}^3$ & 0.8861±0.0073 & 0.9399±0.0075 & 0.849±0.0083 \\
  $A_{\rm cal}^4$ & 0.8775±0.008 & 0.9272±0.0081 & 0.8405±0.0097 \\
  $100\,\epsilon_{\rm Q}^1$ & 0.291±0.025 & 0.319±0.021 & 0.492±0.059 \\
  $100\,\epsilon_{\rm Q}^2$ & 0.402±0.024 & 0.39±0.022 & 0.418±0.06 \\
  $100\,\epsilon_{\rm Q}^3$ & 0.555±0.025 & 0.838±0.021 & 2.096±0.066 \\
  $100\,\epsilon_{\rm Q}^4$ & 0.603±0.033 & 0.912±0.03 & 2.221±0.084 \\
  $100\,\epsilon_{\rm U}^1$ & 0.584±0.027 & 0.74±0.025 & 0.735±0.064 \\
  $100\,\epsilon_{\rm U}^2$ & 0.648±0.025 & 0.748±0.023 & 0.642±0.058 \\
  $100\,\epsilon_{\rm U}^3$ & 0.851±0.027 & 1.238±0.023 & 1.33±0.063 \\
  $100\,\epsilon_{\rm U}^4$ & 0.83±0.035 & 1.174±0.03 & 1.121±0.092 \\
  $\psi_{\rm pol} \, [^\circ]$ & 0.393±0.024 & 0.419±0.021 & -0.188±0.079 \\
  $\beta_{\rm pol}$ & 0.44±0.20 & 0.60±0.28 & 0.51±0.26 \\\toprule
  \multicolumn{4}{c}{\textbf{Prior}}  \\\midrule
  \ & {95\,GHz} & {150\,GHz} & {220\,GHz} \\\midrule
  $A_{\rm cal}^1$ & & 0.9275±0.0094 & \\
  $A_{\rm cal}^2$ & & 0.9095±0.0089 & \\
  $A_{\rm cal}^3$ & & 0.9386±0.0096 & \\
  $A_{\rm cal}^4$ & & 0.927±0.011   & \\
  $\beta_{\rm pol}$ & $\mathcal{U}(0,1)$ & $\mathcal{U}(0,1)$ & $\mathcal{U}(0,1)$ \\
  \bottomrule
\end{tabular}
\caption{Posteriors on systematics parameters given our all-frequency MUSE run (top section) and priors which were input (bottom section). Blank entries or parameters not listed under priors have uniform priors on $(-\infty,\infty)$. Superscripts denote each of the four subfields. Note that no cosmological model is assumed here.}
\label{tab:muse_sys}
\end{table}

\subsubsection{Higher order T-to-P leakage}

\begin{figure}
    \centering
    \includegraphics[width=\columnwidth]{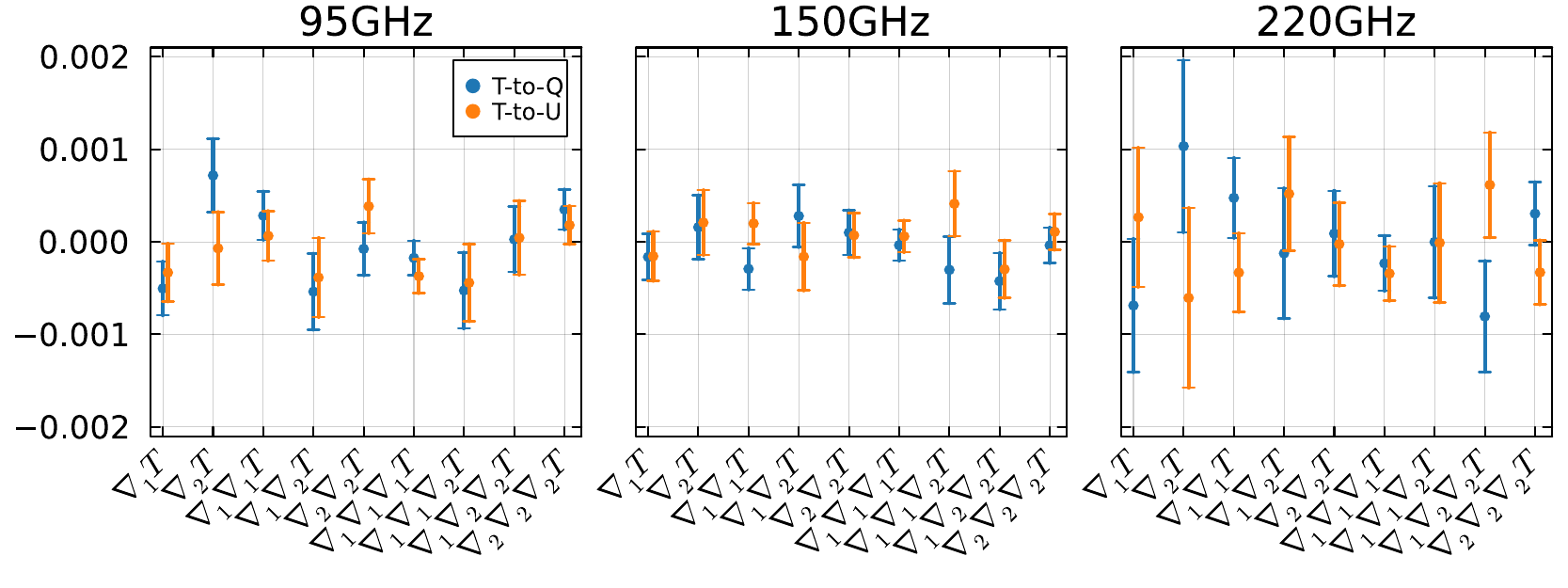}
    \caption{No evidence for dipole, quadrupole, or octopole T$\rightarrow$P leakage. These are MUSE posteriors from a run on all-frequency data that includes these extra leakage templates in addition to the other usual bandpower and systematics parameters. In our baseline run, we thus do not include such higher-order templates.\vspace{-0.2cm}}
    \label{fig:dipole_quadrupole_TP}
\end{figure}

In our baseline model, only monopole T-to-P leakage is modeled. However, it is possible that higher-order leakage exists. To verify our baseline choice, we add leakage templates up to octopole in the MUSE model and infer their leakage amplitudes. These templates corresponds to higher-order spatial gradients of the observed temperature maps. The results are shown in Fig.~\ref{fig:dipole_quadrupole_TP}, where no higher-order T-to-P leakage amplitude parameters are detected at ${>}\,3\,\sigma$.

\subsubsection{MAPs}

\begin{figure*}
    \centering
    \hspace*{0.84in}\includegraphics[width=0.9\textwidth]{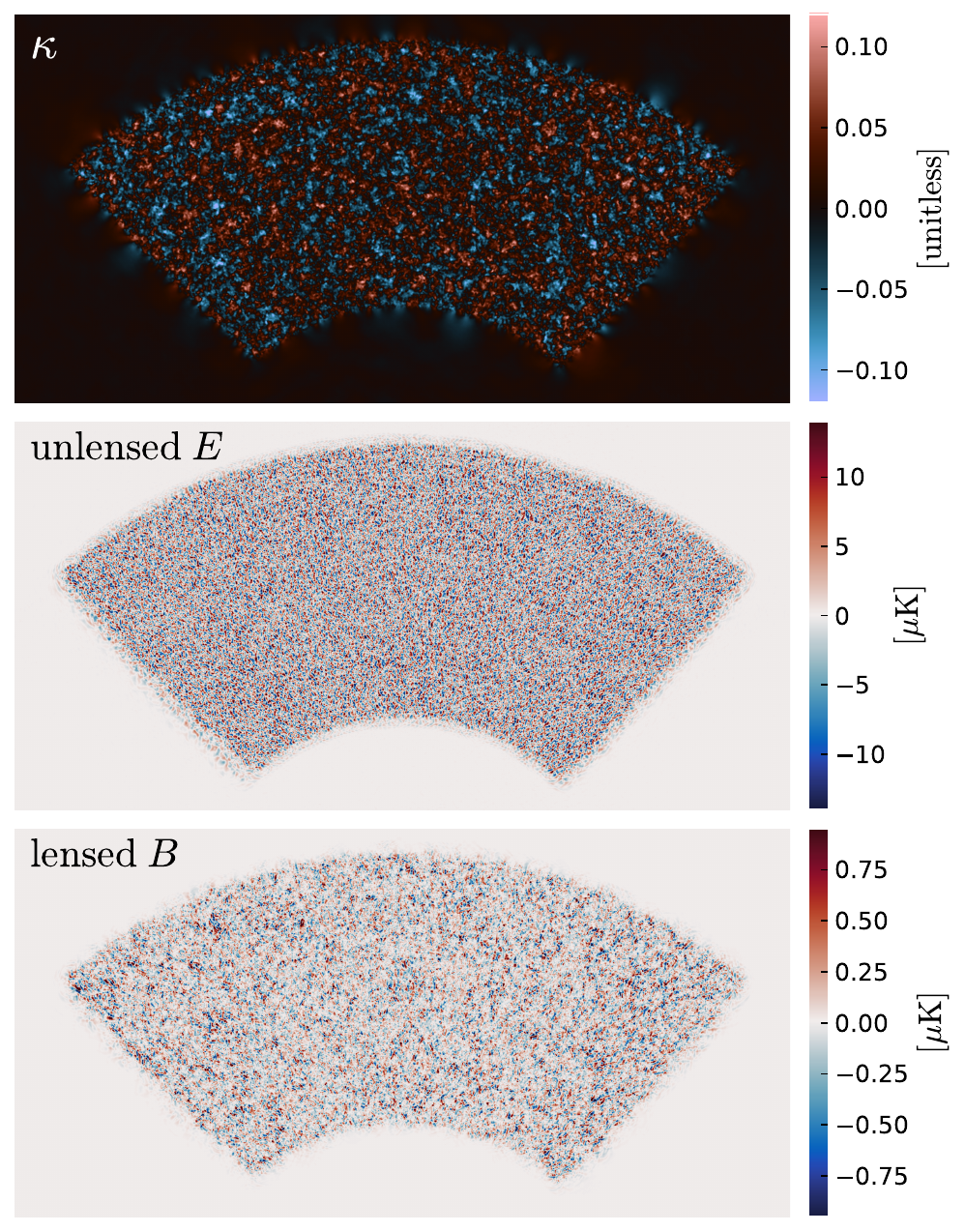}
    \caption{Maximum a posteriori (MAP) maps from 95+150+220\,GHz data at the MUSE estimate of theory and systematics parameters, $\hat \theta$. MAPs correspond to a filtering of the data which maximizes signal relative to noise (akin to a linear Wiener filter, but in the case of the MAP $\kappa$, a non-linear filter). The statistical anisotropy visible particularly in the $E$ map originates from this filtering. No other additional smoothing has been applied, although we have subtracted the simulation-mean of $\kappa$. Except for this subtraction, these MAPs correspond exactly to $\hat z$ which enters the MUSE estimate in Eqn.~\eqref{eq:smap}. We have chosen to show only unlensed $E$ and lensed $B$ {(derived from the $\kappa$ and unlensed $E$ in the top two panels)}, as lensed and unlensed $E$ look nearly identical at the scale of this figure, and unlensed $B$ is effectively zero by assumption.}
    \label{fig:MAPs}
\end{figure*}

We note that prior to unblinding we did look at MAP estimates of the lensing potential and of the lensed and unlensed CMB polarization, and thus present them in this section of the paper. These are shown in Fig.~\ref{fig:MAPs}.

\begin{figure*}
    \centering
    \includegraphics[width=\textwidth]{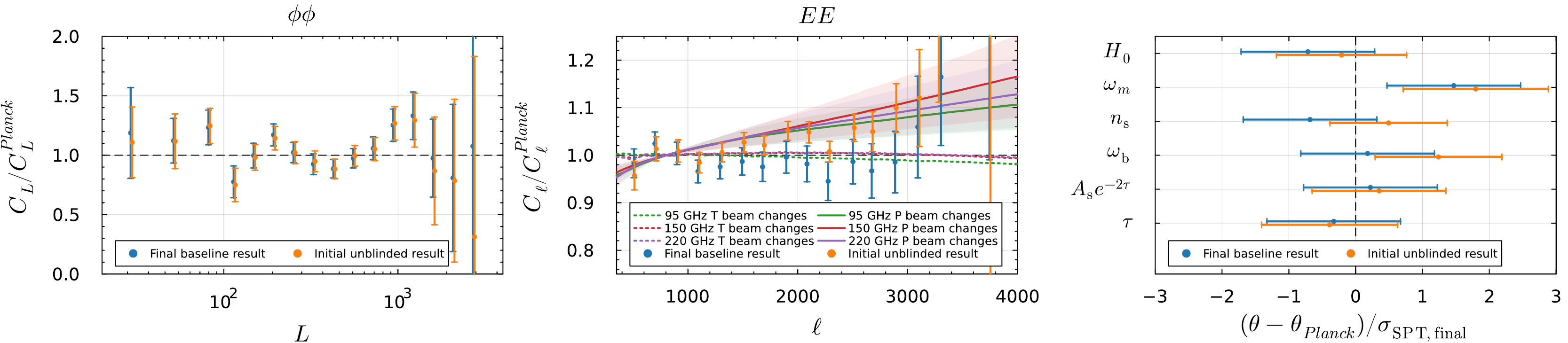}
    \caption{Changes to our results made after unblinding, and their origin in changes to beams. In the left two panels, blue and orange data points show $\phi\phi$ and unlensed $EE$ bandpowers relative to \planck \lcdm theory for our final result and our initial unblinded result, respectively. Colored lines in the $EE$ panel show ratios of temperature (dashed lines) and polarization beams (solid lines) between final and initial unblinded versions. For the polarized beam changes, we also show the posterior uncertainty from the final polarized beam fit with shaded bands indicating $1\,\sigma$. In the right panel, the same two colors show the corresponding changes to cosmological parameter results from the SPT dataset.
    \vspace{-0.4cm}}
    \label{fig:v1_v3_comparison}
\end{figure*}

\subsection{Post-unblinding changes}
\label{sec:post-unblind}

With all of these validation tests passed, we unblinded results, fit cosmological models, and compared to data from other experiments. We found \lcdm was a good fit to our own data and that our $\phi\phi$ spectrum agreed very well with other experiments. However, the $EE$ spectrum disagreed with the \planck theory prediction by a smooth slope (shown in the middle panel of Fig.~\ref{fig:v1_v3_comparison}), leading to an overall $3.5\,{\sigma}$ disagreement across 5 \lcdm parameters (excluding $\tau$) between \planck and SPT datasets. 
The slope was most significant at small scales and would have been undetectable by \planck, ACT, or previous SPT measurements, thus could have indicated a newly discovered departure from \lcdm. We were thus motivated to exhaustively check the robustness of the slope. Doing so, however, we discovered that its likely origin was an overly strong assumption we had been making in our polarization beam modeling.

Prior to unblinding, we had assumed the polarized and temperature beams were the same at the scales used in the analysis. While the temperature beam sidelobes are mapped with high signal-to-noise via observations of bright point sources and planets, we lack sufficiently bright polarized sources to similarly map the sidelobe response of the polarized beams. To check the robustness of the $EE$ slope to this assumption, we devised a model of the polarized beams to realistically account for how much these might differ from the temperature beams. Appendix~\ref{sec:beams} describes the model in more detail, but in summary, we took the polarized beams to be the sum of two components. The first is a central main lobe which can be modeled physically using the basic optics of the telescope, and is expected from physical modeling to be fully polarized and thus the same in polarization as in temperature. This model is fit to match temperature beams themselves at small scales where the model is a good description of the beam. The second component is a diffuse sidelobe stemming from effect such as scattering or reflection from of optical elements in the camera and scattering from panel gaps in the primary mirror, which is not necessarily expected to be fully polarized. We marginalize over the polarization fraction of the sidelobe component with a new parameter at each frequency, $\beta_{\rm pol}^\nu$. In this definition, our earlier assumption had corresponded to $\beta_{\rm pol}^\nu\,{=}\,1$, which makes the polarized beam identical to the temperature beam.

With these free parameters introduced, we re-ran the MUSE fit, finding $\beta_{\rm pol}^\nu\,{=}\,[0.44 \pm 0.20, 0.60 \pm 0.28, 0.51 \pm 0.26]$ at [95, 150, 220]\,GHz. That is, our CMB data alone is able to internally constrain the sidelobe polarization fractions, and it constrains them away from our previously assumed value of unity with mild significance. The ability to constrain these parameters at all arises from inter-frequency differences, as changing the $\beta_{\rm pol}^\nu$ changes the beam at each frequency differently. Indeed, if we re-examine the inter-frequency agreement of our data with the  $\beta_{\rm pol}^\nu$ marginalized, we find slightly improved agreement. This is shown as the orange line in Fig.~\ref{fig:data_inter_freq}. Thus, even though our original blind inter-frequency tests did not show any statistically significant evidence for disagreement, there was indeed some small systematic disagreement hidden below the noise, which is alleviated by this improved beam modeling. 

Regarding the slope in $EE$ relative to \planck \lcdm, the change from $\beta_{\rm pol}^\nu\,{=}\,1$ to the new marginalized posterior values largely removes it. It also increases our $EE$ uncertainties at $\ell\,{>}\,2000$ by as much as a factor of two for some bins, which, however, is highly correlated between bins, with the effect much like an additional relative calibration between high-$\ell$ and low-$\ell$. The impact to our $\phi\phi$ spectrum of this post-unblinding change is negligible, as optimal lensing estimation is largely insensitive to smooth changes to beams. Since much of our cosmological information is coming from lensing, it also means that our \lcdm constraints from SPT data are not that strongly affected; for example we originally found $H_0\,{=}\,67.3\,{\pm}\,0.77$ with the sidelobe polarization fraction fixed to unity and $H_0\,{=}\,66.81\,{\pm}\,0.81$ with it marginalized. A larger impact occurs in the $(n_{\rm s},\Omega_{\rm b}h^2)$ plane, with both parameters shifting just over 1\,$\sigma$ and the posteriors broadening. These changes are all summarized in Fig.~\ref{fig:v1_v3_comparison}. Our final agreement with \planck across 5 \lcdm parameters is $0.5\,{\sigma}$, with the change roughly coming from equal parts central value shifts and posterior broadening in the $(n_{\rm s},\Omega_{\rm b}h^2)$ plane.

We consider this beam model sufficient to capture our polarized beam uncertainty because we find no significant shifts in results if we change the main lobe model to a simple Gaussian, suggesting the exact details of the main lobe modeling are not relevant. Additionally, allowing for a free linear dependence of the sidelobe polarization fraction as a function of angular distance from the peak response does not significantly shift results either, suggesting the model of a constant sidelobe polarization fraction is sufficient. 

With this improvement to the polarized beam modeling, some additional small tweaks mentioned in App.~\ref{app:model} which lead to insignificant changes, and all validation tests passing at the same or better levels than they were prior to unblinding, we adopted this as our final baseline model. All results we will now present in the following section utilize this baseline model.

% \clearpage

\begin{table*}[th!]
\centering
\begin{tabular}{l|l}
\toprule
\textbf{Name} & \textbf{Data Set} \\
\midrule
{SPT} & This work, i.e. SPT-3G 2019/2020 \footnote{The likelihood can be found in \url{https://pole.uchicago.edu/public/data/ge25/index.html} \cite{MUSE2yr}. We refer to this data in the figures as "2yr-main"; the "main" field is the 1500\,deg$^2$ field as defined in \citep{SPT-3G:2024qkd}.} unlensed $EE$ + optimal $\phi\phi$ from 95+150+220\,GHz\\
{SPT$\phi\phi$} & Just the $\phi\phi$ spectrum from the SPT dataset, with systematics and unlensed $EE$ still marginalized over\\
{PlanckT\&E} & \planck 2018 high-$\ell$  $TT,TE,EE$ + low-$\ell$ $TT$ \citep{Planck:2019nip}\\
{Planck$\phi\phi$} & \planck NPIPE PR4 $\phi\phi$\footnote{\url{https://github.com/carronj/planck_PR4_lensing}} \ \citep{Carron:2022eyg}\\
{Planck} & {PlanckT\&E} + {Planck$\phi\phi$} \\
{ACT$\phi\phi$} &  ACT DR6 $\phi\phi$\footnote{\url{https://github.com/ACTCollaboration/act_dr6_lenslike}} \ \citep{ACT:2023kun, ACT:2023dou}\\
{ACT} & ACT DR4 $TT,TE,EE$\footnote{\url{https://github.com/ACTCollaboration/pyactlike}} \citep{aiola2020, ACT:2020frw} + {ACT$\phi\phi$}\\
{WMAP} & WMAP9\footnote{\url{https://github.com/HTJense/pyWMAP}} \ ($\ell_{\rm min}^{\rm TE}=24$, i.e. no $\tau$ information) \citep{Hinshaw2013}\\
{BAO} & 6dF \citep{beutler20116df} + SDSS DR7 MGS \citep{Ross:2014qpa} + SDSS DR12 LRG \citep{BOSS:2016wmc}
+ SDSS DR16 Ly$\alpha$ \citep{eBOSS:2020yzd} + DESI \citep{DESI:2024mwx,DESI:2024lzq, DESI:2024uvr} \\
\bottomrule
\end{tabular}
\caption{Glossary of data set abbreviations used in the text and in figure captions. When combining SPT with WMAP and \planck data, we sum the likelihood directly assuming no correlation, an assumption justified by the small sky area of the SPT \texttt{main} field (1500\,deg$^2$). When combining ACT DR4 with WMAP and \planck data, we followed the procedure in \citep{aiola2020}. When using both \planck NPIPE PR4 lensing and ACT DR6 lensing, we used the built-in likelihood implementation of \texttt{act\_dr6\_lenslike} package with \texttt{variant=actplanck\_baseline} which accounts for correlations between ACT and Planck lensing due to overlapping sky coverage.\footnote{{When combining SPT with other CMB datasets, we ignore any error correlations. We expect these to be very small for two reasons. First, the relatively small area of the SPT-Main survey (at 3.5\% of the full sky) means that the ACT and Planck errors predominantly arise from sources of fluctuation which are not in the SPT-Main survey region. Second, for the lensing likelihood specifically, SPT lensing bandpowers are reconstructed from CMB polarization only, while other CMB lensing reconstructions are predominantly informed by CMB temperature maps.}} For the BAO dataset, we followed the proposal in Sect.~3.3 of \citet{DESI:2024mwx} to replace the DESI BGS and lowest redshift LRG measurements with SDSS MGS ($z_{\rm eff}\,{\sim}\,0.15$) and two SDSS DR12 LRG points ($z_{\rm eff}\,{\sim}\,0.38,0.51$), as well as replacing the DESI Ly$\alpha$ point with joint eBOSS+DESI Ly$\alpha$ BAO. With the exception of Sec.~\ref{sec:Excess}, any CMB dataset combinations we use includes a prior on $\tau$ of $0.051\,{\pm}\,0.006$ \citep{Planck:2020olo}. In Sec.~\ref{sec:Excess}, we instead drop the $\tau$ prior and PlanckT\&E is to be understood as including the Planck low-$\ell$ $EE$ likelihood.}
\label{table:dataset}
\end{table*}

\section{Results}
\label{sec:results}

In this section we present our $\phi\phi$ and unlensed $EE$ bandpowers and resulting cosmological parameter constraints in various model spaces, with and without inclusion of some external datasets. Nomenclature for the external datasets, as well as our SPT-3G dataset, is described in Tab.~\ref{table:dataset}. We follow the convention of naming cosmological parameters in \citep{Planck:2013pxb} (see Table.~1 there for detail). The setup for MCMC analysis is described in Sect.~\ref{sec:cosmo_inference}. For simplicity, we adopt a Gaussian prior, $\tau\,{\sim}\,\mathcal{N}(0.051, 0.006)$, using the result from the joint processing of \planck HFI and LFI data \citep{Planck:2020olo} as baseline unless otherwise specified, instead of using \planck low-$\ell$ $EE$ likelihood or WMAP TE data at $\ell\,{<}\,24$. Unless otherwise specified, all uncertainties quoted in this paper represent the 68\% confidence interval.

\subsection{Bandpowers}

\begin{figure*}[tph!]
    \centering
    \includegraphics[width=\textwidth]{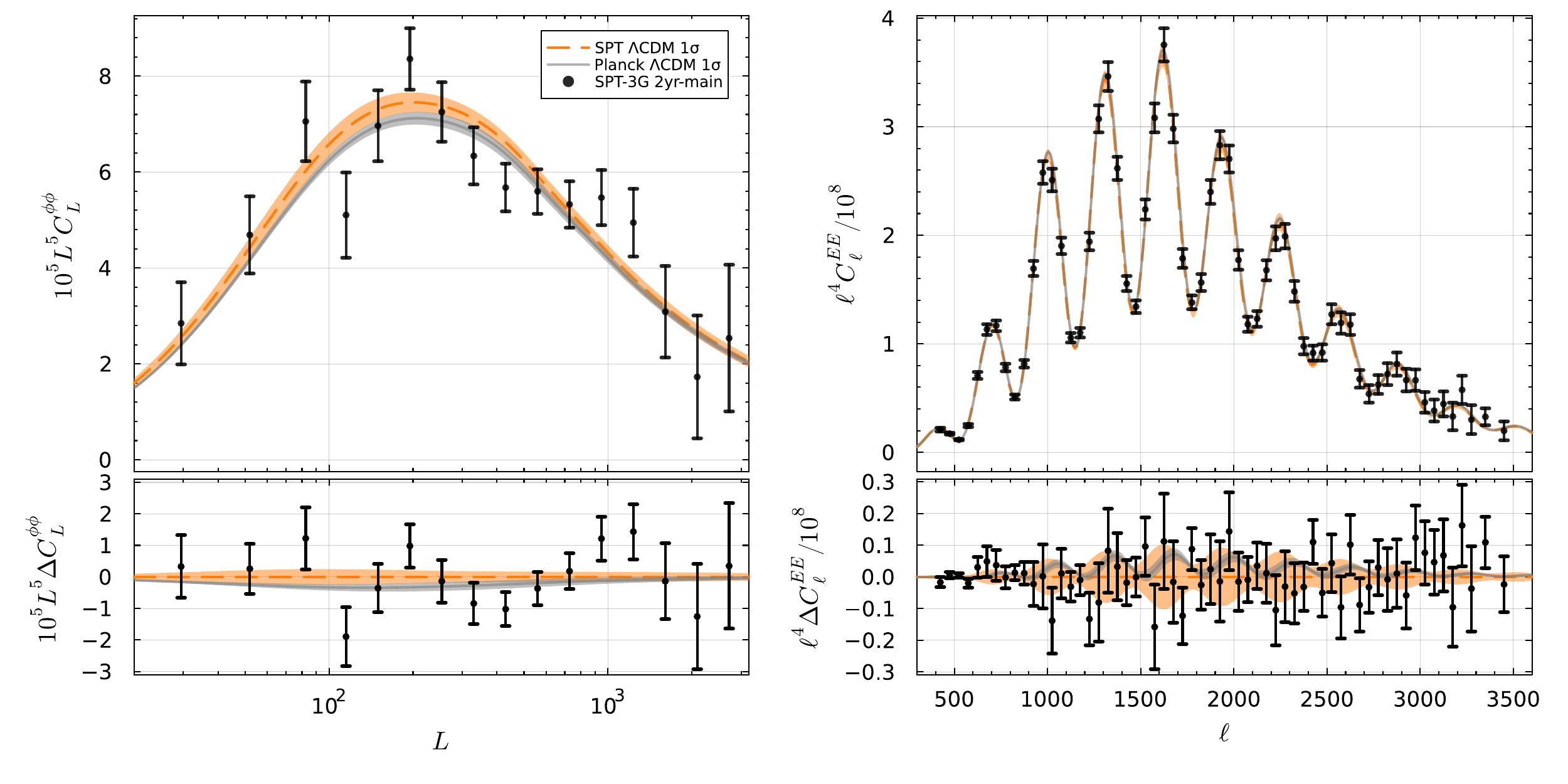}
    \caption{{\it Top panels}: MUSE measurement of the CMB lensing $\phi\phi$ bandpowers and the unlensed $EE$ bandpowers using \mbox{SPT-3G} 2019/20 CMB polarization data. The lensing bandpowers represent a $38\,\sigma$ detection of lensing power. These are the most precise measurements to date at $\ell\,{>}\,2000$ for $EE$ and $L\,{>}\,350$ for $\phi\phi$. {\it Bottom panels}: Residual bandpowers with respect to the SPT best-fit cosmology. The gray and orange bands show the 68\% confidence region of \lcdm model spectra fit to \planck and SPT data, respectively.}
    \label{fig:spectra}
\end{figure*}

In Fig.~\ref{fig:spectra}, we show our inference of $\phi\phi$ and unlensed $EE$ bandpowers. The \lcdm model that best fits these data provides a good fit with $\chi^2\,{=}\,75.8$ given 71 degrees of freedom (76 bandpowers minus 5 cosmological parameters constrained beyond their prior), which corresponds to a PTE of $32.7\%$. By this test, the SPT data are consistent with the standard cosmological model. 
We also show as an orange band the 68\% confidence level predictions of the \lcdm model for the CMB lensing and unlensed $EE$ power spectra given these SPT data. 

\begin{figure*}[bph!]
    \centering
    \includegraphics[width=\textwidth]{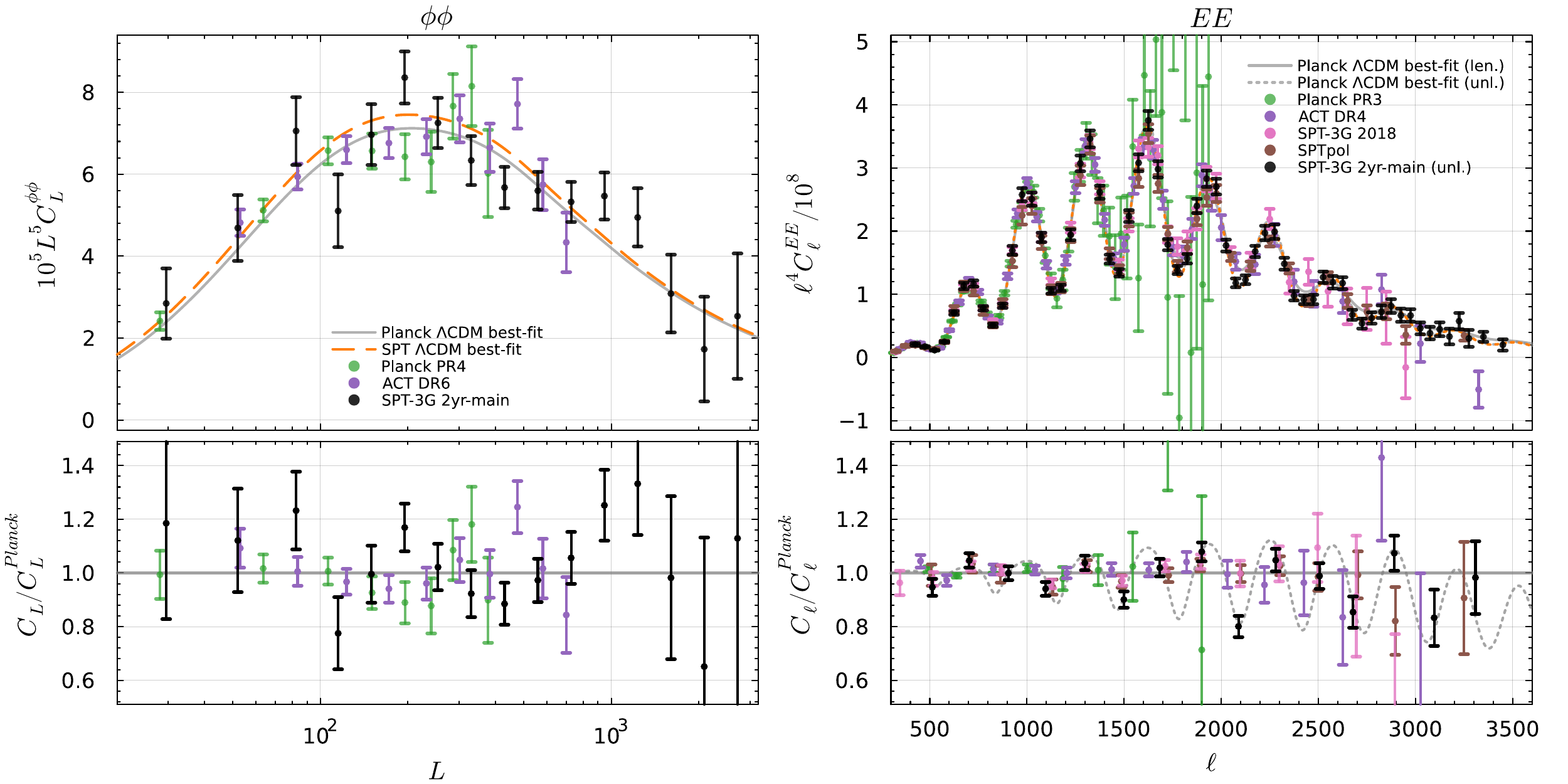}
    \caption{Comparison of SPT bandpowers with existing measurements. {\it Top panels}: CMB $\phi\phi$ and $EE$ bandpowers. {\it Bottom panels}: Bandpowers divided by the Planck best-fit cosmology predictions. Note that the $EE$ bandpowers have been further rebinned as compared to the top panel for clarity. As a reminder, the SPT $EE$ points are inferences of unlensed $EE$. The dotted gray line shows the unlensed $EE$ model prediction. The orange dashed and the solid gray lines are the same as Fig.~\ref{fig:spectra}.}
    \label{fig:bandpowers_muse_vs_others}
\end{figure*}

We can also check if the SPT data are consistent with the \lcdm predictions given Planck data. The 68\% confidence level predictions of the \lcdm model given Planck data are shown as a gray band, with the best-fit \lcdm model given Planck data shown as a black line.   
The $\chi^2$ of the SPT bandpowers given the Planck best-fit \lcdm model is $80.7$, which corresponds to a PTE of $20.2\%$. Thus, the SPT bandpowers agree with the Planck \lcdm best-fit prediction at the $1.3\,\sigma$ level.

In Fig.~\ref{fig:bandpowers_muse_vs_others}, we compare the SPT bandpowers to prior measurements. 
The SPT lensing bandpower measurements extend to $L\,{\sim}\,3000$ and are more precise than prior published measurements at $L\,{>}\,350$.
The unlensed $EE$ bandpowers extend to $\ell\,{\sim}\,3500$ and are more precise than prior published measurements at $\ell\,{>}\,2000$ where we have clear detections of the eighth, ninth and tenth acoustic peaks.

\begin{figure*}
    \centering
    \vspace{1cm} % force onto own page
    \includegraphics[width=\textwidth]{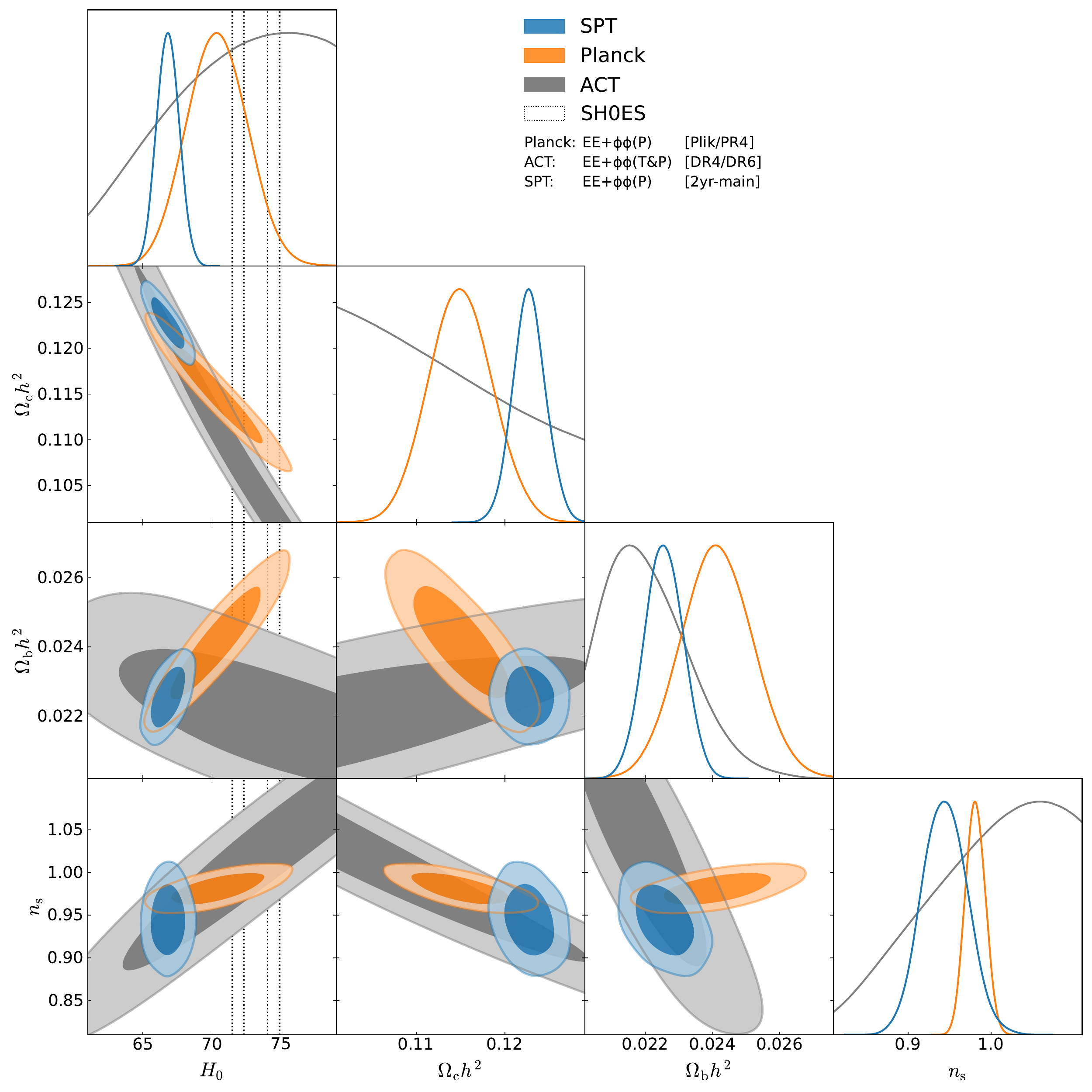}
    \caption{Posterior distributions of \lcdm cosmological parameters derived only from polarization $EE$ spectra and polarization-based lensing reconstruction, or in the case of ACT\footnote{{For the ACT DR4 EE likelihood, we adopted their default prior for \texttt{$yp2\sim[0.5,1.5]$}.}}, also including temperature in the lensing reconstruction. We see tighter constraints on $\Omega_b h^2$, $\Omega_c h^2$, and $H_0$ from the SPT data. We show the SH0ES constraint on $H_0$ for comparison. See also Tab.~\ref{table:dataset} for more details on dataset definitions.}
    \vspace{1cm}
    \label{fig:lcdm_ponly}
\end{figure*}

\subsection{Origins of $\Lambda$CDM parameter constraints}

The SPT bandpowers represent a significant advance in our capacity to constrain cosmological parameters with polarization data alone. In Fig.~\ref{fig:lcdm_ponly} we show constraints on four \lcdm parameters from SPT, ACT, and Planck, assuming the \lcdm model, and limiting to only $EE$ spectra and polarization-based lensing reconstruction, or in the case of ACT, also including temperature in the lensing reconstruction. We see significantly tighter constraints on three of the parameters from SPT, while the higher dynamic range of the Planck $EE$ data (which extends to lower $\ell$ than shown in Fig.~\ref{fig:bandpowers_muse_vs_others}) leads to a tighter constraint on $n_{\rm s}$ for Planck.

Constraints on \lcdm parameters from SPT data and other dataset combinations, now including both temperature and polarization as detailed in Tab.~\ref{table:dataset}, are summarized in Tab.~\ref{table:lcdm_full}. The SPT constraints are comparable to full \planck results in the matter density, $\Omega_{\rm m}h^2$, and angular scale of the sound-horizon, $\theta_\star$. The constraining power on $\Omega_{\rm m}h^2$ from SPT is mainly due to the CMB lensing bandpowers, which are sensitive to the amount of matter in the Universe. This is evidenced by the fact that adding just SPT lensing to WMAP data reduces the error on $\Omega_{\rm m}h^2$ by over a factor of 3 from 0.0045 \citep{Hinshaw2013} to 0.0014, and that further adding SPT $EE$ to this combination leaves the error unchanged. 

That the sensitivity to $\Omega_{\rm m} h^2$ primarily comes from the lensing reconstruction rather than primary power spectra (in this case unlensed $EE$) is a unique feature of these data compared to prior CMB datasets. The sensitivity of primary power spectra to matter density is weakened due to its weighting toward small scales that are sensitive to conditions at horizon crossing; at such early times the matter density is negligible compared to the radiation density. Conversely, the lensing power spectrum, if probed precisely enough, has high sensitivity to $\Omega_{\rm m} h^2$. This comes mainly (but not solely) from how CMB lensing power depends on the scale of matter-radiation equality.  Additionally, the sensitivity increases at the small scales probed here, with $C_L^{\phi \phi}\,{\propto}\,(\Omega_{\rm m} h^2)^2$ at $L\,{\sim}\,200$ and asymptoting to $(\Omega_{\rm m} h^2)^3$ by $L\,{\sim}\,2000$ \citep{Pan:2014xua}. 

The SPT data are also sensitive to the angular scale of the sound horizon, $\theta_\star$. This sensitivity comes from the $EE$ bandpowers, where the peak and trough locations of the $EE$ spectrum are highly sensitive to $\theta_\star$. The importance of the $EE$ bandpowers for this constraint, as opposed to the lensing bandpowers, can be seen in the comparison of $\theta_\star$ constraints from SPT, WMAP+SPT and WMAP+SPT$\phi\phi$; the exclusion of SPT $EE$ significantly degrades the $\theta_\star$ determination. We note that the constraint on $\theta_\star$ is known to be improved by delensing \citep{green2016}, a result which applies to our unlensed spectra inferences here as well, although we do not explicitly attempt to quantify exactly how much improvement we have achieved. 

The constraints on $\tau$ from the SPT dataset are driven almost entirely by Planck low-$\ell$ polarization data, included here as a $\tau$ prior. Although lensing can, in theory, help constrain $\tau$ via degeneracy breaking, the contribution at current noise levels is negligible. The constraint on $\log(10^{10} A_\mathrm{s})$ is similarly largely driven by \planck, since it is degenerate both with $\tau$ and the absolute calibration, the latter which also comes from \planck.

The SPT constraints on $\Omega_{\rm b}h^2$ and $n_{\rm s}$ are weaker than those inferred from \planck. Both $n_{\rm s}$ and $\Omega_{\rm b}h^2$ cause tilts of the CMB $EE$ spectrum, since $\Omega_{\rm b}h^2$ affects the photon diffusion damping at small angular scales and $n_{\rm s}$ affects the shape of the primordial spectrum. Our observations lack the very largest scales probed by \planck, which, logarithmically, contain a significant range of scales useful for constraining a tilt. Additionally, at small scales our uncertainties are increased by polarized beam uncertainty, preventing what could otherwise be a slightly better constraint. Our final result is a 2.5\% determination of the the baryon density.

With the baryon density, matter density, and $\theta_\star$ determined, one has everything necessary, in \lcdm, for calculating both $H_0$ and $\Omega_{\rm m}$ \citep[e.g.][]{Knox:2019rjx}. Adding in $A_{\rm s}$ and $n_{\rm s}$ one can also calculate $\sigma_8$ and $S_8$. We show constraints on these four derived parameters in Tab.~\ref{table:lcdm_full}. 

\subsection{Agreement with other CMB datasets in \lcdm}

Before providing interpretations, it is useful to establish the consistency level of our \lcdm parameter results with other CMB measurements. Such comparison provides tests of the \lcdm model that are complementary to and stronger than the bandpower-level tests above. 

We perform a simple consistency check applicable to uncorrelated datasets by calculating the $\chi^2$ between the two sets of parameters,
\begin{equation}
    \chi^2 = (\vec{x}_1-\vec{x}_2)^T(C_1 + C_2)^{-1}(\vec{x}_1-\vec{x}_2),
    \label{eq:agreement}
\end{equation}
where $\vec{x}$ is a vector of parameter estimates, $C$ are the parameter covariances from each, and the subscripts 1 \& 2 distinguish datasets. We then compute the PTE for the given $\chi^2$ and quote a 1-tailed conversion to number of $\sigma$. Since, as discussed, our estimates of $\tau$ and $A_s$ are highly correlated to Planck, we consider only $\vec{x}\,{=}\,(\theta_\star, \Omega_{\rm b}h^2, \Omega_{\rm c}h^2, n_{\rm s})$. For these parameters, the SPT results are very close to independent of both Planck and ACT due to non-overlapping sky area and multipole ranges. We find \lcdm parameters from SPT and Planck datasets are consistent at $0.8\,\sigma$, and from SPT and ACT datasets at $0.8\,\sigma$. 

\begin{figure*}
    \centering
    \vspace{1cm} % force onto own page
    \includegraphics[width=\textwidth]{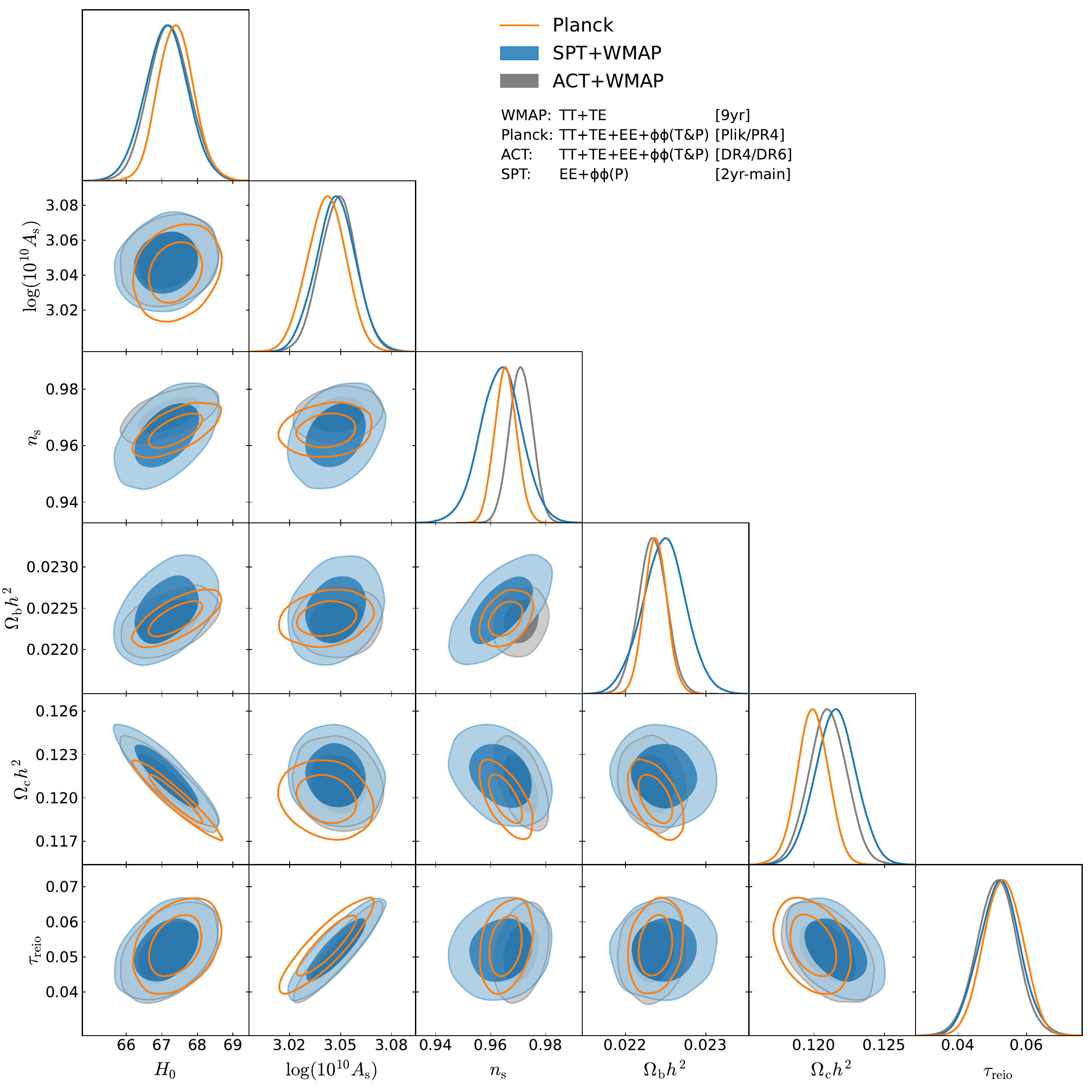}
    \caption{Posterior distributions of \lcdm cosmological parameters derived from SPT+WMAP, ACT+WMAP, and Planck. The details of the datasets are listed in Tab.~\ref{table:dataset}. We see that either ACT or SPT, when combined with WMAP for constraints on larger angular scales, achieve constraints on cosmological parameters with similar constraining power as the constraints from Planck.}
    \vspace{1cm}
    \label{fig:lcdm_triangle}
\end{figure*}

A triangle plot comparing marginal posterior distributions from SPT+WMAP, ACT+WMAP, and Planck is shown in Fig.~\ref{fig:lcdm_triangle}.

\begin{table*}
    \centering
    \begin{tabular}{c|ccccc}
\toprule
    Parameters &       SPT &      Planck &   Planck+SPT &    WMAP+SPT  & WMAP+SPT$\phi\phi$ \\
\midrule

$\log(10^{10} A_\mathrm{s})$ & $3.045\,{\pm}\,0.024$     & $3.042\,{\pm}\,0.011$     & $3.046\,{\pm}\,0.011$     & $3.047\,{\pm}\,0.012$     & $3.042\,{\pm}\,0.013$     \\
 $n_\mathrm{s}$               & $0.944\,{\pm}\,0.027$     & $0.965\,{\pm}\,0.004$     & $0.9647\,{\pm}\,0.0037$   & $0.9637\,{\pm}\,0.0076$   & $0.961\,{\pm}\,0.011$     \\
 $100\theta_\mathrm{MC}$      & $1.04162\,{\pm}\,0.00066$ & $1.04092\,{\pm}\,0.00031$ & $1.04104\,{\pm}\,0.00028$ & $1.04154\,{\pm}\,0.00057$ & $1.0394\,{\pm}\,0.0022$   \\
 $\Omega_\mathrm{b} h^2$      & $0.02255\,{\pm}\,0.00057$ & $0.02237\,{\pm}\,0.00014$ & $0.02239\,{\pm}\,0.00013$ & $0.02248\,{\pm}\,0.00027$ & $0.02226\,{\pm}\,0.00045$ \\
 $\Omega_\mathrm{c} h^2$      & $0.1227\,{\pm}\,0.0018$   & $0.1199\,{\pm}\,0.0011$   & $0.12039\,{\pm}\,0.00094$ & $0.1215\,{\pm}\,0.0014$   & $0.1210\,{\pm}\,0.0014$   \\
 $\tau_\mathrm{reio}$         & $0.052\,{\pm}\,0.006$     & $0.0532\,{\pm}\,0.0056$   & $0.0547\,{\pm}\,0.0058$   & $0.052\,{\pm}\,0.006$     & $0.0515\,{\pm}\,0.0059$   \\
 \midrule
 $H_0$ [km/s/Mpc]                       & $66.81\,{\pm}\,0.81$      & $67.4\,{\pm}\,0.5$        & $67.28\,{\pm}\,0.42$      & $67.1\,{\pm}\,0.6$        & $66.4\,{\pm}\,1.1$        \\
 $100\theta_\star$                   & $1.04181\,{\pm}\,0.00069$ & $1.0411\,{\pm}\,0.0003$   & $1.04125\,{\pm}\,0.00027$ & $1.04173\,{\pm}\,0.00057$ & $1.0396\,{\pm}\,0.0022$   \\
 $\Omega_\mathrm{m} h^2$      & $0.1459\,{\pm}\,0.0019$   & $0.143\,{\pm}\,0.001$     & $0.14340\,{\pm}\,0.00089$ & $0.1446\,{\pm}\,0.0015$   & $0.1439\,{\pm}\,0.0015$   \\
 $\Omega_\mathrm{m}$          & $0.327\,{\pm}\,0.011$     & $0.3147\,{\pm}\,0.0068$   & $0.3169\,{\pm}\,0.0058$   & $0.3211\,{\pm}\,0.0085$   & $0.326\,{\pm}\,0.011$     \\
 $\sigma_8$                   & $0.8143\,{\pm}\,0.0088$   & $0.8104\,{\pm}\,0.0051$   & $0.8137\,{\pm}\,0.0046$   & $0.8177\,{\pm}\,0.0065$   & $0.8118\,{\pm}\,0.0086$   \\
 $S_8=\sigma_8\sqrt{\Omega_m/0.3}$     & $0.850\,{\pm}\,0.017$     & $0.830\,{\pm}\,0.012$     & $0.8363\,{\pm}\,0.0097$   & $0.846\,{\pm}\,0.016$     & $0.847\,{\pm}\,0.016$     \\
\bottomrule
\end{tabular}
\caption{Parameter estimates for various datasets assuming \lcdm. The datasets are defined in Tab.~\ref{table:dataset}.}
\label{table:lcdm_full}
\end{table*}

We now turn to estimates of three parameters of particular contemporary interest in cosmology: $H_0$, $S_8$, and $\Omega_{\rm m}$.

\subsection{The Hubble constant, $H_0$}
\begin{figure}[tbph!]
    \centering
    \includegraphics[width=\columnwidth]{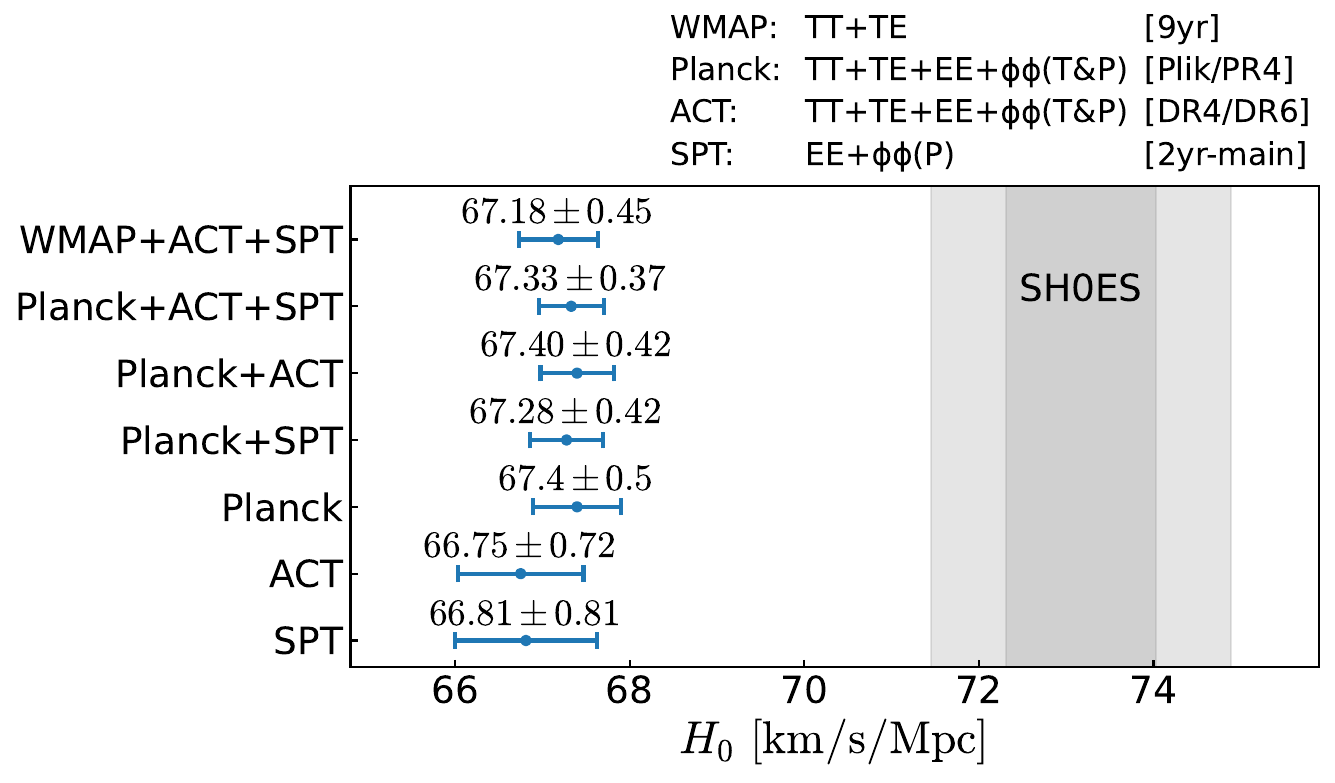}
    \caption{Comparison of the \lcdm constraints on $H_0$ inferred from different CMB observations with the Cepheid-calibrated SNe Ia observations of the SH0ES collaboration \citep{Breuval:2024lsv,Riess:2021jrx}. All CMB \lcdm constraints are significantly lower.}
    \label{fig:lcdm_h0}
\end{figure}

The Hubble constant inferred from \planck\ data, assuming the \lcdm model, has been found to be in tension with the most precise "distance ladder" measurement \citep{Breuval:2024lsv, Riess:2021jrx}, which uses Cepheid-calibrated SNe Ia, at the $5.8\,\sigma$ level.\footnote{See also, e.g. \citet{Freedman:2023jcz, Freedman:2024eph, Riess:2024vfa} for comparison of alternative "rung 2" methods with the Cepheid method.} In Fig.~\ref{fig:lcdm_h0}, we show the Hubble constant inferred from various observations. Using the SPT data alone, we report 
\begin{equation}
    H_0 = 66.81\,{\pm}\,0.81 \ \mathrm{km/s/Mpc}.
\end{equation}
This is consistent with the Hubble constant inferred from Planck, and is in $5.4\,\sigma$ tension with \cite{Breuval:2024lsv}. This confirms previous hints of a tension given only CMB polarization data, where Planck, ACTpol, and SPTpol yielded $H_0\,{=}\,68.7\,{\pm}\,1.3$ \citep{addison2021}, but almost halves the error bar and increases the significance of the tension from less than $3\,\sigma$ to greater than $5\,\sigma$. 

The agreement with previous CMB constraints, that rely heavily on temperature data, is also significant as a passed test of \lcdm. Of course, the continued disagreement with distance ladder methods remains a mystery. We expect that these new SPT data will be a significant hurdle for many of the proposed models \citep{DiValentino:2021izs}, and models yet to come, that address the Hubble tension. 

Combining Planck, ACT, and SPT, we report 
\begin{equation}
    H_0 = 67.33\,{\pm}\,0.37 \ \mathrm{km/s/Mpc},
\end{equation}
which results in tension with the $H_0$ reported in \cite{Breuval:2024lsv} at ${6.2}\,\sigma$.
This is the tightest constraint on $H_0$ inferred from CMB observations assuming the \lcdm model. We also note that the combination WMAP+ACT+SPT achieves $H_0=67.18\,{\pm}\,0.45 \ \mathrm{km/s/Mpc}$, a result with precision nearly the same as that of the \planck result and nearly independent of \planck data. Overall, inferences of the Hubble constant from a wide range of CMB observations are in good agreement. While the potential for confirmation bias to drive this agreement could be a concern, we highlight that both the unlensed $EE$ and $\phi\phi$ spectra which enter our SPT result here were obtained blindly, and while there were some changes to $H_0$ after we unblinded, these changes were small compared to the magnitude of the tension.

\subsection{The matter density, $\Omega_{\rm m}$}

\begin{figure}[tbph!]
    \centering
    \includegraphics[width=\columnwidth]{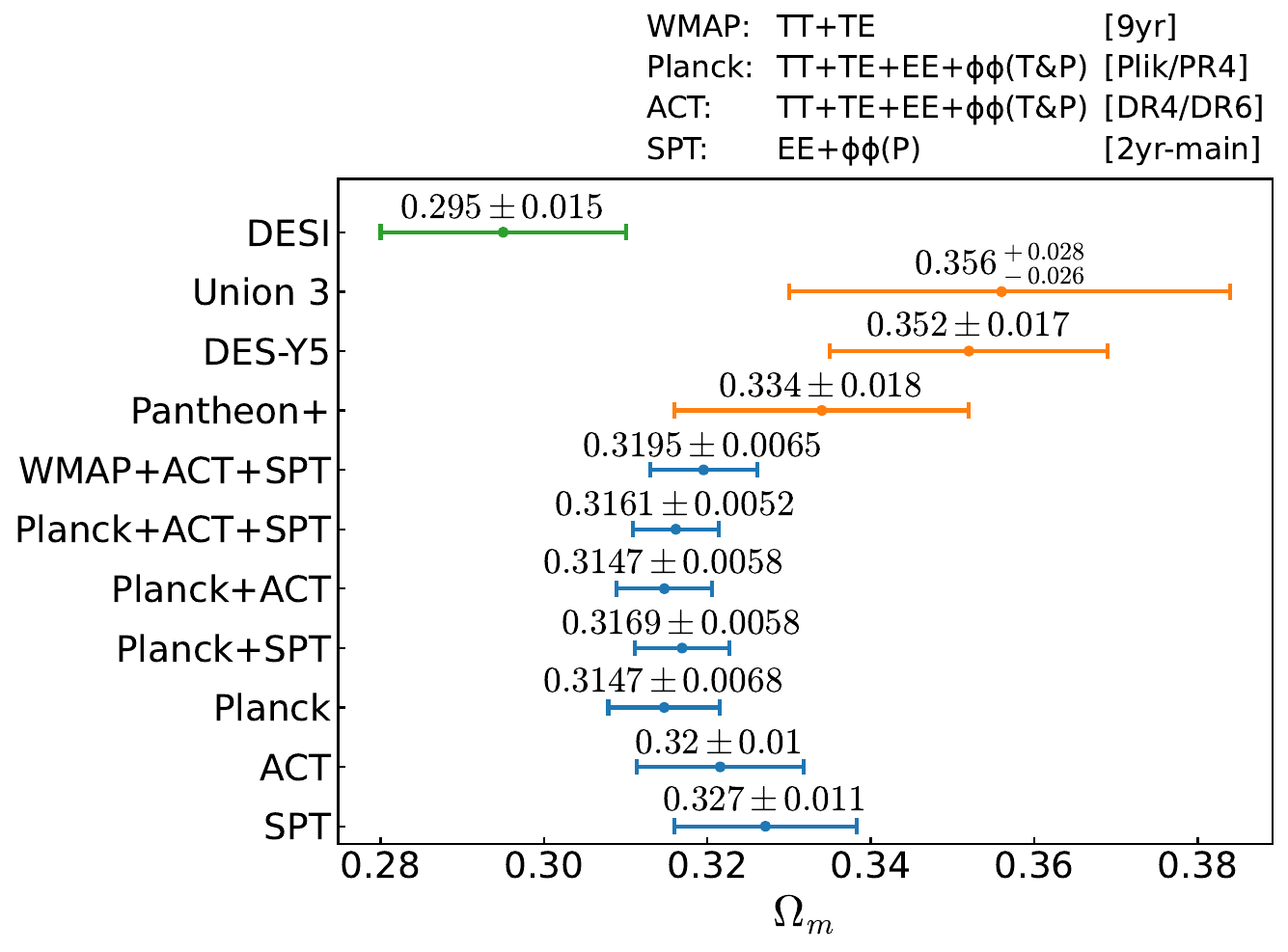}
    \caption{Comparison of the \lcdm constraints on $\Omega_{\rm m}$ inferred from different CMB observations (blue) with those from SNe observations (orange) and DESI BAO (green). The SNe observations include Pantheon+ \citep{Brout:2022vxf}, DES-Y5 \citep{Abbott:2024agi} and Union 3 \citep{Rubin:2023ovl}. The biggest discrepancy is between DESI BAO and the DES-Y5 determination of $\Omega_{\rm m}$, a $2.5\,\sigma$ difference.}
    \label{fig:lcdm_Om}
\end{figure}

There has recently emerged some degree of tension between $\Omega_{\rm m}$ as inferred from BAO, and values inferred from supernova surveys \cite{DESI:2024mwx}. We summarize the situation in Fig.~\ref{fig:lcdm_Om}. The biggest discrepancy is between DESI BAO and the DES-Y5 determination of $\Omega_{\rm m}$ which yields a $2.5\,\sigma$ difference. Inferences from CMB data, with consistency among different observations, are more precise than those from the supernovae and BAO, lying in between the low BAO value and the higher supernova values. From SPT data alone we infer
\begin{equation}
    \Omega_{\rm m}= 0.327\,{\pm}\,0.011.
\end{equation}
This is the highest central value of the inferences from different choices of CMB dataset, though consistent with all of them, and with an uncertainty only 65\% greater than the \planck uncertainty.

\subsection{The structure amplitude parameter, $S_8$}

\begin{figure}[tbph!]
    \centering
    \includegraphics[width=\columnwidth]{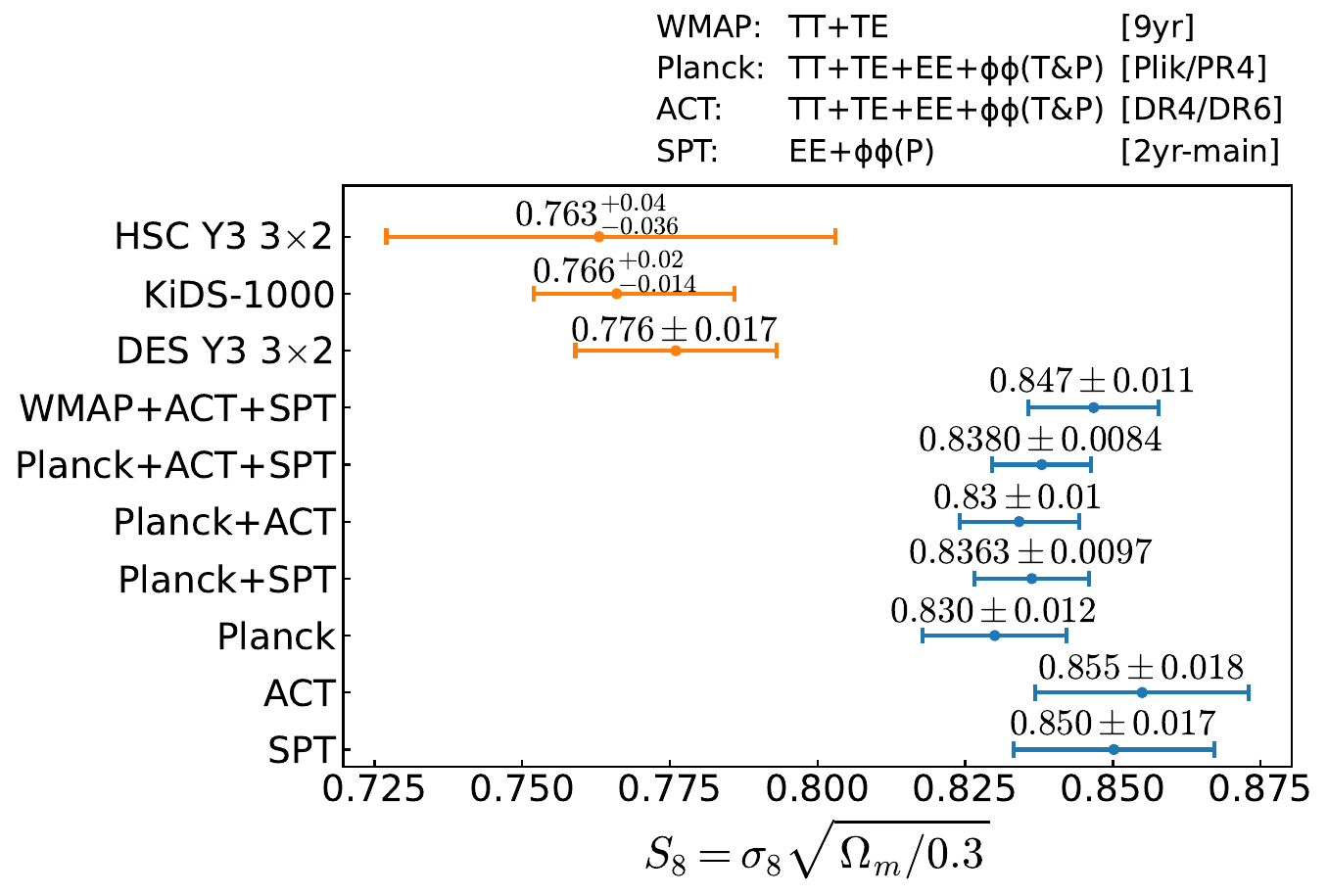}
    \caption{Comparison of the \lcdm constraints on $S_8$ inferred from different CMB observations (blue) with those from galaxy surveys of HSC Y3 , KiDs-1000 and DES-Y3 (orange). The estimate from Planck+ACT+SPT differs with the one from HSC-Y3 at $2\,\sigma$, the one from KiDs at $3.3\,\sigma$, and with the one from DES-Y3 at $3\,\sigma$.}
    \label{fig:lcdm_s8}
\end{figure}

The amplitude of structure measured from galaxy weak lensing surveys also shows mild tension with the \lcdm inference using CMB data. In Fig.~\ref{fig:lcdm_s8}, we show the value of $S_8\,{\equiv}\,\sigma_8\sqrt{\Omega_\mathrm{m}/0.3}$ inferred from various observations. From SPT, we find
\begin{equation}
    S_8 =0.850\,{\pm}\,0.017,
\end{equation}
which is in agreement with the result from Planck at a similar level of precision. Our result is in $2\,\sigma$, $3.2\,\sigma$, and $3\,\sigma$ disagreement with HSC-Y3 \citep{Miyatake:2023njf}, KiDs-1000 \citep{Heymans:2020gsg} and DES-Y3 \citep{DES:2021wwk}, respectively. From the joint fit of Planck+ACT+SPT, we get 
\begin{equation}
    S_8 = 0.8380\,{\pm}\,0.0084,
\end{equation}
which is at $1.8\,\sigma$, $3.3\,\sigma$, and $3.3\,\sigma$ tension with HSC-Y3, KiDs and DES-Y3 results.
We also note that the constraint on $S_8$ from a combination of WMAP+ACT+SPT is consistent with the $S_8$ inferred from Planck alone, which provides a nearly-independent check of the Planck result.

\subsection{Amplitude of nonlinear structure growth}

One proposed phenomenological resolution to the $S_8$ tension is a suppression of the matter power spectrum at non-linear scales, potentially due to baryonic feedback or dark matter structure growth which differs from the standard picture \citep{Amon:2022azi, Preston:2023uup}. The CMB lensing reconstruction we present in this paper is the most precise measurement of the CMB lensing spectrum at the smaller angular scales where non-linear effects are more important, and can be used to further constrain the space of possible models which resolve this tension. To do so, we introduce a new parameter, $A_{\rm mod}^{\rm CMB}$, equivalent to the $A_{\rm mod}$ parameter of \cite{Amon:2022azi} under the Limber approximation, but here measured at CMB lensing scales and redshifts. We implement this by directly scaling the non-linear correction to the linear-theory CMB lensing spectrum,
\begin{equation}\label{eq:Amod}
    C_{L}^{\phi\phi} = C_{L,\rm lin}^{\phi\phi} + A_\mathrm{mod}^{\rm CMB}(C_{L,\rm nonlin}^{\phi\phi}-C_{L,\rm lin}^{\phi\phi}),
\end{equation}
where $C_{L,\rm lin}^{\phi\phi}$ and $C_{L,\rm nonlin}^{\phi\phi}$ are the linear and non-linear CMB lensing spectra, the latter computed via our default \texttt{Halofit}. This modified lensing spectrum is then also propagated to changes in the lensed  primary spectra. 

As a small aside, we have verified that constraints on $A_{\rm mod}^{\rm CMB}$ are insensitive to the choice of nonlinear and baryonic feedback models. Given that $A_{\rm mod}^{\rm CMB}$ is the parameter most sensitive to the non-linear prediction, our constraints on other cosmological parameters in this paper are insensitive as well.

\begin{figure}
    \centering
    \includegraphics[width=0.8\columnwidth]{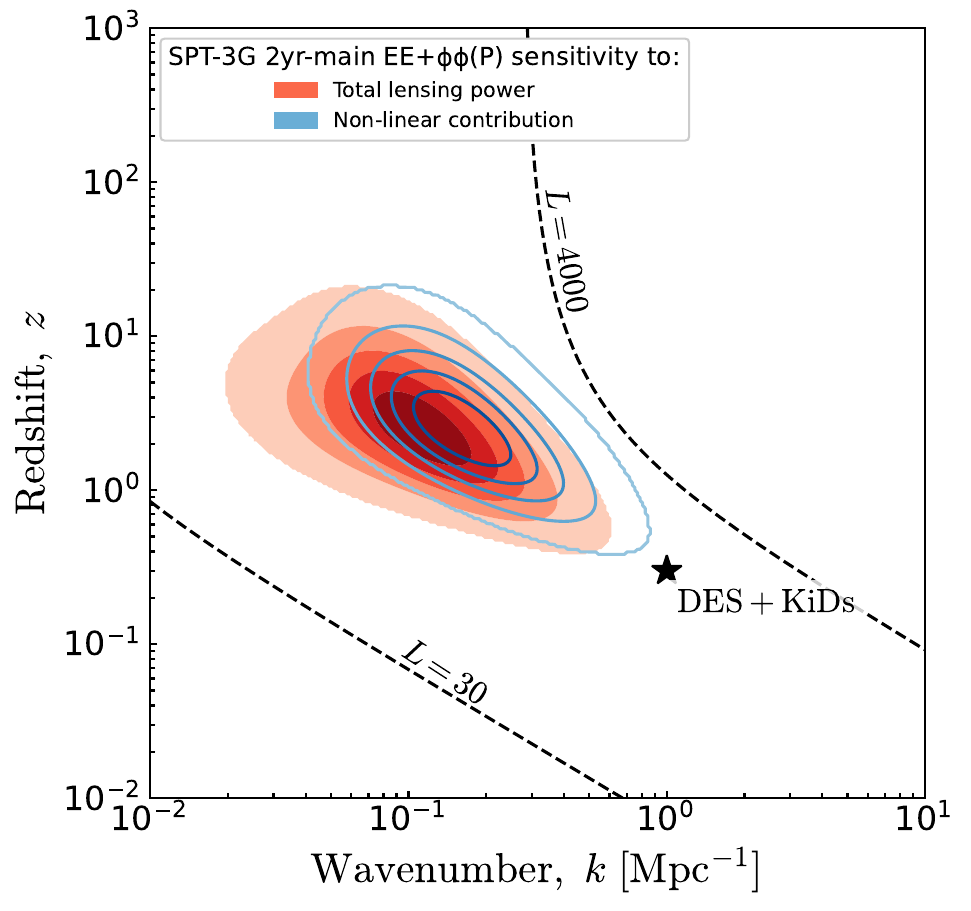}
    \includegraphics[width=0.95\columnwidth]{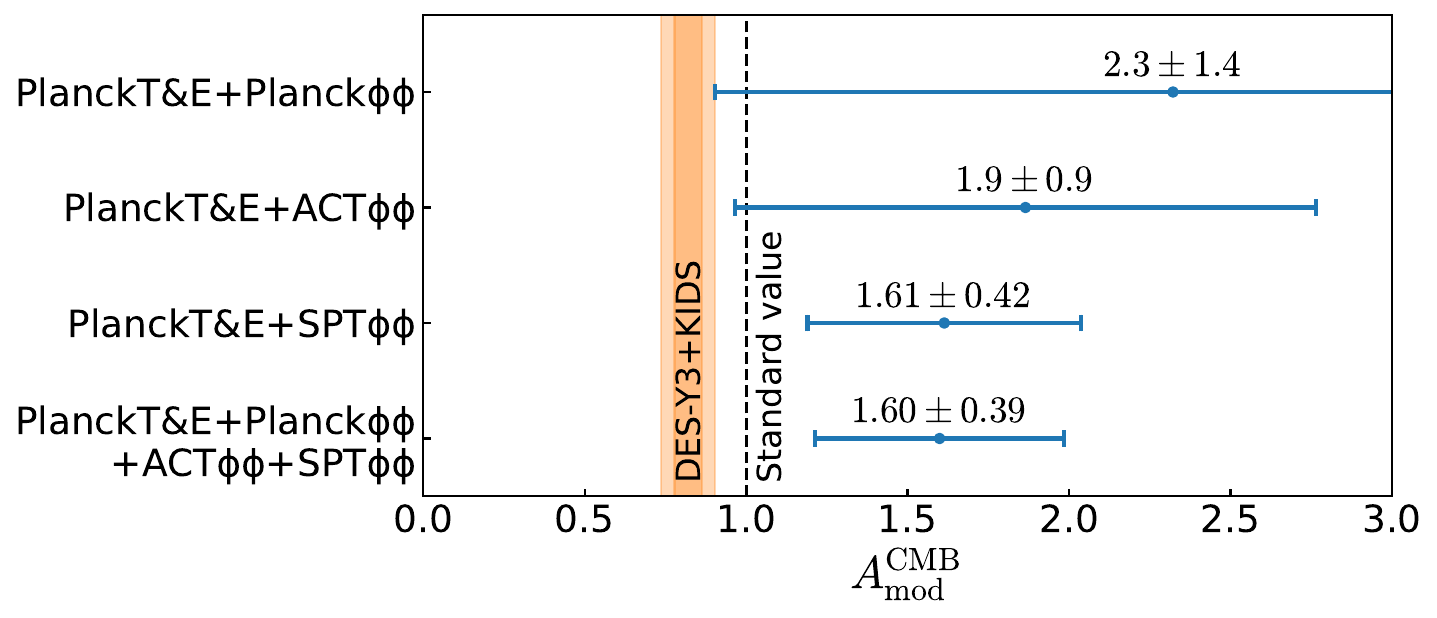}
    \caption{{\it Top panel:} Redshifts and scales to which our lensing measurements are sensitive. Colored regions are iso-contours in the contribution to the square root of the Fisher information from logarithmic bins in $k$ and $z$. The outermost contour corresponds to 10\% of the peak. Red is for the Fisher information in the total lensing power, whereas blue is for the Fisher information just in $A_{\rm mod}^{\rm CMB}$ (and is weighted more towards higher $L$). Under the Limber approximation, a given $(k,z)$ contributes to a single CMB lensing multipole, $L$, and dashed lines denote this mapping for scales of $L\,{=}\,30$ and {$L\,{=}\,4000$}. {\it Bottom panel:} Constraints on $A_\mathrm{mod}^{\rm CMB}$, defined as the amplitude of non-linear contributions to the CMB lensing spectrum (Eqn.~\ref{eq:Amod}), from \planck primary CMB measurements (labeled PlanckT\&E) in combination with lensing measurements from \planck, ACT DR6, and the SPT dataset presented here. The orange band shows the 68\% and 95\% constraints of $A_{\rm mod}$ inferred from DES-Y3 and KiDs datasets in \citep{Preston:2023uup}.}
    \label{fig:A_mod}
\end{figure}

Fig.~\ref{fig:A_mod} shows results of fitting the \lcdm{+}$A_{\rm mod}^{\rm CMB}$ model to various CMB datasets. We take the \planck primary CMB measurements as a baseline, which tightly limit the linear contribution to lensing, then explore the impact to constraints on the non-linear contribution, $A_{\rm mod}^{\rm CMB}$, from adding lensing measurements from \planck, ACT, or the SPT results from this work. As expected from the spectra shown in Fig.~\ref{fig:bandpowers_muse_vs_others}, there are significant improvements in constraining the non-linear contribution from \planck to ACT, and again from ACT to SPT. The \planck primary CMB together with SPT detect a non-zero $A_{\rm mod}^{\rm CMB}$ at ${>}\,3\,\sigma$ for the first time. 

All measurements are consistent with the standard value $A_{\rm mod}^{\rm CMB}\,{=}\,1$, and the tightest combination of all lensing measurements yields $A_{\rm mod}^{\rm CMB}\,{=}\,1.60\,{\pm}\,0.39$. The $A_{\rm mod}$ inferred from DES-Y3 and KiDs is $A_{\rm mod}\,{=}\,0.820\,{\pm}\,0.042$ \citep{Preston:2023uup}. This latter result measures the amplitude of nonlinear structure growth at redshift $z\,{\sim}\,0.3$ and scales of $k\,{\sim}\,1\,{\rm Mpc^{-1}}$, whereas our CMB lensing measurements are sensitive to higher redshifts and larger scales. Using a Fisher calculation based on our actual achieved errorbars, in the top panel of Fig.~\ref{fig:A_mod} we have computed the range of $(k,z)$ which contribute most to our constraints on total lensing power, as well as to just the non-linear contribution to the lensing power, i.e. to $A_{\rm mod}$.

This plot, taken together with our $A_{\rm mod}$ constraints, suggests (at ${\sim}\,2\,\sigma$) that if the $S_8$ tension is to be explained by some non-standard physics suppressing the non-linear matter power spectrum, these effects have not started at the higher redshift and larger scales of CMB lensing as compared to galaxy lensing, and evolve in the decade of separation in $(k,z)$ before suppressing the galaxy lensing power by ${\sim}\,20\%$ at later times and smaller scales. We note that with even better future measurements of small-scale CMB lensing, sensitivity to $A_{\rm mod}^{\rm CMB}$ will push to smaller scales; e.g., significantly tightening up measurement of lensing power in a broad band centered on {$L\,{=}\,4000$} will lead to sensitivity to galaxy weak lensing scales, although at $z \simeq 1$ to 2 rather than $z=0.3$.

{

\subsection{Massive neutrinos}\label{sec:mnu}
The \lcdm model, as usually understood, includes a strong assumption about the spectrum of neutrino masses, namely that the sum is 0.058 eV. This particular value follows from assuming that the lightest neutrino mass is effectively zero, that normal ordering applies, and that the usual interpretation of atmospheric and solar neutrino oscillations is correct. Here we relax this assumption and study constraints on the sum of neutrino masses, $\Sigma m_\nu$. For a review of neutrino physics and its cosmological significance see \cite{Lesgourgues:2013sjj}.

Relative to the case of massless neutrinos, neutrino rest masses increase the mean energy density and therefore the expansion rate. To keep the angular size of the sound horizon fixed, the cosmological constant must then decrease, with a result that the expansion rate is boosted at $z > 1$ and decreased at $z < 1$ \cite{Pan:2015bgi}. The increased expansion rate at $z> 1$ reduces growth on scales below the neutrino free-streaming length. Increasing mass also decreases the neutrino free-streaming length. On scales above the free-streaming length, neutrinos can cluster, approximately canceling the impact on growth of the increased expansion rate. For neutrino masses consistent with the data we consider, the free-streaming lengths are on very large scales so that the clustering of neutrinos only impacts the CMB lensing power spectrum at $L \lesssim 200$ \cite{Loverde:2024nfi}. For this reason, nearly all of the constraining power of CMB data on neutrino masses comes from the changes to the expansion rate, rather than the changed free-streaming scale \cite{Bertolez-Martinez:2024wez}. This is not to say that CMB lensing does not contribute to the constraints, as it is indeed sensitive to the reduction in growth that occurs as a result of the increased expansion rate.

In Table~\ref{tab:lcdm_ext}, we show the constraints on $\Sigma {m_\nu}$ from various datasets. Combining the CMB observations of Planck, SPT, and ACT, we report the following constraint on neutrino mass
\begin{equation}
    \Sigma {m_\nu}\,{<}\,0.20\,\mathrm{eV}\,\mathrm{(95\% \, C.L.)}.
\end{equation}
Additionally including BAO measurements yields
\begin{equation}
    \Sigma {m_\nu}\,{<}\,0.075\,\mathrm{eV}\,\mathrm{(95\% \, C.L.)},
\end{equation}
which puts extreme pressure on the inverted hierarchy minimal mass sum of about 0.098 eV.

These constraints on $\Sigma m_\nu$ should be interpreted with some caution. The need for that caution is well-illustrated by Figure 3 in \citet{Loverde:2024nfi}. Here they show that, due to constraints from primary CMB data on $\theta_{\rm s}$, $\omega_{\rm bc}$, and $r_{\rm d}$ (where $\omega_{\rm bc} = \omega_b + \omega_c$), and constraints from BAO data on $\Omega_{\rm m}$ and $\omega_{\rm m}r_d^2$ (where here $\omega_{\rm m} = \omega_b + \omega_c + \omega_\nu$ and likewise for $\Omega_{\rm m}$), the preference of the combined constraints is for $\Omega_{\rm m}\,{<}\,  \Omega_{\rm bc}$; i.e., for an (unphysical) negative neutrino mass. Further, the Pantheon+SH0ES constraint in this plane is completely at odds with both the CMB and BAO (either DESI or SDSS) constraints. \citet{Loverde:2024nfi} argue that the main cause of DESI BAO data leading to tighter constraints on $\Sigma m_\nu$ is the central value of their constraints in the $\Omega_{\rm m}$ - $\omega_{\rm m}r_d^2$ plane, rather than a decrease in the uncertainty in this plane. 

Similar reason for caution is given by \citet{Craig:2024tky} and \citet{Green:2024xbb}. They point out that there is some weak ($2.5 \sigma$) evidence for an excess in lensing power present in CMB data, relative to \lcdm predictions given CMB and BAO data. This is, again, opposite the expected influence of an increase to $\Sigma m_\nu$ from its minimal (\lcdm) value of 0.058 eV. We turn to this question of excess lensing power in the next subsection.
}
%\end{comment}

%
\subsection{Excess lensing power?}
\label{sec:Excess}

As mentioned, \citet{Craig:2024tky} pointed out that the lensing power inferred from the Planck lensing reconstruction and from the ACT DR6 lensing reconstruction are somewhat in excess of \lcdm predictions given DESI BAO and primary CMB data. Here we investigate this claim, first with $L$-independent lensing template parameters and then with the phenomenological "negative neutrino mass" parameter used in \citep{Craig:2024tky}, $\Sigma \tilde m_\nu$, which can be thought of as an $L$-dependent lensing template parameter.

To conduct our analysis, we define three $L$-independent lensing template parameters: 
\begin{enumerate}
    \item $A_{\rm recon}$ scales the \lcdm model lensing power used to predict the reconstructed lensing power, 
    \item $A_{\rm 2pt}$ scales the model lensing power used to compute any lensed power spectra (2-point correlation functions), and 
    \item $A_{\rm lens}$ scales both of these model power spectra; i.e., it is what we call $A_{\rm recon}$ and $A_{\rm 2pt}$ when we force them to be equal to each other.  
\end{enumerate} 
Note that what we call $A_{\rm 2pt}$ has often been called $A_{\rm L}$. 

We consider four model spaces, each an extension of \lcdm. They are \lcdm{+}$A_{\rm recon}$, \lcdm{+}$A_{\rm 2pt}$, \lcdm{+}$A_{\rm lens}$, and \lcdm{+}$A_{\rm recon}${+}$A_{\rm 2pt}$. Taking $\theta$ to be the \lcdm parameters, the extensions to the model spaces that include these template parameters are done via
\begin{equation}
C_L^{\phi \phi}(A_X,\theta) = A_X C_L^{\phi \phi}(\theta),
\end{equation}
where $X$ = recon, 2pt, or lens and $C_L^{\phi \phi}(\theta)$ on the right hand side is the \lcdm model evaluated at $\theta$.\footnote{Recall that in this paper the \lcdm model has $\Sigma m_\nu$ fixed to 0.058 eV.} To assist in comparison with \cite{Craig:2024tky,Green:2024xbb}, in this subsection only we take "PlanckT\&E" to include the Planck low-$\ell$ $EE$ likelihood and remove the $\tau$ prior, and we restrict our use of ACT data to ACT DR6 lensing only. Our choice of BAO data can be found in Table~\ref{table:dataset}. 

\begin{figure}[tpbh]
    \centering
    \includegraphics[width=\columnwidth]{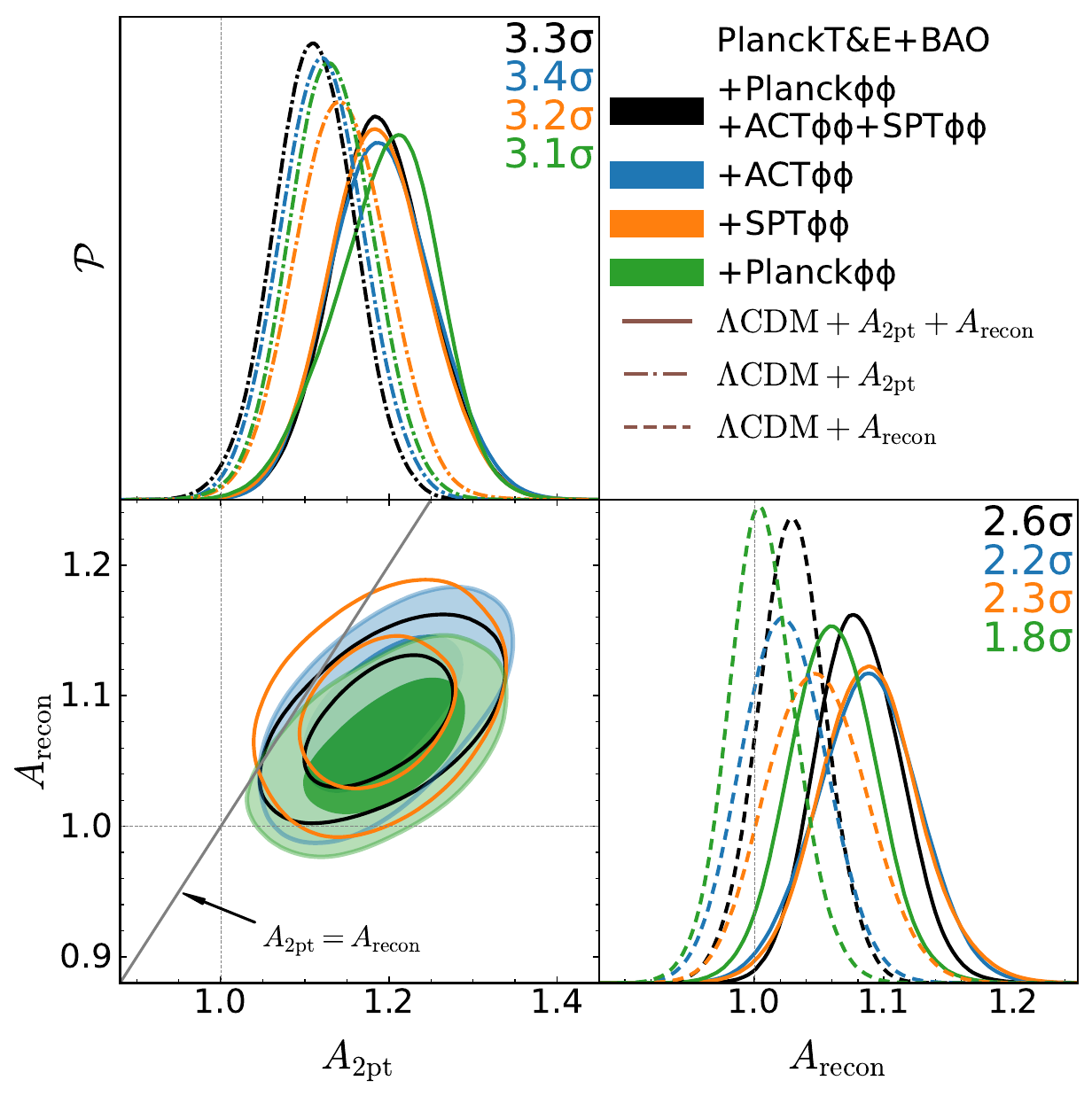}
    \caption{Posteriors of the lensing power phenomenological scaling parameters $A_{\rm 2pt}$ and $A_{\rm recon}$. Different combinations of CMB lensing data added to the Planck primary CMB and BAO are shown in different colors. The 2D contours and solid lines show the marginal posterior from the \lcdm{+}$A_\mathrm{2pt}+A_\mathrm{recon}$ model; the dot-dashed and dashed lines show the results from \lcdm{+}$A_\mathrm{2pt}$ and \lcdm{+}$A_\mathrm{recon}$ respectively. The $\sigma$ values give the level at which $A_{\rm recon}$ ($A_{\rm 2pt}$) values less than 1 are excluded, for the case of marginalization over $A_{\rm 2pt}$ ($A_{\rm recon}$); i.e., for the solid curves.}
    \label{fig:lcdm_A2ptArecon}
\end{figure}

In the dash-dotted curves of the top panel of Fig.~\ref{fig:lcdm_A2ptArecon} we see the well-known "A lens" or "$A_L$" anomaly, which in our language is an $A_{\rm 2pt}$ anomaly driven by Planck PR3 $TT$ data \cite{Planck:2018vyg}. This is a preference, in the lensed $TT$, $TE$, and $EE$ spectra, for excess lensing power beyond what one gets from the \lcdm model. In the lower left panel we can see that if both $A_{\rm recon}$ and $A_{\rm 2pt}$ are allowed to vary freely, the 95\% confidence regions for the four considered data combinations have central values greater than 1 in both dimensions and no overlap with the \lcdm point of $A_{\rm recon}\,{=}\,A_{\rm 2pt}\,{=}\,1$.

Turning to the question of an excess of reconstructed lensing power, we see in the solid lines of the right panel that if we marginalize over $A_{\rm 2pt}$ (solid lines) then there is some preference for $A_{\rm recon}\,{>}\,1$. The significances of these preferences are shown as ranging from 1.8\,$\sigma$ to 2.6\,$\sigma$. From these significance levels and from the blue, orange, and green curves we see that we get a very similar result whether we use Planck, ACT DR6, or SPT lensing reconstructions, with the ACT DR6 and SPT curves nearly overlapping. If we remove DESI BAO data\footnote{When removing DESI we also add back in BAO data we had removed (see Tab.~\ref{table:dataset}), namely the SDSS DR16 LRG, ELG, and QSO BAO \cite{eBOSS:2020yzd} and the DR12 BAO bin with $z_{\rm eff}\,{=}\,0.61$ \citep{BOSS:2016wmc}.} (a case not shown in the figure), the preferences for $A_{\rm recon}\,{>}\,1$ persist, although with somewhat reduced significance ranging from 1.4\,$\sigma$ to 2.2\,$\sigma$. 
The preference for $A_{\rm recon}\,{>}\,1$ is weakened much more if we fix $A_{\rm 2pt}$ to 1 rather than marginalize over it (dashed curves); any evidence for excess reconstructed lensing power is then quite weak. 

This dependence of the $A_{\rm recon}$ posteriors on the treatment of $A_{\rm 2pt}$ (fixing to 1 or marginalizing) makes sense given the correlation we see in the lower left panel. The lower the assumed $A_{\rm 2pt}$, the lower the distribution of $A_{\rm recon}$ values. Physically, this correlation emerges because as $A_{\rm 2pt}$ is increased, the \lcdm parameters used to fit the Planck T\&E data adjust in a way that decreases lensing power (both $\Omega_{\rm m}h^2$ and $A_s$ decrease). With the \lcdm lensing power decreased, a given level of reconstructed CMB lensing power needs a larger $A_{\rm recon}$ to fit it. 

Readers who are familiar with \cite{Green:2024xbb} may be surprised at this sensitivity of $A_{\rm recon}$ to choice of treatment of $A_{\rm 2pt}$ (marginalizing it or fixing it to 1), given arguments made in that paper about the {\it un}importance of the 2-point lensing anomaly to their conclusion of excess lensing power. But these two different results do not contradict each other. \citet{Green:2024xbb} define a slightly $L$-dependent lensing template parameter, $A_L(\Sigma \tilde m_\nu)$, and an $L$-independent template parameter, $B_{\rm lens}$, which are approximately related to our parameters by $A_{\rm recon}\,{=}\,A_L(\Sigma \tilde m_\nu)$ and $A_{\rm 2pt}\,{=}\,A_L(\Sigma \tilde m_\nu)B_{\rm lens}$.\footnote{This is approximate because 1) we have ignored the slight $L$-dependence of $A_L(\Sigma \tilde m_\nu)$ and 2) their template is defined with respect to \lcdm models with $\Sigma m_\nu = 0$ eV.} The resulting correlation between $B_{\rm lens}$ and $\Sigma \tilde m_\nu$ is much smaller than the correlation we see in Fig.~\ref{fig:lcdm_A2ptArecon}, which explains why they see the posterior for $\Sigma \tilde m_\nu$ not change much if they switch from fixing $B_{\rm lens} = 1$ to marginalizing over it.

\begin{figure}[tpbh!]
    \centering
    \includegraphics[width=\columnwidth]{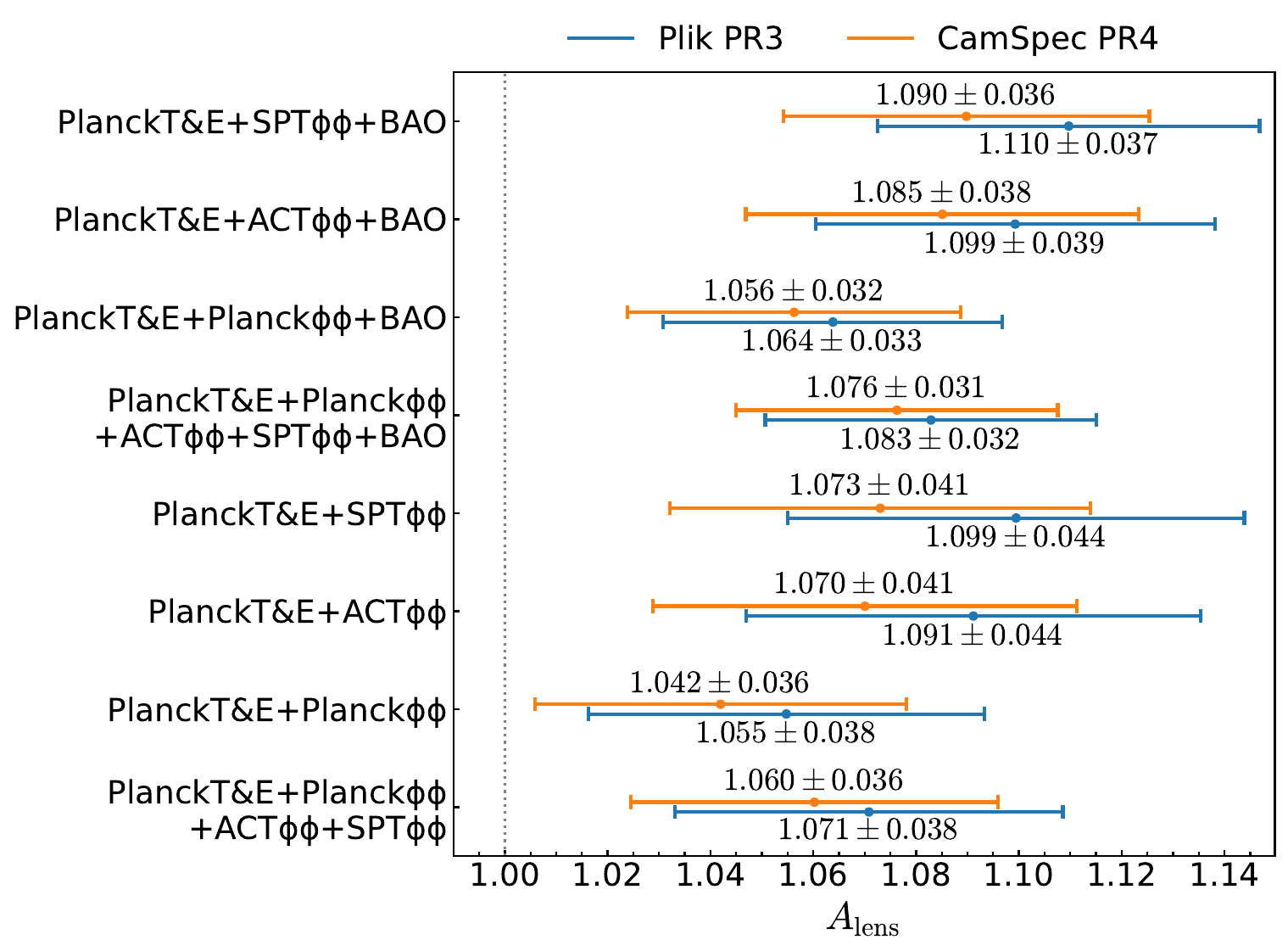}
    \caption{Constraints on the lensing power phenomenological scaling parameter $A_{\rm lens}$ (68\% C.L.) using Planck PR3 \texttt{Plik} T\&E likelihood (in blue) and \texttt{CamSpec} PR4 T\&E likelihood (in orange).}
    \label{fig:lcdm_Alens}
\end{figure}

We have a strong theoretical prior that the lensing power one would infer from lensed CMB spectra (given the right cosmological model) should be equal, within the expected uncertainties, to what one would infer in a CMB lensing reconstruction.\footnote{Although there are models in which this is violated too; see \citet{Craig:2024tky}.} So, even though the bulk of the probability lies to one side of the $A_{\rm recon}\,{=}\,A_{\rm 2pt}$ line, it is worth exploring what we find if we force these two parameters to be equal. This is the model space we explore with an even greater variety of dataset combinations in Fig.~\ref{fig:lcdm_Alens}, and it is also very similar to the model space chiefly studied by \citet{Green:2024xbb} (with their $B_{\rm lens}=1$). 

One of the dataset variations we consider throughout Fig.~\ref{fig:lcdm_Alens} is the replacement of the Planck PR3 Plik T\&E likelihood with the Camspec PR4 T\&E likelihood \cite{Rosenberg:2022sdy}.\footnote{This replacement does not change the low-$\ell$ $TT$ and $EE$ likelihood, which remain PR3.} The Planck PR4 maps use more timestream data than does PR3 and the Camspec PR4 T\&E likelihood uses more sky than does the Plik PR3 T\&E likelihood. If the $A_{\rm 2pt}$ anomaly is driven by some fluctuations that drive the spectra away from the mean (in a manner similar to enhanced lensing) then we would expect the support for it to go down as we add in more data. For Plick PR3 we have, from $TT$,$TE$,$EE$+lowE, $A_L=1.180\,{\pm}\,0.065$ \cite{Planck:2018vyg}. With the additional PR4 data, we instead have $A_L = 1.095\,{\pm}\,0.056$ \cite{Rosenberg:2022sdy}. Note that there is one other PR4 T\&E likelihood \cite{Tristram:2023haj}, in which they find an even lower $A_L = 1.039\,{\pm}\,0.052$. 

The lower half of the entries in Fig.~\ref{fig:lcdm_Alens} do not include any BAO data. We see that without the BAO data any conclusion of $A_{\rm lens}\,{>}\,1$ by more than $2\,\sigma$ is not very robust, occurring only for one of the dataset combinations.  With the addition of BAO data (top half of the figure), such a conclusion becomes much more robust. The combinations that include Planck T\&E, BAO, and all three lensing reconstructions become robust to the difference between PR3 and PR4. With PR3 specifically we find PlanckT\&E+BAO+Planck$\phi\phi$+ACT$\phi\phi$+SPT$\phi\phi$ yields
{
\begin{equation}
    A_{\rm lens} = 1.083\,{\pm}\,0.032
\end{equation}
}
and a $2.7\,\sigma$ exclusion of $A_{\rm lens}\,{<}\,1$. Note that the constraints are weakened by removing DESI from the BAO data set. Also, if one does so, the results become dependent again on the choice of Planck T\&E. Removing DESI (in the same manner as done above) and switching from Planck PR3 T\&E to Planck PR4 T\&E we find {$A_{\rm lens}\,{=}\,1.064\,{\pm}\,0.032$}.

We have checked that switching to the $L$-dependent lensing template used in \cite{Craig:2024tky,Green:2024xbb} in place of $A_{\rm lens}$ delivers a similar result. Their template depends on their neutrino-mass-like parameter $\Sigma \tilde m_\nu$. For PlanckT\&E(PR3)+BAO+Planck$\phi\phi$+ACT$\phi\phi$+SPT$\phi\phi$, we find
\begin{equation}
    \Sigma \tilde{m}_\nu = -0.122\,{\pm}\,0.072\,\mathrm{eV}
\end{equation}
and a $2.7\,\sigma$ exclusion of $\Sigma \tilde m_\nu\,{>}\,0.058\,\mathrm{eV}$.

We note that this evidence for excess lensing power arises from 3 types of constraints: 1) lensing power measurements, 2) inferences of \lcdm parameters (other than $\tau$) from the combination of CMB and BAO data, and 3) inference of $\tau$ from large angular-scale CMB polarization. We have seen that the preference for $A_{\rm lens} > 1$ is robust to selection of different datasets for (1).  Of particular relevance for this paper, the preference is present when the \planck and ACT lensing measurements are replaced with the SPT one. The SPT lensing is inferred entirely from polarization, making it much less susceptible to possible biases from extragalactic foregrounds. Regarding (2), we see that a more than $2\,\sigma$ preference for $A_{\rm lens}\,{>}\,1$ requires either BAO data (including DESI) or PR3 T\&E. The next release of BAO data from DESI should be informative, as will the forthcoming lensed $TT$, $TE$, and $EE$ spectra from SPT-3G. Note that $\tau$ (item 3 above) is
inferred from \planck low-$\ell$ $EE$ data. The importance of $\tau$ comes from its implications for $A_{\rm s}$ as the combination $A_{\rm s} e^{-2\tau}$ is very well determined by CMB data: a higher $\tau$ would lead to higher $A_{\rm s}$ and hence a higher lensing power prediction. Independent and competitive constraints on $\tau$ may come via the kinematic SZ effect \citep{Alvarez:2020gvl,raghunathan2024}.

The conclusion that there is an excess of lensing power also relies on assumption of the \lcdm\ model.  \citet{Craig:2024tky} introduced a number of changes to the model that could possibly restore concordance. \citet{Lynch:2024hzh} noted that non-standard recombination can simultaneously reduce the $H_0$ tension and lead to a higher predicted lensing power. The DESI collaboration points out that with free $w_0$ and $w_a$ the constraints on $\Sigma m_\nu$ are significantly relaxed \citep{DESI:2024mwx}.

\subsection{Other \lcdm extensions}

\begin{table*}[tbph]
    \centering
    \begin{tabular}{c|ccccc}
    \toprule
Ext. Parameter& WMAP+SPT             & Planck                & Planck+SPT           & Planck+ACT+SPT        & Planck+ACT+SPT+BAO\\
\midrule
 $\Omega_k$    & $-0.021^{+0.014}_{-0.011}$   & $-0.0070\,{\pm}\,0.0056$    & $-0.0098\,{\pm}\,0.0052$    & $-0.0076^{+0.0053}_{-0.0047}$      & $0.0014\,{\pm}\,0.0014$    \\
\hline
 $w_0$         & $-1.42^{+0.29}_{-0.47}$    & $-1.48^{+0.20}_{-0.37}$       & $-1.52^{+0.18}_{-0.35}$       & $-1.52^{+0.19}_{-0.35}$       & $-1.075\,{\pm}\,0.046$   \\
\hline
 $w_{0}$       & $-1.10^{+0.50}_{-0.69}$    & $-1.18^{+0.46}_{-0.62}$       & $-1.16^{+0.48}_{-0.60}$       & $-1.18^{+0.48}_{-0.58}$       & $-0.65\,{\pm}\,0.23$ \\
 $w_{a}$       & $-1.1^{+0.6}_{-1.9}$       & $-1.06^{+0.61}_{-1.94}$       & $-1.18^{+0.52}_{-1.82}$       & $-1.21^{+0.57}_{-1.79}$       & $-1.15^{+0.71}_{-0.59}$      \\
\hline
 $N_{\rm eff}$ & $2.95\,{\pm}\,0.33$    & $2.89\,{\pm}\,0.18$      & $2.85\,{\pm}\,0.16$      & $2.66\,{\pm}\,0.14$    & $2.86\,{\pm}\,0.13$\\
\hline
 $\Sigma m_\nu$        & $<0.38\,\mathrm{eV}$       & $<0.31\,\mathrm{eV}$          & $<0.17\,\mathrm{eV}$          & $<0.20\,\mathrm{eV}$           & $<0.075\,\mathrm{eV}$        \\
\hline
 $N_{\rm eff}$ & $2.95\,{\pm}\,0.32$    & $2.89\,{\pm}\,0.19$      & $2.84\,{\pm}\,0.17$      & $2.67\,{\pm}\,0.14$    & $2.83\,{\pm}\,0.13$\\
 $\Sigma m_\nu$& $<0.38\,\mathrm{eV}$     & $<0.31\,\mathrm{eV}$        & $<0.17\,\mathrm{eV}$        & $<0.21\,\mathrm{eV}$        & $<0.061\,\mathrm{eV}$      \\
 \hline
 $N_\mathrm{eff}$    & $2.79\,{\pm}\,0.52$    & $2.8\,{\pm}\,0.3$        & $2.77\,{\pm}\,0.26$      & $2.60\,{\pm}\,0.23$    & $2.89\,{\pm}\,0.23$     \\
 $Y_\mathrm{P}$      & $0.256\,{\pm}\,0.028$  & $0.248\,{\pm}\,0.018$    & $0.250\,{\pm}\,0.016$    & $0.246\,{\pm}\,0.014$  & $0.241\,{\pm}\,0.015$   \\
\bottomrule
\end{tabular}
\caption{Summary of constraints on selected extensions to the \lcdm model. The errors indicate the 68\% confidence region, and the upper limits on the sum of neutrino masses are 95\% upper limits.}
\label{tab:lcdm_ext}
\end{table*}

\begin{figure*}[tbph]
    \centering
    \includegraphics[width=\linewidth]{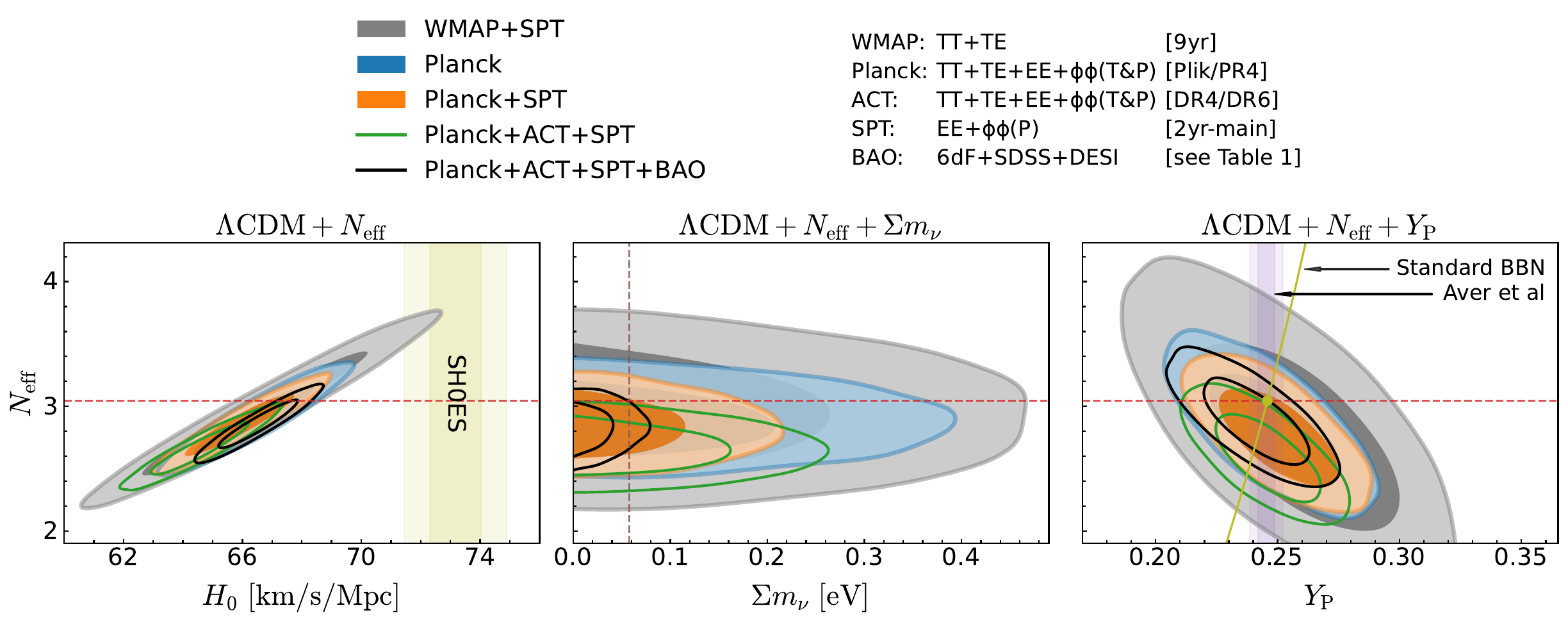}
    \caption{Marginal posterior distributions of the $\Lambda$CDM extension models involving $N_{\rm eff}$. The horizontal red dashed line shows $N_{\rm eff}=3.044$. The yellow vertical band in the left panel shows the constraint on $H_0$ from \citet{Breuval:2024lsv}. The vertical brown dashed line shows $\Sigma m_\nu=0.058\,\mathrm{eV}$. The purple vertical band shows the inference of the primordial helium fraction from \citet{Aver:2020fon}. 
}
    \label{fig:lcdm_ext_all}
\end{figure*}

We also explore \lcdm extensions with curvature, the density of light relics, primordial abundance of helium, and a time-varying dark energy equation of state. Constraints on additional parameters are summarized in Table \ref{tab:lcdm_ext}. 

For the \lcdm extension with free curvature and for the ones with changing dark energy equation of state, 
we find that all CMB datasets are consistent within $2\,\sigma$ with the fiducial values of the \lcdm model.
The constraints from different CMB datasets are consistent. Adding the BAO data significantly tightens the constraints. This is due to the BAO sensitivity to the expansion rate and the distance-redshift relation over redshifts when the dark energy begins to dominate the density.

The energy density of additional relativistic particles other than photons is parameterized by the effective number of neutrino species, $N_{\rm eff}$, defined via 
\begin{equation}
    \rho_{\rm rad} = \rho_\gamma\left[1+N_{\rm eff}\frac{7}{8}\left(\frac{4}{11}\right)^{4/3}\right],
\end{equation}
where $\rho_\gamma$ is the photon energy density. Assuming the standard cosmological model, $N_{\rm eff}=3.044$ \citep{Bennett:2020zkv, Akita:2020szl, Froustey:2020mcq}.
The CMB power spectrum is sensitive to $N_{\rm eff}$ \citep{Bashinsky:2003tk, Baumann:2015rya, Follin:2015hya, Ge:2022qws, Shao:2024mag}. 

By increasing the density of light relics, one increases the expansion rate during the radiation-dominated era. To keep the precisely measured angular scale of the sound horizon unchanged, the expansion rate after last scattering also increases \citep[e.g.][]{Hou:2011ec}. Therefore, there is a positive correlation between $H_0$ and $N_{\rm eff}$. The marginal posterior distribution of $N_{\rm eff}$ and $H_0$ is shown in the left panel of Fig.~\ref{fig:lcdm_ext_all}. Combining \planck, SPT, ACT, and BAO, we get
\begin{alignat}{2}
    &N_{\rm eff} = 2.83\,&&{\pm}\,0.13 \\
    &H_0 = 66.52\,&&{\pm}\,0.87\, {\rm km/s/Mpc}
\end{alignat}
where $N_{\rm eff}$ is consistent with 3.044 from the standard model prediction, but $H_0$ is still in $5.4\,\sigma$ tension with the SH0ES \citep{Breuval:2024lsv} measurement.

One can further extend the model to allow for a free $\Sigma m_\nu$ as well. The marginal posterior of $N_{\rm eff}$ and $\Sigma m_\nu$ is shown in the middle panel of Fig.~\ref{fig:lcdm_ext_all}. There is nearly no degeneracy between $N_{\rm eff}$ and $\Sigma m_\nu$. Adding SPT data to \planck improves the constraints on  $N_{\rm eff}$ slightly and $\Sigma m_\nu$ by about a factor of 2 as discussed in  Sect.~\ref{sec:mnu}. The lensing information in SPT data is useful for constraining the neutrino mass sum, as discussed in Sect.~\ref{sec:mnu}, and the improvement of the $N_{\rm eff}$ constraint is mostly from the more precisely measured $EE$ bandpowers. 

The density of light relics is also partially degenerate with the primordial fraction of baryonic mass in helium, $Y_{\rm P}$. Both $N_{\rm eff}$ and $Y_{\rm P}$ affect the photon diffusion damping of the CMB power spectra at small angular scales. In the right panel of Fig.~\ref{fig:lcdm_ext_all}, we show the marginal posterior of $N_{\rm eff}$ and $Y_{\rm P}$. Compared to the constraints using only \planck data, the uncertainty is reduced by $\sim\,41\%$ with the addition of SPT data.\footnote{ {Although prior work has
shown that de-lensing can improve constraints on $N_{\rm eff}$ and $Y_{\rm P}$ \citep{abazajian2016, green2016}, we suspect that the improvement we see here is largely unrelated to the unlensed nature of our bandpowers; i.e., it would be quite similar even if our bandpowers were lensed bandpowers.}}
%This improvement is primarily due to the addition of high signal-to-noise data of $EE$ and $\phi\phi$. The reduction in uncertainty is larger than the improvement comparing the forecast constraints using delensed over lensed CMB spectra for future CMB observations with lower noise levels than this work \citep{abazajian2016, green2016}. A detailed investigation will need one swap the unlensed EE bandpowers with lensed EE from the same data. We leave this for a future work.} 
When we fit jointly for $N_{\rm eff}$ and $Y_{\rm P}$ using \planck, ACT, SPT, and BAO measurements, we find values of $N_{\rm eff}$ that are consistent with the standard model prediction of 3.044, $N_{\rm eff} - Y_{\rm P}$ contours that are consistent with BBN predictions, and $Y_{\rm P}$ constraints that are consistent with direct measurements of HII regions in metal-poor galaxies \citep{Aver:2020fon}.

\section{Conclusions}
\label{sec:conclusions}

In this paper, we presented an optimal joint inference of the CMB lensing potential power spectrum and the unlensed CMB $EE$ power spectrum, derived entirely from CMB polarization maps. These maps were made from observations of 1500\,deg$^2$ taken during the SPT-3G 2019 and 2020 Austral winter observing seasons. The inference was performed using a Bayesian framework called the Marginal Unbiased Score Expansion (MUSE) method, which also enables propagation of systematic uncertainties into the estimated bandpower uncertainties. The inferred bandpowers have smaller uncertainties than previously published results at $\ell\,{>}\,2000$ for the $EE$ spectrum and at $L\,{>}\,350$ for the lensing spectrum, with a $38\,\sigma$ detection of lensing power.

After checking that the \lcdm model provides a good fit to the SPT data (combined with a prior on $\tau$), we reported constraints on the \lcdm parameters. We found $H_0=66.81\pm 0.81 \ \mathrm{km/s/Mpc}$, a significantly tighter constraint than from prior polarization-only measurements. It is consistent with the Planck $TT/TE/EE$ plus Planck lensing constraints assuming \lcdm, and in 5.4\,$\sigma$ tension with the most precise distance ladder measurement \cite{Breuval:2024lsv}. The SPT constraints on $\Omega_c h^2$ and $\Omega_b h^2$ are also tighter than from prior polarization-only or polarization plus CMB lensing measurements. We also found $S_8 = 0.850\pm 0.017$, consistent with the result from \planck data and with comparable uncertainty.

We estimated \lcdm parameters for several other CMB datasets and checked for consistency between these estimates. These consistency tests are stringent tests of the \lcdm model. All differences investigated were within expectations given the measurement uncertainties and the \lcdm model. The comparison between SPT and Planck provides a particularly interesting test of the \lcdm model, because the SPT constraints are more heavily weighted toward small-scale polarization power, and especially lensing power, than is the case for \planck. 
 
 We estimated \lcdm parameters from SPT data in combination with other CMB datasets. When combined with primary CMB and lensing from \planck and ACT, we find $H_0=67.33\pm0.37 \ \mathrm{km/s/Mpc}$, a 6.2\,$\sigma$ tension with the distance ladder \cite{Breuval:2024lsv},  and $S_8 = 0.8380\,{\pm}\,0.0084$, in $3.3\, \sigma$ tension with $S_8$ from the 3 x 2pt analysis of KiDs-1000 \citep{Heymans:2020gsg} and also from the 3 x 2pt analysis of DES-Y3 \citep{DES:2021wwk}.

Motivated by the work of \citet{Amon:2022azi} and \citet{Preston:2023uup} regarding the $S_8$ tension, we investigated the sensitivity of our lensing measurement to non-linear corrections. By introducing the $A_{\rm mod}^{\rm CMB}$ parameter that scales the amplitude of non-linear corrections to the CMB lensing spectrum, we found $A_{\rm mod}^{\rm CMB}\,{=}\,1.60\,{\pm}\,0.39$. This is the first ${>}\,3\,\sigma$ "detection" of the non-linear correction for CMB lensing, and is ${\sim}\,2\,\sigma$ higher than the value needed for resolving the $S_8$ tension between galaxy lensing and CMB measurements. Our data clearly do not favor any suppression of the non-linear contribution, though the uncertainties are still large. A suppressed $S_8$ could arise at lower redshifts and from slightly smaller scales than the ones we are probing. 

The CMB lensing potential is also sensitive to the neutrino mass sum, $\Sigma m_\nu$. We investigated constraints on $\Sigma m_\nu$ with the SPT lensing spectrum. In the \lcdm{+}$\Sigma m_\nu$ model, we find $\Sigma m_\nu\,{<}\,0.20\,\mathrm{eV \ (95\%\, C.L.)}$ from the joint fit to \planck, ACT, and SPT data. Adding BAO data, the constraint is further tightened to $\Sigma m_\nu\,{<}\,0.075\,\mathrm{eV \ (95\%\, C.L.)}$, placing pressure on the inverted hierarchy minimal mass sum of $0.098\ \mathrm{eV}$.

With the addition of BAO data to the CMB datasets, the strongest challenge to the standard model we see in our results comes from an investigation of lensing power amplitudes. We introduced $A_{\rm 2pt}$ to scale the \lcdm model CMB lensing spectrum used to calculate lensed model $TT/TE/EE$ spectra and $A_{\rm recon}$ to scale that same \lcdm model CMB lensing spectrum, but for comparison to the reconstructed CMB lensing spectrum instead. With both of these parameters free, the \lcdm point in the $A_{\rm recon}, A_{\rm 2pt}$ plane is outside the 95\% confidence region for a variety of dataset combinations. After marginalization over $A_{\rm 2pt}$, Planck lensing, ACT DR6 lensing, and SPT lensing all favor $A_{\rm recon}\,{>}\,1$ at $\sim\,2\,\sigma$. 

We also set $A_{\rm recon}\,{=}\,A_{\rm 2pt}$ and called the single template parameter $A_{\rm lens}$. We investigated the posterior distribution of $A_{\rm lens}$ for a variety of datasets. For our most comprehensive combination of datasets we find $A_{\rm lens}\,{=}\,1.083\,{\pm}\,0.032$ and a $2.7\,\sigma$ exclusion of $A_{\rm lens}\,{<}\,1$.  
Following \citet{Craig:2024tky} and \citet{Green:2024xbb}, we found a similar constraint on their neutrino-mass-like $L$-dependent lensing template parameter, $\Sigma \tilde m_\nu$. For the same dataset we found a $2.7\,\sigma$ exclusion of $\Sigma \tilde m_\nu\,{>}\,0.058$ eV. We also saw that these constraints are weakened by dropping DESI BAO, or by switching from Planck PR3 T\&E data to Planck PR4 T\&E data. Doing both we find $A_{\rm lens}\,{=}\,1.064\,{\pm}\,0.032$.

We also discussed the cosmological inference with SPT CMB lensing potential bandpowers and the unlensed $EE$ bandpowers on other one- and two-parameter extensions of the \lcdm model. These model extensions include varying the amount of spatial curvature, the dark energy equation of state and its evolution, the density of light relics, and the primordial fraction of baryonic mass in helium. The constraints on the extended parameters are summarized in Tab.~\ref{tab:lcdm_ext}. Combining the SPT results with \planck, ACT, and BAO, we saw no significant preference for any of these extensions. 

Although the 2019-2020 SPT-3G CMB polarization data analyzed here primarily constrain $H_0$ and $S_8$, SPT-3G constraints on other \lcdm parameters and extensions will be improved significantly in the near future. The addition of temperature information to both the power spectrum and lensing analyses will significantly improve the constraints on the inferred cosmological parameters, in particular the effective number of neutrino species, $\Sigma m_\nu$, and curvature \citep{SPT-3G:2024qkd}. With polarized beam uncertainties identified as an important contribution to our systematic error budget, we have planned additional deep observations of polarized point sources to better map out the polarized beam response and reduce this source of uncertainty in future analyses. The 1500\,deg$^2$ field has since been observed for three more seasons. The noise level of the full 2019-2023 1500\,deg$^2$ data is expected to be $1.7$ times lower than the 2019-20 observations, and the figure of merit for \lcdm parameters is expected to be 2 times better \citep{SPT-3G:2024qkd}. The addition of ${\sim}\,8500$ deg$^2$ more sky mapped with SPT-3G by the end of 2024 will strengthen constraints even further, eventually surpassing those from \planck \citep{SPT-3G:2024qkd}. The analysis of these data, particularly the ultra-deep 1500\,deg$^2$ data, will benefit even more from optimal analyses such as the MUSE method demonstrated here. This work has demonstrated the strength of joint Bayesian CMB lensing potential bandpower reconstruction and unlensed CMB $EE$ bandpower estimation, in particular its ability to propagate systematic uncertainties, and paves the way for future analyses of SPT-3G observations and similarly deep datasets in the future. 
\vspace{2cm}

%%%%%%%%%%%%%%%%%%%%% Ackn., bib, appendix %%%%%%%%%%%%%%%%%%%%%
\begin{acknowledgements}
 The South Pole Telescope program is supported by the National Science Foundation (NSF) through awards OPP-1852617 and OPP-2332483. Partial support is also provided by the Kavli Institute of Cosmological Physics at the University of Chicago. This research used resources of the National Energy Research Scientific Computing Center (NERSC), a DOE Office of Science User Facility supported by the Office of Science of the U.S. Department of Energy under Contract No. DE-AC02-05CH11231, the Infinity Cluster hosted by Institut d’Astrophysique de Paris and the Peloton Cluster of the High Performance Computing Core Facility at the University of California, Davis. We acknowledge the computing resources provided on Swing and Crossover, high-performance computing clusters operated by the Laboratory Computing Resource Center at Argonne National Laboratory. The UC Davis group acknowledges support from Michael and Ester Vaida. Argonne National Laboratory’s work was supported by the U.S. Department of Energy, Office of High Energy Physics, under contract DE-AC02-06CH11357. The Paris group has received funding from the European Research Council (ERC) under the European Union’s Horizon 2020 research and innovation program (grant agreement No 101001897), and funding from the Centre National d’Etudes Spatiales. The Melbourne authors acknowledge support from the Australian Research Council’s Discovery Project scheme (No. DP210102386). Work at Fermi National Accelerator Laboratory, a DOE-OS, HEP User Facility managed by the Fermi Research Alliance, LLC, was supported under Contract No. DE-AC02-07CH11359. The SLAC group is supported in part by the Department of Energy at SLAC National Accelerator Laboratory, under contract DE-AC02-76SF00515.
 %We also acknowledge support from the Argonne  Center for Nanoscale Materials.

\end{acknowledgements}

\bibliographystyle{prd}
\bibliography{misc,marius,marius2}

\onecolumngrid
\appendix

\section{Modeling details} \label{app:model}

In this appendix, we give full details about our simulation and posterior models, proceeding in the order in which each piece is used to generate a simulation. In cases where the posterior model is different from the simulation model, each is described individually.

\subsection{Lensed CMB}
\label{sec:lensedcmb}
\paragraph*{Simulation model}
In the simulation model, the lensed CMB is computed on the curved sky. Spherical harmonic coefficients for the unlensed CMB and for the lensing potential are sampled from a normal distribution with variance given by the appropriate $C_\ell$ values. We use the \textsc{Ducc} library to perform an inverse spherical harmonic transform of the coefficients to obtain simulated \healpix \footnote{\url{http://healpix.sourceforge.net}} \citep{Gorski:2004by, Zonca2019} $N_{\rm side}\,{=}\,2048$ maps. The lensing operation is computed with \textsc{Lenspyx} \citep{Reinecke:2023gtp}, taking as input the unlensed CMB maps, $f$, and the gravitational lensing potential maps, $\phi$. The lensed map is then projected to the Lambert projection using Non-Uniform FFTs (NFFTs), as described in the next section.

For the MUSE covariance calculation, we also require the Jacobian of the lensed map with respect to bandpower parameters. Since here we use external libraries for SHTs and curved-sky lensing, we must define the Jacobians for these operations for the autodifferentiation (AD) system. The Jacobian of a spherical harmonic transformation (SHT), $f\,{=}\,\mathbb{Y} a$ and the Jacobian with respect to $f$ of the lensing operation, $\tilde f\,{=}\,\mathbb{L}(\phi) f$, are trivially just $\mathbb{Y}$ and $\mathbb{L}$ themselves since they are linear operators. The Jacobian of lensing with respect to $\phi$ is derived from
\begin{align}
    f(x + \nabla (\phi + \epsilon)) = f(x + \nabla \phi) + \nabla f(x + \nabla \phi) \cdot \nabla \epsilon + O(\epsilon^2),
\end{align}
where $x$ is a pixel location and $\epsilon$ is a small perturbation to the lensing potential. The second term demonstrates that the Jacobian is the operator $\nabla f(x + \nabla \phi) \cdot \nabla$, and involves the spatial gradient of a lensed map, which itself is computed with \textsc{Lenspyx}. With these AD rules implemented, our curved-sky lensing model is fully compatible with the MUSE covariance calculation.

\paragraph*{Posterior model}
In the posterior model, we assume the flat-sky approximation, so simulated maps are generated from white-noise maps FFT-convolved with appropriate kernels. We implement lensing with \textsc{LenseFlow} \citep{Millea:2017fyd}, which is GPU-compatible and automatically differentiable up to second-order as needed by the MUSE covariance calculation.

\paragraph*{Bandpower parameters}
Both simulation and posterior models take as input a theory unlensed CMB spectrum and gravitational lensing potential spectrum. We introduce free parameters which we will infer, $A_b^{\rm EE}$ and $A_b^{\phi\phi}$, which control the amplitude of these spectra relative to a fiducial model within each bin. Our $EE$ bandpowers have linearly spaced bins with $\Delta\ell\,{=}\,50$ between 350 and 3500, and our $\phi\phi$ bandpowers have 12 logarithmically spaced bins between 20 and 3000. We take the fiducial model to be the \planck best-fit spectrum, but this introduces no dependence on \planck, rather is just a choice of definition of the bandpower amplitude parameters. Within a bin, the parameters scale the theory spectrum uniformly across $\ell$'s. One can show that with this choice, our inferred $A_b^{\rm EE}$ and $A_b^{\phi\phi}$ parameters correspond to a constraint on an inverse-variance weighted average of the theory spectrum across the bin. Since this inverse-variance weighting arises only implicitly, we also construct explicit bandpower window functions (used in parameter estimation) from an interpolation of the diagonal of the MUSE covariance, $\Sigma_{\rm MUSE}$, to every $\ell$. The tests on simulations in Sec.~\ref{sec:endtoend}, which demonstrate no bias and correct scatter of cosmological parameters derived from simulations, provide verification of this procedure.

\subsection{NFFT-based projections}
Within our model, we perform interpolation between maps in different projections in four different places: 1) as mentioned, in the simulation model, the lensed CMB is projected from \healpix $N_{\rm side}\,{=}\,2048$ to a Lambert pixelization, 2) in both simulation and posterior models, the transfer function model involves projecting from a Lambert to an Equirectangular pixelization, applying some weights, then projecting back to Lambert, 3) when computing the transfer function, a set of end-to-end pipeline simulations are compared to a direct projection of the \healpix $N_{\rm side}\,{=}\,8192$ maps which are input to the end-to-end simulations, and 4) to model deflections due to relativistic aberration.

To ensure easy differentiability, GPU compatibility, and the best possible accuracy, we perform these interpolations using Non-Uniform FFTs (NFFTs). NFFT interpolation implicitly uses information from all pixels to compute the interpolated value at a given location, as opposed to other typically-used interpolations such as bilinear interpolation, which uses only the nearest pixel neighbors. It is exact for a band-limited signal (i.e. when there is no power beyond the Nyquist frequency), and remains very accurate for nearly band-limited signals such as ours, where power beyond the Nyquist frequency is highly suppressed by the beam. 

We use the CPU and GPU compatible NFFT library \texttt{NFFT.jl} \citep{knopp2023nfft}. An interpolation is performed by a pair of forward and adjoint NFFTs, $\mathcal{F}$, with appropriate grid-points, $\gamma$, which describe the relative location of the pixel centers in the map projections being interpolated between,
\begin{align}
    f_1 = \mathcal{F}(\gamma_1) \mathcal{F}(\gamma_2)^\dagger f_2.
\end{align}
The adjoint of the projection operator, which arises in automatic differentiation, is then simply
\begin{align}
    \big(\mathcal{F}(\gamma_1) \mathcal{F}(\gamma_2)^\dagger\big)^\dagger = \mathcal{F}(\gamma_2) \mathcal{F}(\gamma_1)^\dagger.
\end{align}

\begin{figure}
    \centering
    \includegraphics[width=0.4\columnwidth]{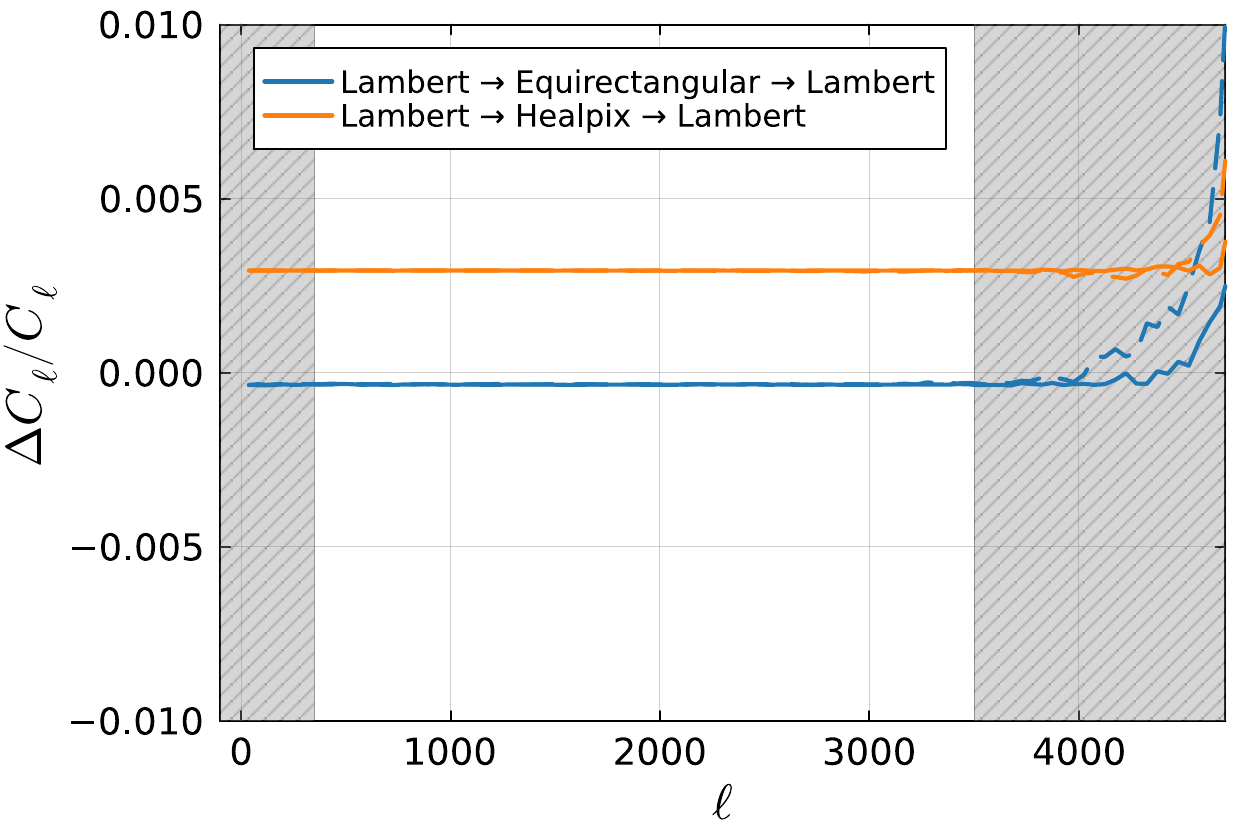}
    \includegraphics[width=0.4\columnwidth]{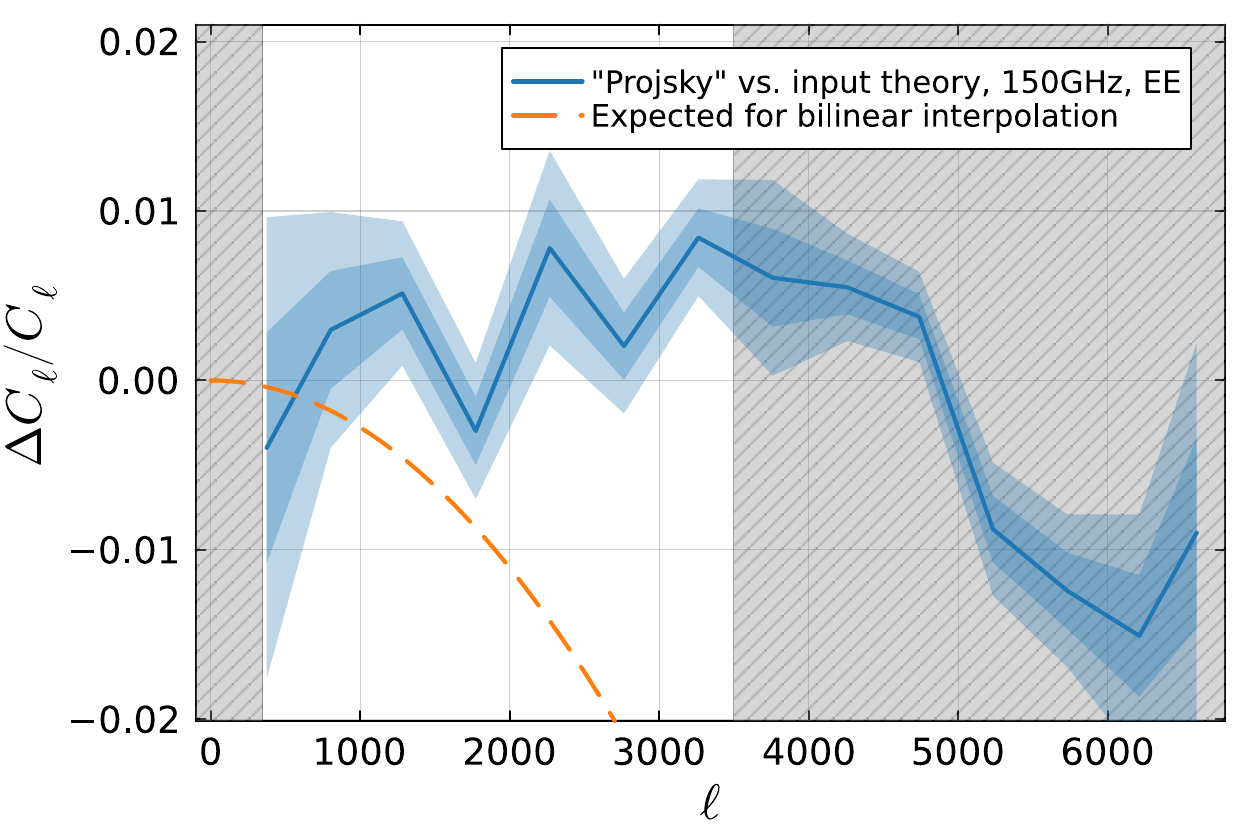}
    \caption{\textit{(Left)} Change to the $EE$ (solid) and $BB$ (dashed) power spectrum of a lensed CMB simulation when using NFFTs to project a pair of Q/U maps to Equirectangular or \healpix $N_{\rm side}=2048$ pixelizations and back. This demonstrates negligible modeling error in our simulation and posterior models due to these projections. \textit{(Right)} Comparison of the pseudo-powerspectra of NFFT Lambert $2.25^\prime$ projections of the \healpix $N_{\rm side}=8192$ maps which are input to mock observation (labeled ``projskies'') vs. the input theory which generated those \healpix maps. Blue bands represent 1 and 2 $\sigma$ Monte Carlo error from the finite number of simulations used to compute the blue curve. This confirms the ``projskies'' sufficiently recover the input theory so as not to bias the transfer function calculation, where they are used. For reference, this plot also shows the expected change to the power spectra if bilinear interpolation was instead used, which shows much worse performance than NFFT interpolation.}
    \label{fig:projection}
\end{figure}

We verify these projections are accurate enough in Fig.~\ref{fig:projection}. For use cases (1) and (2), in the left panel, we generate a simulated lensed CMB map in the Lambert pixelization. We then project to equirectangular and back to Lambert, as well as to \healpix $N_{\rm side}\,{=}\,2048$ and back to Lambert. We then examine the change to the spectrum of the original map. Fig.~\ref{fig:projection} demonstrates this change is ${\lesssim}\,0.3\%$ up to the maximum multipole used in the analysis, $\ell\,{<}\,3500$. For use case (3), in the right panel, we compare the spectrum of the projected \healpix $N_{\rm side}\,{=}\,8192$ maps to the input theory spectrum, again finding ${\lesssim}\,0.5\%$ changes up to the maximum multipole. These differences are well below our uncertainties; for example typical $EE$ bandpower errors are at the 5 to 10\% level. 

\subsection{Relativistic aberration}
\label{sec:aberration}

Relativistic aberration distorts the CMB maps due to our proper motion relative to the CMB rest frame \cite{Jeong:2013sxy}. This is a small but non-negligible effect, which we expect could bias $\theta_s$ by as much as ${\sim}\,0.3\,\sigma_{\theta_s}$ if unaccounted for. One method for modeling aberration is to apply a first-order correction to the power spectrum \cite{Jeong:2013sxy}, 
\begin{align}
    C_\ell \rightarrow C_\ell + C_\ell \frac{d \log C_\ell}{d \log \ell} \beta \langle \cos \psi \rangle,
    \label{eq:aberration}
\end{align}
where $\beta$ is our peculiar Lorentz factor and $\langle \cos \psi \rangle$ measures the average angle to the direction of proper motion for pixels in the field. We note that this correction applies to a lensed spectrum, and does not apply to the unlensed spectra inferred here. In particular, some a priori unknown part of the aberration deflection is interpreted in the posterior as very long wavelength gravitational lensing deflection, and then is automatically removed when producing inferences of the unlensed spectrum. Eqn.~\eqref{eq:aberration} is thus an upper bound on the impact aberration might have on our particular analysis, but it could be much smaller. Without knowing the exact magnitude, we choose to simply include the aberration effect directly in the map-level model. Since the effect is small, it is sufficient to only include it in the simulation model to prevent biases. 

The deflection angle, $\Delta \psi$, is given by 
\begin{align}
    \Delta \psi = \psi - \cos^{-1}\left( \frac{\cos\psi - \beta}{1 - \beta\cos\psi} \right)
\end{align}
where $\psi$ is the angle on the sphere to the direction of proper motion. We assume a fixed velocity of $\beta\,{=}\,0.00128$ and a direction of $(l,b)\,{=}\,(264^\circ, 48^\circ)$ \citep{Planck:2018nkj}, ignoring next-order effects from uncertainties on these values. The deflection is performed via NFFTs. 

We note that we do not model the CMB dipole also induced by our proper motion, as this is filtered out by timestream and map-level high-pass filters. 

\subsection{Beams}
\label{sec:beams}

Next, the CMB maps are convolved with a model for the polarized instrument beams at each frequency, $\mathbb{B}^\nu_P(\beta_n, \beta^\nu_{\rm pol})$. We assume a rotationally symmetric convolution kernel, performed on the Lambert maps with FFTs in both simulation and posterior models. The shape of the convolution kernel in real space, $B(\theta)$, or in multipole space, $B_\ell$, is controlled by free parameters which are marginalized over in the analysis. These parameters include a set of temperature beam eigenmode amplitudes, $\beta_n$, and the polarization fraction of the beam sidelobes at each frequency, $\beta^\nu_{\rm pol}$. We first describe the temperature beams, then how we derive the polarization beams from them. More detail on the beam determination will be presented in Huang et al., (in prep).

\subsubsection{Temperature beams}
The first step in determining our polarized beams is to determine the temperature beams. At low radii (${\lesssim}\,2$\,arcmin), temperature beams are determined from observations of several bright AGN in our observing field. The AGN, however, do not have sufficient brightness to fully map out the beam at larger scales (referred to as the beam ``sidelobes''), where instead we use observations of Saturn. In turn, observations of Saturn cannot be used at small scales, since its intensity saturates detectors and changes the response as compared to typical CMB observations.  Therefore, we stitch the AGN and Saturn maps into a composite beam map, which uses the AGN measurements at low radii, and Saturn for the sidelobes.

To create the composite map, we must account for several systematic effects.
First, Saturn is very slightly extended.  Before stitching, we convolve the AGN maps with a disk of 8.55\,arcsec (the average angular diameter of Saturn during our observations).  We later deconvolve the same disk from the final beam.  The SPT-3G detectors have a finite response time, which spreads signals along the scan direction.  While the response time is constant, the on-sky scan speed varies with declination.  Therefore, we deconvolve the map-space effect of the time constants from each map (both for individual AGN and Saturn), and then reconvolve with the mean effect of the time constant over the SPT-3G winter field footprint.  We also attempt to remove background signals (primarily the CMB) from both the Saturn maps and AGN maps.  Since Saturn moves across the sky, we are able to subtract direct observations of the backgrounds from the Saturn maps.  To do this, we mock observe \planck maps for each of our Saturn observations, and subtract the mock-observed maps from our Saturn maps.  For the AGN maps, we are unable to do this.  However, we are only concerned with small scale ($\lesssim\,7$\,arcmin) information in the AGN maps.  On these scales, the CMB varies very little, and can be modeled with only low-order modes.  Therefore, we fit a simple background model to each AGN consisting of a slope in radius and an offset.  We fit these parameters (using the Saturn maps as a reference for the beam) at 95 and 150 GHz, since the CMB is expected to be common between bands, and the 220 GHz maps are significantly noisier (both from instrumental noise, and higher astrophysical backgrounds).  We subtract the background model from all three bands.

Finally, we apply a very basic model of intra-band frequency dependence to the beam.  We model the beam at each band as a Gaussian made of two components, one which is fixed across the band, and one which scales with frequency.  We model the frequency scaling as $\sigma(\nu) \propto 1 / \nu$, which is the expected behavior for a diffraction-limited beam.  We use the same constant of proportionality across all three bands.  In order to fit these parameters, we assume that the 95 GHz beam is diffraction limited (i.e. the frequency independent component is non-existent).  We approximate the ``effective beam frequency'' for each band using the SPT-3G mean bandpass functions, and the source spectrum.  Then, we fit the three remaining parameters (the constant of proportionality in the frequency dependent portion, and the two frequency independent portions of the 150 and 220 GHz beams) using observations of the brightest AGN.  This simple model allows us to approximate the change in the beam by calculating the effective beam frequency for different source spectra, and convolving the beam maps (or multiplying their spectra) with the appropriate Gaussian.  We apply this model to transform the final beam from the mean AGN spectrum to the CMB spectrum, and to transform the Saturn maps to match the AGN spectrum before stitching.

With the systematic effects accounted for, the stitching process is relatively simple.  For each pair of AGN and Saturn maps, we fit for an amplitude to account for the difference in intensity between the sources.  As discussed above, we cannot use the innermost portion of the Saturn maps.  The AGN maps also become noise-dominated at large distances from the source, so we perform this fit over an annulus with inner radii of (2.25, 1.5, 1.0) arcmin, and outer radii of (3.5, 3.0, 2.5) arcmin at (95, 150, 220) GHz.  Finally, each composite map is constructed by linearly transitioning from the AGN map to the Saturn map over the same annulus.

We make composite maps from all pairs of AGN and Saturn maps.  We also vary several systematic parameters: the choice of stitching radii, uncertainty in the time constants, the choice of the background fitting region.  For each of these variations, we create the full set of composite maps.  Then, we construct cross spectra of composite maps such that none of the component maps are the same.  If we included the same component map more than once, we would incur a noise bias in the beam.  To measure the final beam, we take the mean of these cross spectra, and apply the two final effects described above (specifically, deconvolving the Saturn disk, and converting from the AGN spectrum to the CMB spectrum).

The beam uncertainty is measured from the cross spectra.  By including both different Saturn and AGN maps, as well as the systematic variations, we capture both the map noise, systematic uncertainty from our analysis choices, and systematic differences between the AGN.  Using the Monte-Carlo method, this uncertainty is propagated to a covariance between temperature $B^\nu_\ell$'s at each frequency, $\nu$, and multipole, $\ell$. Finally, we add one additional uncertainty directly to the covariance matrix, which accounts for uncertainty in the detector bandpasses and the AGN spectral index.  This covariance is decomposed into eigenmodes, and in the analysis the amplitude of 5 eigenmodes, $\beta_n$ is free to vary jointly with the bandpower and other systematics parameters. We note that these are eigenmodes of the joint covariance between frequencies and multipoles, so each $\beta_n$ changes the beam simultaneously at each frequency (in different ways at each frequency). The total uncertainty in the determination of the temperature beams is 0.2\% at $\ell\,{=}\,3000$, and is sub-dominant to the uncertainty on polarized beams which we now describe. 

\subsubsection{Polarization beams}

\begin{figure*}
    \centering
    \includegraphics[width=\textwidth]{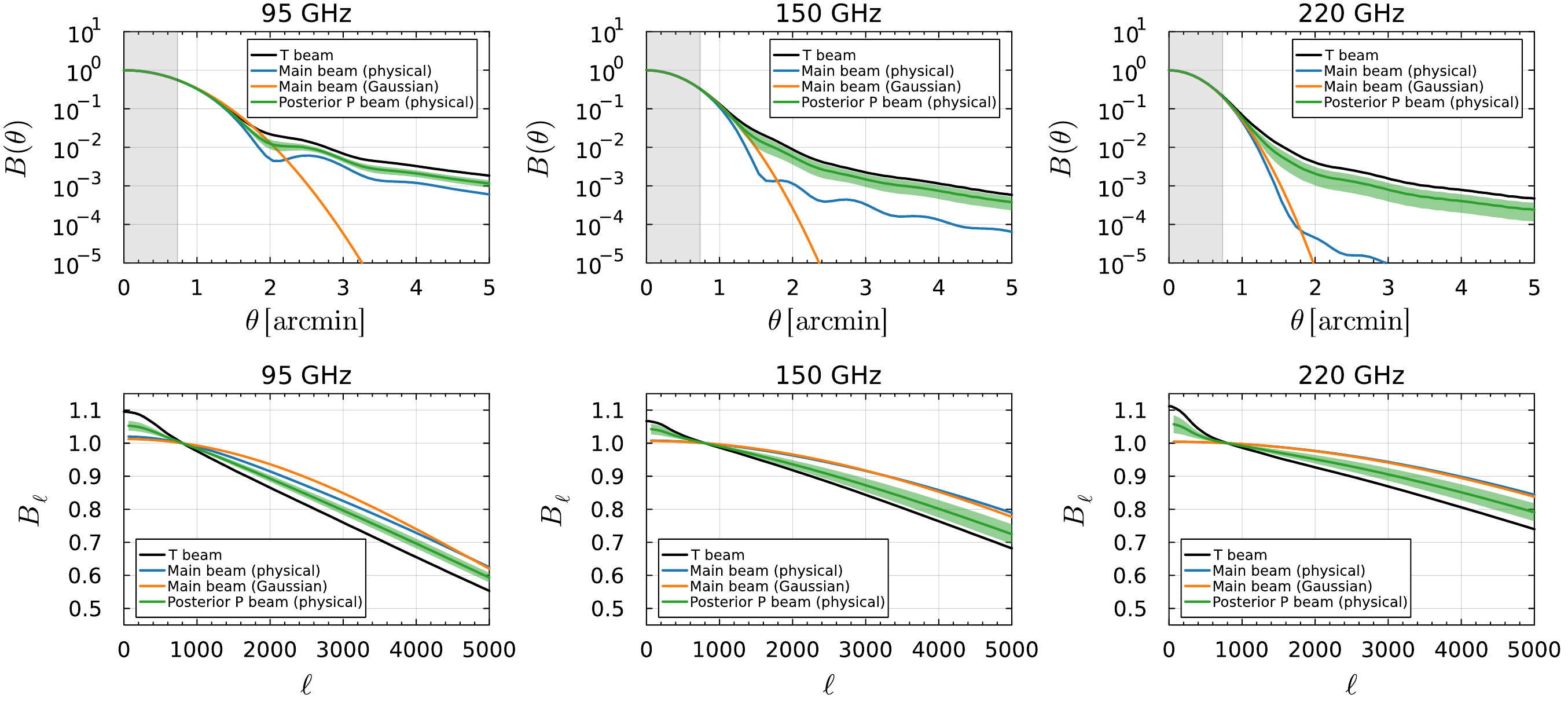}
    \caption{Beams used in this analysis, in real space (top row) and multipole space (bottom row), at each frequency (each column). Black lines show the temperature beams. Blue and orange lines show our best fits to the main beam, where the fitting region is denoted by the gray shaded region. We used the physical main beam for our final result, but we also verified that using the Gaussian main beam leads to no significant shifts in results (see Sect.~\ref{sec:post-unblind}). Green lines and green bands represent the posterior mean and $1\,\sigma$ uncertainty on the polarized beams from the MUSE analysis, which includes marginalization over $\beta_n$, $\beta_{\rm pol}$, and all other systematics and bandpower parameters.}
    \label{fig:beam_fits}
\end{figure*}

While the temperature beams are mapped out to very high signal-to-noise at angular scales relevant to this analysis via observations of bright point sources, we lack sufficiently bright polarized sources to similarly map the sidelobe response of the polarized beams. Prior to unblinding, we had assumed the polarized and temperature beams were identical at the scales used in the analysis. However, the temperature beam has significant diffuse sidelobes due to diffraction and scattering that may not be fully polarized. To account for this, we devised a model of the beams which allows the polarization of the beam sidelobes to vary. This model assumes the  beams are made up of a fully polarized main central beam, and a diffuse beam sidelobe which may be partially unpolarized.

We consider four components in the model for the main beam 1) Gaussian illumination pattern of the primary by the detector lenslets, 2) truncation of that illumination pattern by the Lyot stop, 3) averaging the frequency dependent beam over the detector bandpass weighted by a CMB spectrum, and finally 4) Gaussian broadening of the diffracted beam due to a frequency independent geometric aberration.

We model the electric field illumination pattern on the primary mirror as a truncated Gaussian:
\begin{equation}
  V(r, \nu) = \exp{\left(\frac{-r^2}{2 \left(\sigma_0 \frac{\nu_0}{\nu}\right)^2 }\right)} \Pi(R)
\end{equation}
where $\sigma_0$ is the width of the Gaussian beam from the lenslets at some arbitrary fiducial frequency $\nu_0$, and $\Pi(R)$ is a disk of radius $R$:
\begin{equation}
  \Pi(R) = 
  \begin{cases}
    1 & r \leq R \\
    0 & r > R
  \end{cases}
\end{equation}
For SPT-3G, $R \approx 3.75$ m. The illumination on the sky can then be computed from the Fraunhofer diffraction equation,
\begin{align}
  U(\rho, z, \nu) &= 2 \pi \int_0^\infty V(r, \nu) J_0\left(\frac{2 \pi r \rho}{c / \nu z}\right) r dr
  \\
  U(\theta, \nu) &= 2 \pi \int_0^\infty V(r, \nu) J_0\left(\frac{2 \pi r \theta}{c / \nu}\right) r dr
\end{align}
where $z$ is the distance to the image plane, $\rho$ is the distance from the origin on the image plane, and $J_0$ is the zero-th order Bessel function. In the second line, we have taken the small angle approximation $\rho / z = \theta$ to put this in terms of angle on the sky. The beam on the sky is the intensity which is the square of the electric field pattern, 
\begin{align}
B(\theta, \nu)_{\rm ideal} = U(\theta, \nu)^2
\end{align}

The beam for our wide frequency bands is then integrated over the bandpass and source spectrum. We integrate over the beam intensity, since the electric field phases will decohere at any significant frequency separation. Adjacent frequencies where phase coherence matters have very nearly the same illumination pattern, because both the spectra of our sources and bandpasses vary slowly. Thus we have,
\begin{equation}
  B(\theta)_{\rm ideal} = \frac{\int_{\rm bandpass} I(\nu) t(\nu) B(\theta, \nu) d\nu}{\int_{\rm bandpass} I(\nu) t(\nu) d\nu},
\end{equation}
where $t(\nu)$ is the bandpass function, and $I(\nu) = \frac{dB}{dT}|_{T=T_{\rm CMB}}$ is the observed spectrum.

Finally, we need to account for the effect of geometric aberrations.
This is simply a matter of convolving $B(\theta)_{\rm ideal}$ with a Gaussian, a representation of geometric aberrations, so the final beam is,
\begin{equation}
  B(\theta) = B(\theta)_{\rm ideal} \otimes \exp{\left( -\frac{\theta^2}{2 \sigma_{\rm geom}^2} \right)}
\end{equation}
In principle, we expect $\sigma_{\rm geom}$ to be the same for all three bands.

This model for the main beam has five free parameters, $\sigma_0$, $\sigma_{\rm geom}$, and an overall amplitude at each frequency, $A_\nu$, and predicts the main beam simultaneously in each of our bands. We fit this model to the observed temperature beams as determined in the previous section, limiting to an inner radius of $\theta\,{<}\,0.75\,{\rm arcmin}$ where we observe a good fit, suggesting the main beam dominates and the fit is not biased by the presence of temperature beam sidelobes. For this fit, we construct a $\chi^2$ of the form $\sum_\nu \int{\rm d}\theta \, (B_{\rm model}^\nu(\theta, A, \sigma_0, \sigma_{\rm geom}) - B_{\rm observed}^\nu(\theta))^2$ which ignores uncertainties in the temperature beams, since we expect the systematic uncertainty from our choice of physical model to dominate to the total uncertainties in fit. 

With this fiducial main beam determined, we denote the difference between this and the temperature beam as the sidelobes, and model the polarized beams as the sum of the fixed main beam and the sidelobes, the latter scaled by some unknown scale-independent polarization fractions, $\beta^\nu_{pol}$. That is, the polarized beams used in our analysis are,
\begin{equation}
    \mathbb{B}^{\nu}_{P}(\beta_n, \beta^\nu_{\rm pol}) = \mathbb{B}^\nu_{\rm main} + \beta^\nu_{\rm pol} \left(\mathbb{B}^\nu_{T}(\beta_n) - \mathbb{B}^\nu_{\rm main}\right)
\end{equation}
where the temperature beam at each frequency is
\begin{equation}
    \mathbb{B}^{\nu}_{T}(\beta_n) = \mathbb{B}_0^\nu + \beta_1 \mathbb{B}_1^\nu + \beta_2 \mathbb{B}_2^\nu + ...,
\end{equation}
where $\mathbb{B}_0$ is the best-fit beam and $\mathbb{B}_n$ is the $n$-th eigenmode of the beam covariance. The $\beta_n$ is the amplitude of the n-th eigenmode, which is the free parameter. The $\mathbb{B}_n$ is normalized such that $\beta_i$ has a unit normal prior.

\subsection{Absolute calibration}
\label{sect:calibration}

In the analysis, we apply calibration factors, $A_{\rm cal}^{\nu,i}$, to each frequency map, $\nu$, and subfield, $i$, independently. We assume a Gaussian prior on the $A_{\rm cal}^{150,i}$ based on calibrating our maps to \planck, and leave uniform priors for $A_{\rm cal}^{95,i}$ and $A_{\rm cal}^{220,i}$. App.~\ref{sec:external_sys} describes our choice of prior for the 150 GHz calibration.

\subsection{Global polarization angle}

 To model a systematic error on the global polarization angle calibration, we rotate the global polarization angle of the entire field by an angle, $\psipol^\nu$, at each frequency, $\nu$. The rotation is given by,
\begin{equation}
    \mathbb{R}(\psi) \begin{pmatrix} Q \\ U \end{pmatrix} = \begin{pmatrix}
        \cos(2\psi) & -\sin(2\psi) \\
        \sin(2\psi) & \cos(2\psi)
    \end{pmatrix}
    \begin{pmatrix} Q \\ U \end{pmatrix}.
\end{equation}

\subsection{Transfer function} \label{sec:tf}

\begin{figure}
    \centering
    \includegraphics[width=0.95\columnwidth]{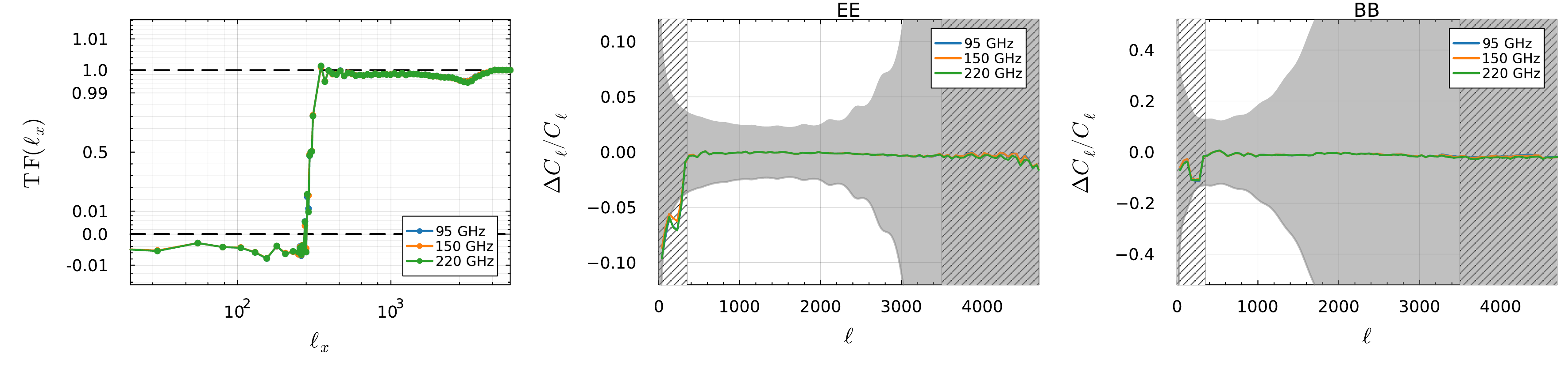}
    \caption{\textit{Left panel:} The filtering applied in the scan direction as part of the transfer function model. This function is fit on simulations at the spline interpolation points denoted with markers. Note the non-linear y-axis scaling. \textit{Right two panels:} Comparison of the $EE$ and $BB$ power spectra of an end-to-end pipeline simulation of the CMB as compared to the model transfer function applied to the same input sky which generated the simulation. Across the multipole range used for the analysis, the model is accurate to ${\lesssim}0.2\%$ in $EE$ and ${\lesssim}2\%$ in BB, both less than 10\% of our uncertainties (denoted as gray bands).}
    \label{fig:transfer_function}
\end{figure}

%Next, we apply the transfer function model, $\mathbb{TF}$. 
The transfer function accounts for the filtering of the timestreams during mapmaking described in Sec.~\ref{sec:data_processing}. In our transfer function model $\mathbb{TF}$, we must approximate the impact of this filtering directly to a Lambert-projected map, since our model is not at the level of timestreams. The non-trivial procedure of sequentially deprojecting templates while masking point sources still corresponds to a linear operator, but not one which is diagonal in any easily accessible basis. 

Our model begins by first reprojecting the map to an Equirectangular projection using NFFTs, such that individual scans are straight lines in the horizontal direction. We then create a single map-level deprojection template by multiplying the point source mask assumed in mapmaking and applying a 1D filtering in the horizontal direction using FFTs where the 1D filter is a free function which we will fit for using simulations. This template is subtracted from the entire map (including the masked regions), before projecting back to Lambert. The operation can be represented as:
\begin{equation}
    \mathbb{TF} = \mathbb{P}^{-1} \cdot \left[\mathbbm{1} - \left(\mathbbm{1}-{\rm TF}(\ell_x)\right) \cdot \mathbb{M}_{\rm ptsrc}\right] \cdot \mathbb{P}
\end{equation}
where $\mathbb{P}$ is the NFFT projection from Lambert to Equirectangular, $\mathbb{M}_{\rm ptsrc}$ is the point source mask and ${\rm TF}(\ell_x)$ is the fitted FFT kernel in the horizontal direction. 

To fit for the appropriate function, ${\rm TF}(\ell_x)$, we compare 1) mock observations of the signal, $m$, which simulate the entire timestream filtering and mapmaking process, and 2) direct projections of the simulated signal maps which were fed into the mock observations, $p$. The relation between these maps are
\begin{equation}
    m = \mathbb{PWF}_{2.25^\prime} \cdot \mathbb{TF} \cdot \mathbb{PWF}_{8192}^2 \cdot p.
\end{equation}
The factor of the \healpix $N_{\rm side}=8192$ pixel window function is due to the interpolation of $p$ onto mock timestreams during the simulation process, and does not arise in the real data. The factor of Lambert $2.25^\prime$ pixel window function, $\mathbb{PWF}_{2.25^\prime}$ arises from the binning of these timestreams into $2.25^\prime$ resolution maps. We then minimize the function $\lVert m_{\rm true} - m_{\rm model} \rVert^2$ by fitting ${\rm TF}(\ell_x)$ at a set of spline points. We find that a single pair of $m$ and $p$ simulations is sufficient to obtain a stable estimate, and have verified the result changes negligibly between a few different simulations.

The left panel of Fig.~\ref{fig:transfer_function} shows the fitted ${\rm TF}(\ell_x)$. As expected, it is roughly zero until around $\ell_x\,{=}\,300$, then transitions to unity at higher multipoles. We choose to allow some small negative values as it gives a better overall fit with no other observed impact. The right two panels of the figure then show the comparison of $EE$ and $BB$ spectra of the mock observations with those computed by applying the best-fit model transfer function to $p$.  Across the multipole range used for the analysis, the model is accurate to ${\lesssim}0.2\%$ in $EE$ and ${\lesssim}2\%$ in BB, both less than 10\% of our uncertainties (which are denoted as gray bands).

\subsection{Temperature-to-polarization leakage}
T-to-P leakage can be caused by a variety of sources.
Gain mismatch between detector pairs leaks a scaled copy of the temperature map into polarization (``monopole'' leakage), whereas differential detector pointing and beam ellipticity introduce copies of the first and second derivatives of the temperature map \citep[``dipole'' and ``quadrupole'' leakage;][]{hu.etal03,henning18}.
We find no evidence of dipole or quadrupole leakage in the SPT-3G data (Figure \ref{fig:dipole_quadrupole_TP}), so we only correct for monopole leakage.

In the MUSE simulation and posterior model, we construct the T-to-P leakage as $t_{\rm Q}^\nu = (T^\nu, 0)$ and $t_{\rm U}^\nu = (0, T^\nu)$, where $T^\nu$ is the temperature map at frequency $\nu$. The temperature maps are made following the same map-making procedure in Sect.~\ref{sec:data_processing}. The amplitudes of the T-to-P templates, $\epsilon_{\rm Q}^\nu$ and $\epsilon_{\rm U}^\nu$, are free parameters. Note that we used independent leakage parameters per frequency per subfield to be consistent with the choice of calibration parameters.

\subsection{Masking}
\label{sec:masking}

We apply a pixel mask ($\mathbb{M}_{\rm pix}$) in the model, which includes a border mask and a point source mask. The border mask is set by zeroing out the areas with weights less than 30\% of the median, padding the border of the mask with zero within the radius of 120\,arcmin and apodizing with a cosine apodization function of 60\,arcmin radius. The point source mask contains emissive sources with brightness greater than 50 mJy and clusters with signal-to-noise ratio greater than 15 at 150GHz.

We also apply a trough mask ($\mathbb{M}_{\rm trough}$) to exclude specific modes from analysis. As discussed in Sect.~\ref{sec:tf}, the pixel mask aliases the power to higher $\ell_x$ modes in Equirectangular projection due to the TOD filtering, and smears out the power when the maps are projected to Lambert. Therefore, we cut out the modes with $\ell_x<400$ in Equirectangular projection to avoid this aliasing power. We also masked out the region that has been contaminated in the observation by a narrow-band signal around $1.1\,\mathrm{Hz}$. 

The mid-pass filter in the Fourier space ($\mathbb{M}_{\rm fourier}$) constrains the modes used in the analysis. We include modes between $350\leq \ell \leq 4000$. 

The total mask used in the analysis is $\mathbb{M} = \mathbb{M}_{\rm fourier} \cdot \mathbb{M}_{\rm trough} \cdot \mathbb{M}_{\rm pix}$, that is, $\mathbb{M}_{\rm pix}$ is applied first. The mask is also applied to the real data map before passed down into the analysis.

\subsection{Monte-carlo correction}

With the aforementioned modeling ingredients specified, we can compare the pseudo power spectrum of the signal part of $d^\nu$ between simulation-model simulations, as in Eqn.~(\ref{eq:sim_data_model}), and mock observations. We find a remaining 0.3\% discrepancy at $\ell\lesssim 1250$ in the $EE$ spectrum between mock observations and MUSE simulations. We believe this is due to residual transfer function modeling errors. Before unblinding, we implement this as an empirical multiplicative Monte Carlo (MC) correction to the $A_b^{\rm EE}$ parameters, with a shape fit from these simulations. We note that this correction is small relative to our uncertainties and corresponds to a 0.01\,$\sigma$ shift in $H_0$ or a 0.07\,$\sigma$ shift in $\Omega_{\rm b}h^2$.

\subsection{Noise modeling}
\paragraph*{Simulation model}
In the simulation model, we use sign-flip noise realizations generated empirically from the real data. This guarantees the noise in the simulation model has the same statistical properties at the true data noise, up to the small correlation between sign-flip realizations, which is not large enough to appreciably impact our final results. 

\paragraph*{Posterior model}
In the posterior model, we must be able to compute the posterior probability function, thus must explicitly specify a random distribution for the noise. We chose a Gaussian model with a covariance derived from the sign-flip noise realizations themselves. We model the noise as diagonal in Fourier space with isotropic noise spectra, 
\begin{equation}
    N_\ell = A (\ell/3000)^\alpha,
\end{equation}
where $A$ and $\alpha$ are parameters to fit to 20 sign-flip noise realizations across the region $500 \leq \ell \leq 3000$, separately at each frequency. We find this gives an acceptable fit, and note that any deviations of this model from true noise distribution will only lead to slight loss of optimality, not any bias. 

\section{Calibration priors and alternative systematics estimates}
\label{sec:external_sys}

Here, we describe the priors on $A_{\rm cal}^{150,i}$ which enter the MUSE analysis, and the alternative estimates of $A_{\rm cal}^{95,i}$, $A_{\rm cal}^{220,i}$, $\epsQ$, $\epsU$, and $\psi_{\rm pol}$ which are compared against the MUSE results.

In practice, calibrating to \planck is performed as two steps. First, we determine a calibration factor for temperature maps in each subfield, $T_{\rm cal}^i$. Then, we determine a correction on top of this for polarization maps, $P_{\rm cal}$, which is assumed the same across all subfields. By our convention, these factors multiply the SPT data, rather than multiplying the theory model like the MUSE definition of $A_{\rm cal}^{\nu,i}$. The relation is thus
\begin{equation}
    A_{\rm cal}^{150,i} = 1/(T_{\rm cal}^{150,i} P_{\rm cal}^{150}),
\end{equation}

The temperature calibration is determined by computing the cross-spectrum of the full-depth 150\,GHz SPT-3G data with the full-depth 143\,GHz \textit{Planck} map passed through the SPT-3G observing pipeline normalized by the cross-spectrum of two half-depth SPT-3G maps,
\begin{equation}
    T_{{\rm cal, \,\rm external}}^{150}  = \frac{{\rm SPT}^{150}_{\rm full} \times {\rm \textit{Planck}}^{143}_{\rm full}}{{\rm SPT}^{150}_{\rm half1} \times {\rm SPT}^{150}_{\rm half2}}.
\end{equation}\label{eq:acal_150}
The uncertainty on $T_{\rm cal, \, external}$ is computed from the scatter in the quantity when it is computed \textcolor{red}{using 100} different subsets of SPT observations entering the SPT maps. After unblinding, we also realized we had neglected to fold in a $0.25\%$ uncertainty \citep{planckcollaboration.etal19v} on the \textit{Planck} absolute calibration \citep{Planck:2018lkk}, which leads so insignificant changes, but is fixed in our final results. 

Although not used in MUSE, we also obtain temperature calibrations of 95 and 220GHz SPT-3G maps by cross-correlating against the 150GHz map (this yields tighter constraints than calibrating these against \planck). The calibrations are
\begin{align}
    T_{{\rm cal, \,\rm internal}}^{95}  & = 
        \frac{{\rm SPT}^{150}_{\rm half1} \times {\rm SPT}^{150}_{\rm half2}}{{\rm SPT}^{95}_{\rm half1} \times {\rm SPT}^{150}_{\rm half2}} \ T_{{\rm cal, \,\rm external}}^{150}, \notag \\[0.5em]
    T_{{\rm cal, \,\rm internal}}^{220}  & = 
        \frac{{\rm SPT}^{150}_{\rm half1} \times {\rm SPT}^{150}_{\rm half2}}{{\rm SPT}^{220}_{\rm half1} \times {\rm SPT}^{150}_{\rm half2}} \ T_{{\rm cal, \,\rm external}}^{150}.
    \label{eq:tcal}
\end{align}

We recalibrate the $Q$ and $U$ maps by this factor, then correct them for temperature-to-polarization leakage and polarization angle calibration as described below. Then, we determine $P_{\rm cal}^{150}$ by cross-correlating the SPT 150\,GHz polarization maps to \planck 143\,GHz polarization maps in a manner analogous to the absolute temperature calibration in Eqn.~\eqref{eq:tcal}. The estimated $P_{\rm cal}^{150}$ is then corrected by an additional factor of $\sqrt{0.966}$ to match the final recalibration factor applied to the 143\,GHz \textit{Planck} polarization bandpowers \citep[Eqn.~45,][]{planckcollaboration.etal19v}, and additionally has a $0.5\%$ uncertainty folded in corresponding to the uncertainty on the \textit{Planck} polarization calibration. Similarly used only for comparison, we then determine $P_{\rm cal}^{95}$ and $P_{\rm cal}^{220}$ by internally cross-correlating the SPT maps. 

Next, we describe the alternative estimate of temperature-to-polarization leakage and global polarization angle correction which is used above in determining the calibration factors and is compared against the MUSE results in Fig.~\ref{fig:systematics_comparison_pipelines}. The main difference from the MUSE estimate is that these are performed at the power spectrum rather than at the map level, and are done with an iterative procedure which assumes a theory cosmological model rather than simultaneously and jointly with all systematics and bandpower parameters.

Independent estimates of the monopole leakage coefficients, $\epsilon^{Q,TT}$,$\epsilon^{Q,TQ}$,$\epsilon^{U,TT}$, are formed by fitting a template to the $C_\ell^{TQ}$ and $C_\ell^{TU}$ cross-spectra using cross-spectra of the half-depth data maps and cross-spectra from simulations that have no monopole leakage: 
\begin{align}
    C_{\ell, \rm template}^{TQ} & = \epsilon^{Q,TT} C_{\ell, \rm data}^{TT} + \epsilon^{Q,TQ} C_{\ell, \rm sim}^{TQ}, \\
    C_{\ell, \rm template}^{TU} & = \epsilon^{U,TT} C_{\ell, \rm data}^{TT} + \epsilon^{U,TU} C_{\ell, \rm sim}^{TU}.
\end{align}
In practice, $C_{\ell, \rm sim}^{TU}$ scatters around null, and therefore $\epsilon^{U,TU}$ can take arbitrary values. For this reason, the second term is assumed to be 0 and only fits to $\epsilon^{Q,TT}$,$\epsilon^{Q,TQ}$,$\epsilon^{U,TT}$ are carried out \citep{SPT-3G:2021eoc}.

The observed (and possibly rotated) $Q$ and $U$ are related to the primordial \textcolor{red}{(and lensed)} $\tilde{Q}$ and $\tilde{U}$ by:
\begin{align}
    \qty (Q + iU) & = (\tilde{Q} + i\tilde{U}) e^{i2\psi_{\rm pol}},
    \intertext{which introduces artificial $TB$ and $EB$ correlations:}
    C_\ell^{TB}   & = - \sin(2\psi_{\rm pol}) \tilde{C_\ell}^{TE}                                          \\
    C_\ell^{EB}   & = \frac{1}{2} \sin \qty(4 \psi_{\rm pol}) (\tilde{C_\ell}^{BB} - \tilde{C_\ell}^{EE}),
\end{align}
where $\psi_{\rm pol}$ is an rotation angle between 0 and $2\pi$.

A joint fit to the T-to-P and polarization rotation parameters is performed by generating 100 ($\epsilon^{Q,TT}$,  $\epsilon^{Q, TQ}$, $\epsilon^{U, TT}$, $\psi_{\rm pol}$) Latin hypercube training points, which are applied to the data maps (first T-to-P deprojection, then polarization rotation), before computing the resulting $TB$ and $EB$ data spectra.
An emulator is then trained using the \texttt{CosmoPower} \citep{mancini.etal21} package, 
which allows us to predict $C_{\ell,{\rm template}}^{TQ}$, $C_{\ell,{\rm template}}^{TU}, C_{\ell}^{TB}$, and $C_{\ell}^{EB}$ at arbitrary points in the full parameter space. An MCMC chain is then run to find the best-fit parameters that produces $C_{\ell}^{TQ}$ and $C_{\ell}^{TU}$ spectra that match with data as well as $\psi_{\rm pol}$ that results in the $C_{\ell}^{TE}$ and $C_{\ell}^{EB}$ that is consistent with null.

\section{Covariance shrinkage} 

\label{app:shrinkage}

We develop and use a new procedure to reduce Monte Carlo error in the MUSE covariance. As a reminder, the covariance is given by
\begin{align}
    \Sigma = H^{-1} J \left(H^{-1}\right)^\dagger = H^{-1} \langle s \, s^\dagger \rangle \left(H^{-1}\right)^\dagger,
\end{align}
where $s \equiv \smap - \langle \smap \rangle$ [see also Eqns.~(\ref{eq:muse_cov}-\ref{eq:H})]. The $H$ matrix originates from second order derivatives of the posterior function and hence contains very little MC error. The $J$ matrix, however, is a pure covariance of some finite set of samples of $s$ and can contain significant MC error. 

Under a singular value decomposition of $H = U S V^\dagger$, the covariance, $\Sigma$, can be rewritten as, 
\begin{align}
     V S^{-1} U^\dagger \big \langle s \, s^\dagger \rangle U S^{-1} V^\dagger = V S^{-1/2}  \Big\langle (S^{-1/2} U^\dagger s) \, (s^\dagger U S^{-1/2}) \Big\rangle S^{-1/2} V^\dagger
     \label{eq:cov_svd}
\end{align}
Note that in the case of a Gaussian latent space, $H\,{=}\,J$, hence the quantity in angle brackets in Eqn.~\eqref{eq:cov_svd} is the identity matrix. Even for the non-Gaussian latent space of the lensing problem, we have verified on simplified cases with thousands of samples of $s$, and for our exact problem with hundreds of samples, that this matrix remains extremely close to diagonal. This motivates a method of reducing MC error by applying a shrinkage estimator to compute the covariance in angle brackets in Eqn.~\eqref{eq:cov_svd}. Specifically, we shrink towards a diagonal target covariance with arbitrary diagonal entries. We find that this drastically reduces MC error of the covariance, and importantly, of its inverse, which is much more prone to MC error. Note that naively applying a diagonal shrinkage estimator to $J\,{=}\,\langle s \, s^\dagger \rangle$ directly would not work, since nothing guarantees the $J$ matrix is close to diagonal nor would there be any {\it a priori} known shrinkage target. Instead, we are essentially applying the shrinkage estimator to a preconditioned version of $J$,  where the preconditioning is based on $H$, which itself contains almost no MC error and provides excellent preconditioning.

\end{document}